\documentclass[pre,onecolumn,superscriptaddress,longbibliography,nofootinbib]{revtex4-1}
\usepackage{amsmath,amssymb,graphicx,cases}
\usepackage{epsfig}
\usepackage{textcomp}
\usepackage{epstopdf}
\usepackage{float}
\usepackage{braket}
\usepackage[normalem]{ulem}
\usepackage{enumerate}
\usepackage{enumitem}

\usepackage{ae,aecompl}
\usepackage{lmodern}

\usepackage{xcolor,cancel}
\definecolor{mygreen}{rgb}{0.0,0.55,0.4}

\newcommand{\stkout}[1]{\ifmmode\text{\sout{\ensuremath{#1}}}\else\sout{#1}\fi}

\usepackage[colorlinks=true, urlcolor=blue, anchorcolor=blue, citecolor=blue,filecolor=blue,linkcolor=blue,menucolor=blue]{hyperref}

\newcommand{\aalph}{\alpha} 

\newif\ifimportant
\importantfalse

\begin{document}
	\title{Entropy production and its large deviations in an active lattice gas}
\author{Tal Agranov}
\affiliation{DAMTP, Centre for Mathematical Sciences, University of Cambridge,
Wilberforce Road, Cambridge CB3 0WA, United Kingdom}
\author{Michael E. Cates}
\affiliation{DAMTP, Centre for Mathematical Sciences, University of Cambridge,
Wilberforce Road, Cambridge CB3 0WA, United Kingdom}
\author{Robert L. Jack}
\affiliation{DAMTP, Centre for Mathematical Sciences, University of Cambridge,
Wilberforce Road, Cambridge CB3 0WA, United Kingdom}
\affiliation{Yusuf Hamied Department of Chemistry, University of Cambridge, Lensfield
Road, Cambridge CB2 1EW, United Kingdom}

	\begin{abstract}
	Active systems are characterized by a continuous production of entropy at steady state. We study the statistics of entropy production within a lattice-based model of interacting active particles that is capable of motility-induced phase separation. Exploiting a recent formulation of the exact fluctuating hydrodynamics for this model, we provide analytical results for its entropy production statistics in both typical and atypical (biased) regimes. This complements previous studies of the large deviation statistics of entropy production in off-lattice active particle models that could only be addressed numerically. Our analysis uncovers an unexpectedly intricate phase diagram, with five different phases arising (under bias) within the parameter regime where the unbiased system is in its homogeneous state. Notably, we find the concurrence of first order and second order nonequilibrium phase transition curves at a bias-induced tricritical point, a feature not yet reported in previous studies of active systems.
	\end{abstract}
	
	\maketitle
	\section{Introduction}\label{intro}
The hallmark of active matter systems is the presence of a non-equilibrium drive that acts at the single-particle level \cite{marchetti_hydrodynamics_2013,bechinger_active_2016,fodor_statistical_2018}. This active drive continuously dissipates heat into the environment, so that active systems show a nonvanishing entropy production rate even at steady state. Notably, these far-from-equilibrium steady states display a host of nontrivial collective behavior that is not permitted in equilibrium systems. One of the most striking and well studied of those is Motility Induced Phase Separation (MIPS), where a system composed of self-propelled particles phase separates into dense and dilute phases even when the inter-particle interactions are purely repulsive \cite{cates_motility-induced_2015}. 

It is  now established that the entropy production rate (EPR) is a key observable in the understanding of such collective phenomena and it has been the subject of many recent studies \cite{caprini_entropy_2019,dabelow_irreversibility_2019,ganguly_stochastic_2013,shankar_hidden_2018,fodor_how_2016,mandal_entropy_2017,li_steady_2021,caballero_stealth_2020,fodor_irreversibility_2022,obyrne_time_2022,nardini_entropy_2017,borthne_time-reversal_2020,nardini_entropy_2017,solon_generalized_2018}. This is partly because the EPR quantifies the actual level of irreversibility in a system once the local rules for activity are combined with interactions between particles. For instance if activity is modelled by propulsive forces, but particles are jammed and unable to move, the active forces do no work and the EPR vanishes. More generally it has proven useful \cite{obyrne_time_2022,fodor_irreversibility_2022} to define the informatic entropy production rate (IEPR), defined by the log probability ratio of forward and time reversed dynamics. In most settings, the dynamics of interest is coarse-grained so that the IEPR quantifies the amount of irreversibility that is detectable at coarse-grained level. Note that the IEPR is bounded above by the microscopic EPR, with equality assured only if one avoids coarse-graining entirely, and keeps track of all microscopic variables that are present. 

The IEPR enables us to quantitatively distinguish the emergent dynamics of active systems from their passive counterparts. Its relation to collective phenomena in active matter is two-fold. Firstly, measuring the IEPR was found useful in identifying where activity comes to play in the collective behavior of active systems, such as MIPS \cite{fodor_irreversibility_2022}. 
Secondly, a series of recent studies found that tuning the value of IEPR in active systems, by applying a dynamical bias, can cause  phase transitions, leading to novel states with a rich and interesting phenomenology, which may suggest new design and control strategies for active matter \cite{cagnetta_large_2017,whitelam_phase_2018,tociu_how_2019,gradenigo_first-order_2019,nemoto_optimizing_2019-1,cagnetta_efficiency_2020,chiarantoni_work_2020,fodor_dissipation_2020,yan_learning_2022,grandpre_entropy_2021,keta_collective_2021}. Such a bias allows one to explore sub-ensembles of dynamical trajectories for which the IEPR is larger or smaller than usual. This exploration is achieved by analysing the rate function for large deviations of the IEPR; this can show singularities that are identified as non-equilibrium phase transitions. The same biasing strategy can of course be used to look at large deviations of other physical quantities such as currents. (This is done for the specific model explored below in \cite{agranov_exact_2021} and \cite{agranov_macroscopic_2022}.)

Our present understanding of the nonequilibrium phase transitions that arise on biasing IEPR is very far from complete. So far, these transitions of biased active systems were uncovered primarily within numerical studies; there has been no comprehensive analytical account. Numerically it is hard to resolve the details (for instance the order of particular transitions) even for simple models, and a complete mapping of the phase diagram is typically not possible. 
Indeed, numerically studying large systems under a macroscopic bias, especially near phase transitions, is nontrivial and requires advanced sampling techniques \cite{yan_learning_2022}. 
The general question of what to expect for such a phase diagram therefore remained unanswered. Furthermore, the relation of the phases and transitions of the biased system to those appearing in the unbiased case such as MIPS, is still largely unknown. 

For this reason, a fully-worked out analysis of large IEPR deviations for a nontrivial model of active many-body dynamics is likely to be particularly valuable. However, in almost all such models there is little prospect of analytic progress: particles are strongly interacting and the system is far from equilibrium, which hinders analytical characterisation of fluctuations. Up until now, analytical progress was limited to either phenomenological coarse-grained theories, or field-theoretical descriptions that employ mean-field approximations and gradient expansions, see for example \cite{peshkov_nonlinear_2012,bertin_mesoscopic_2013,wittkowski_scalar_2014,nardini_entropy_2017,solon_generalized_2018-1,solon_generalized_2018,tjhung_cluster_2018}.
While these approaches proved most successful in accounting for the typical behavior, they leave the full statistical description, and in particular, the account of large macroscopic deviations which are at the center of the present work, mostly unattended. 

Another limitation of those approaches is that they do not admit a direct comparison of the IEPR derived at the coarse grained or hydrodynamic scale with the full EPR of the microscopic dynamics defined at particle level. To do this one needs both quantities to be calculable within the same model. 
An exact comparison of this type was so far limited to the case of noninteracting particles only \cite{pietzonka_entropy_2018}.

In this paper we provide the first analytically exact account of the IEPR distribution in a well-chosen model of many interacting active particles of the type capable of undergoing MIPS. We start by deriving exact expressions for both the coarse-grained IEPR and the microscopic EPR evaluated over an arbitrary system path. This offers the first exact comparison of the two in an interacting active system. For reasons we explain, they are found to coincide, creating an unexpectedly close link between the macroscopic IEPR and the microscopic active work dissipated in this particular model \cite{seifert_stochastic_2012,pietzonka_entropy_2018}.

Next we turn to study the probability distribution of the IEPR at large times, which obeys a large deviation principle and is described by a suitable rate function. We provide a complete mapping of the dynamical phase diagram which includes several previously reported phases \cite{cagnetta_large_2017,whitelam_phase_2018,nemoto_optimizing_2019-1,chiarantoni_work_2020,fodor_dissipation_2020,yan_learning_2022,grandpre_entropy_2021,keta_collective_2021}.   In particular, our analysis distinguishes two kinds of stationary inhomogeneous states: phase separated states with sharp interfaces, and states with smoothly modulated density.  Arguments have been presented for both such states~\cite{nemoto_optimizing_2019-1,fodor_irreversibility_2022}, but previous analyses were not able to distinguish them.  Our analytically exact framework allows us to fully work out the intricate details of the transitions to these phases,  where we find a concurrence of first and second order phase transition lines that meet at a nonequilibrium
tricritical point. Moreover, we uncover further novel phases that were not reported before. We argue that these various phases should be generic in that they should also arise if the statistics are biased with other observables of interest, provided that these observables share the same symmetries as those of the IEPR. We also establish the relation between the MIPS criticality in the unbiased system and the transitions of the biased one. As was argued before \cite{nemoto_optimizing_2019-1,fodor_irreversibility_2022}, when the system approaches MIPS criticality, the onset of the transition to the smoothly modulated phase happens at vanishingly small IEPR bias, so that the bias-induced and motility-induced instabilities merge. A similar phenomena was reported in the study of current fluctuations in \cite{agranov_macroscopic_2022}.
 
Our analytical breakthrough is possible thanks to recent progress in establishing the exact fluctuating hydrodynamics of a lattice-based model for repulsively interacting active particles in one dimension. For this active lattice gas model, the noiseless coarse-grained hydrodynamics was established first in~\cite{kourbane-houssene_exact_2018-1}. Later, in \cite{agranov_exact_2021} and \cite{agranov_macroscopic_2022} this was complemented with an exact large deviation framework for macroscopic fluctuations.  (We rely heavily on this framework below.) The active lattice gas model shares many important features, including MIPS \cite{cates_motility-induced_2015}, with previously studied active particle models that are off-lattice and/or in higher dimensions. However, one crucially important difference is that the microscopic rates are scaled with system size. Consequently, in contrast to other standard models of active particles, the particle magnetization density (the excess of right- over left-moving particles) enters as a slow hydrodynamic field (a different scaling would remove the slow magnetization). This slow field plays a major role in allowing the analytic progress that we report below, essentially by ensuring that the system remains mean-field-like, even in proximity to second-order phase transitions. We believe that this is a small price to pay for obtaining a set of exact results that can then serve as a benchmark for numerical studies, and future approximate theories, involving models whose microscopic motivation is more concrete and whose critical behavior is more complex. In particular this choice of model allows an approximation-free analysis of the phase diagram under bias (Fig.~\ref{phasefig} below). This is already so complicated that were it to emerge from a less exact theory, one would need to study its robustness to the specific approximations made.

The rest of the paper is organized as follows: In Sec.~\ref{model} we review the active lattice gas model, its corresponding fluctuating hydrodynamics description and the associated MIPS. Then, in Sec.~\ref{secentro} we present the IEPR functional and its expected value. In Sec.~\ref{over} we present the large deviation framework that accounts for the IEPR fluctuations and overview our main results, focusing for simplicity on positive deviations. In Sec.~\ref{hum} we derive the IEPR fluctuations for the homogeneous phase. Sec.~\ref{exp} explains how this derivation helps expose the other, inhomogeneous, phases whose derivation is detailed in Secs.~\ref{pstb} and \ref{sm}. Sec.~\ref{negative} concludes the mapping of the phase diagram and presents the final expression for the large deviation function describing macroscopic IEPR fluctuations. Here we also exploit a detailed fluctuation theorem that extends all the results so far reported to negative IEPR values. Sec.~\ref{mipsrel} considers the relation of the biased phase transitions to the MIPS criticality.
Sec.~\ref{conc} present a comparison  to previous studies of IEPR in active systems. We conclude and remark on future directions in Sec.~\ref{conc2}.

\section{Active lattice gas model}\label{model}	

\subsection{Definition}

The model we study was first introduced in \cite{kourbane-houssene_exact_2018-1}, and we repeat its definition here. It consists of a one dimensional periodic unit lattice of $L\gg1$ sites where each site $i$ can either be occupied by a~$+$~particle ($\sigma_i^+ = 1$ and $\sigma_i^- = 0$), a~$-$~particle ($\sigma_i^+ = 0$ and $\sigma_i^- = 1$), or be empty ($\sigma_i^+ = 0$ and $\sigma_i^- = 0$). 
The number of particles in the system is $N=\rho_0 L$ where $\rho_0$ is the mean density.

\begin{figure} [t!]
	.\includegraphics[width=0.9\linewidth]{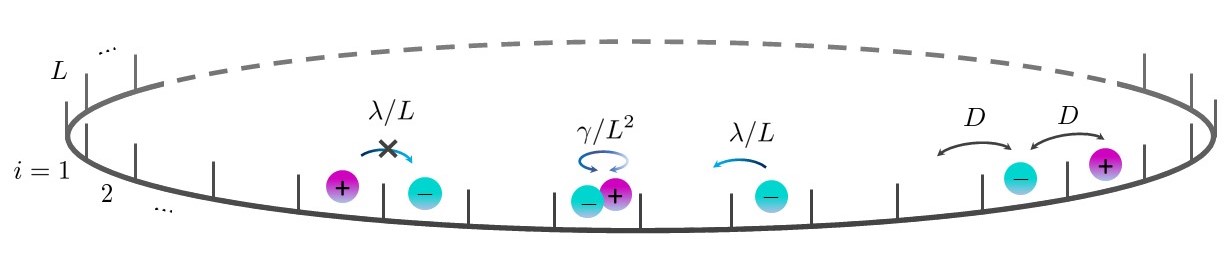}
	\caption{Schematic representation of the microscopic dynamics of the active lattice gas, as described in the text.}
	\label{schem}
\end{figure}

The dynamics of the model is  composed of three rules (see Fig.~\ref{schem}):
\begin{enumerate}[label=(\roman*)]
	\item Symmetric diffusion: A pair of neighboring sites exchange their state with rate $D$.
	\item Self propulsion: A $+$ ($-$) particle jumps to the right (left) neighboring site with rate $\lambda/L$, provided the target site is empty.
	\item Tumbling: A $+$ ($-$) particle converts into a $-$ ($+$) particle with rate $\gamma/L^2$.
\end{enumerate}
The scaling of the rates with $L$ ensures that in the hydrodynamic limit ($L\to \infty$ at fixed $\rho_0$) all processes occur on diffusive time scales. Indeed, the time it takes for particles to traverse a macroscopic system of size $L$ either via diffusive motion, or on account of self-propulsion, scales as $L^2$ which is also the time scale for tumbling events. 

We identify two relevant length scales in this system, which are the persistence length of active motion in the dilute limit, $\hat{\ell}_\mathrm{P} = \lambda L / \gamma$, and the diffusive length $\hat{\ell}_\mathrm{D} =L\sqrt{ D/\gamma}$, which is the typical distance that a particle diffuses, before it tumbles.  One observes that both these length scales are proportional to $L$ so they are macroscopic, due to the $L$-dependence of the rates.
These $L$-dependent rates may not be the obvious choices for a physical model:  one typically expects the tumbling time to be faster, so that the length scales $\hat{\ell}_\mathrm{P},\hat{\ell}_\mathrm{D} $ remain microscopic (of the order of the particle size). However, the choice of $L$-dependent rates allow for many exact results~\cite{kourbane-houssene_exact_2018-1} that would otherwise not be achievable, including those described below.  Interpretation of such exact results can then be used to understand and interpret the results of more realistic physical models.

\subsection{Hydrodynamic equations}
\label{sec:hydro-eq}

To access hydrodynamic behaviour of this model, define a hydrodynamic co-ordinate $x$ such that the
	position of site $i$ is $x=i/L$, which lies in the unit interval $[0,1)$.  The local density in the vicinity of this point is obtained by a mesoscopic coarse-graining:
\begin{equation}
\rho_{\pm}(x,t)=\frac{1}{2L^{\delta}}\sum_{|i-Lx|<L^{\delta}}\sigma_i^{\pm} \:,\label{den}
\end{equation}
where the exponent $\delta$ can take any value in the range  $0<\delta<1$.  The mean density is then
\begin{equation}
\rho_0=\int_0^1\rho(x,t)dx \; .
\end{equation}
We write $\hat{t}$ for the time variable of the microscopic system: since the system is diffusive, the corresponding hydrodynamic time variable is $t=\hat{t}/L^2$.  Define also the density and polarisation (or magnetisation) fields:
	\begin{align}
	\rho & = \rho_+ + \rho_- \nonumber\\
	m & = \rho_+ - \rho_- \; .
	\end{align}
	(The dependence of these fields on $(x,t)$ is left implicit, for a compact notation.)
	
	It was shown in \cite{kourbane-houssene_exact_2018-1} that this model reaches `local equilibrium' over mesoscopic time and length scales. By virtue of this essential property, the density fields \eqref{den} obey hydrodynamic equations:	
\begin{eqnarray}
\partial_t \rho&=& -\partial_x J_{\rho}\label{eq:rho}\\
\partial_t m&=& -\partial_x J_{m}- 2\gamma K, \label{eq:m}
\end{eqnarray} 
where $J_{\rho}, J_m$ are the conservative fluxes and $K$ is the net rate of change of the $\rho_-$ density due to tumbling reactions. (Note that $K$ is used in other works to denote activity, which is not to be confused with the net tumbling rate here.) The fluxes $J_{\rho,m}$ and the net tumbling rate $K$ fluctuate around their typical (most likely) values, which are:
\begin{align}\label{mean1}
\bar{J}_{\rho}&=-D\partial_x\rho+\lambda m(1-\rho) \nonumber\\ 
\bar{J}_{m}&=-D\partial_xm+\lambda\rho(1-\rho)\nonumber\\ 
\bar{K}&= m.
\end{align}
(As above, the dependence of these mean fluxes and rates on the hydrodynamic fields is left implicit, for compactness.)

The deterministic hydrodynamics given by these expressions was studied in \cite{kourbane-houssene_exact_2018-1} where a regime of MIPS was found.  Steady states of the model
obey the time-independent equations
\begin{eqnarray}
0&=&(\ell\partial_x)\rho-\text{Pe}\,m(1-\rho)\nonumber\\
0&=&(\ell\partial_x)^2m-\text{Pe}\,(\ell\partial_x)\left[\rho(1-\rho)\right]-2m,\label{sta}
\end{eqnarray}
where 
\begin{align}
\ell & = \hat{\ell}_\mathrm{D}/L= \sqrt{D/\gamma} 
\nonumber \\
\text{Pe} & = \hat{\ell}_\mathrm{P}/\hat{\ell}_{\rm D}=   \lambda/\sqrt{D\gamma} \; .
\end{align}
This $\ell$ is a rescaled version of the diffusive length, and the P\'eclet number $\text{Pe}$ is the ratio of the persistence length to the diffusive one.
If Pe and $\rho_0$ are sufficiently large, the hydrodynamic equations support non-homogeneous steady states, corresponding to MIPS.
	
	Two comments are in order.  First, while this definition for the P\'eclet number provides a measure for the self-propulsion strength, it is fundamentally different from the common definition in studies of interacting active particles, which compares the persistence length with the particle size.  That definition would be meaningless in this model, because the persistence length ${\hat\ell}_{\rm P}$ is macroscopic.  Hence we choose the definition of Pe as the ratio between the persistence length and the diffusive length, which are both  macroscopic.  
	
	Second, gradients in \eqref{sta} always appear together with factors of $\ell$.  The result is that any inhomogeneous steady states will have spatial modulations whose length scale are determined by this parameter (or equivalently by the diffusive length $\hat{\ell}_{\rm D}$).  In particular, MIPS appears in this model~\cite{kourbane-houssene_exact_2018-1}  as a coexistence of high- and low-density phases, separated by narrow domain walls.  In this work, we assume that ${\ell}\ll 1$ so that the interfacial width for MIPS is indeed narrow when compared to the system size, although it is still macroscopic.  (Anticipating results below, note that biased  ensembles can support phases with a smooth density modulation: in these cases, narrow interfaces do not appear, even as $\ell\to0$.)

\begin{figure} [t!]
	.\includegraphics[scale=0.5]{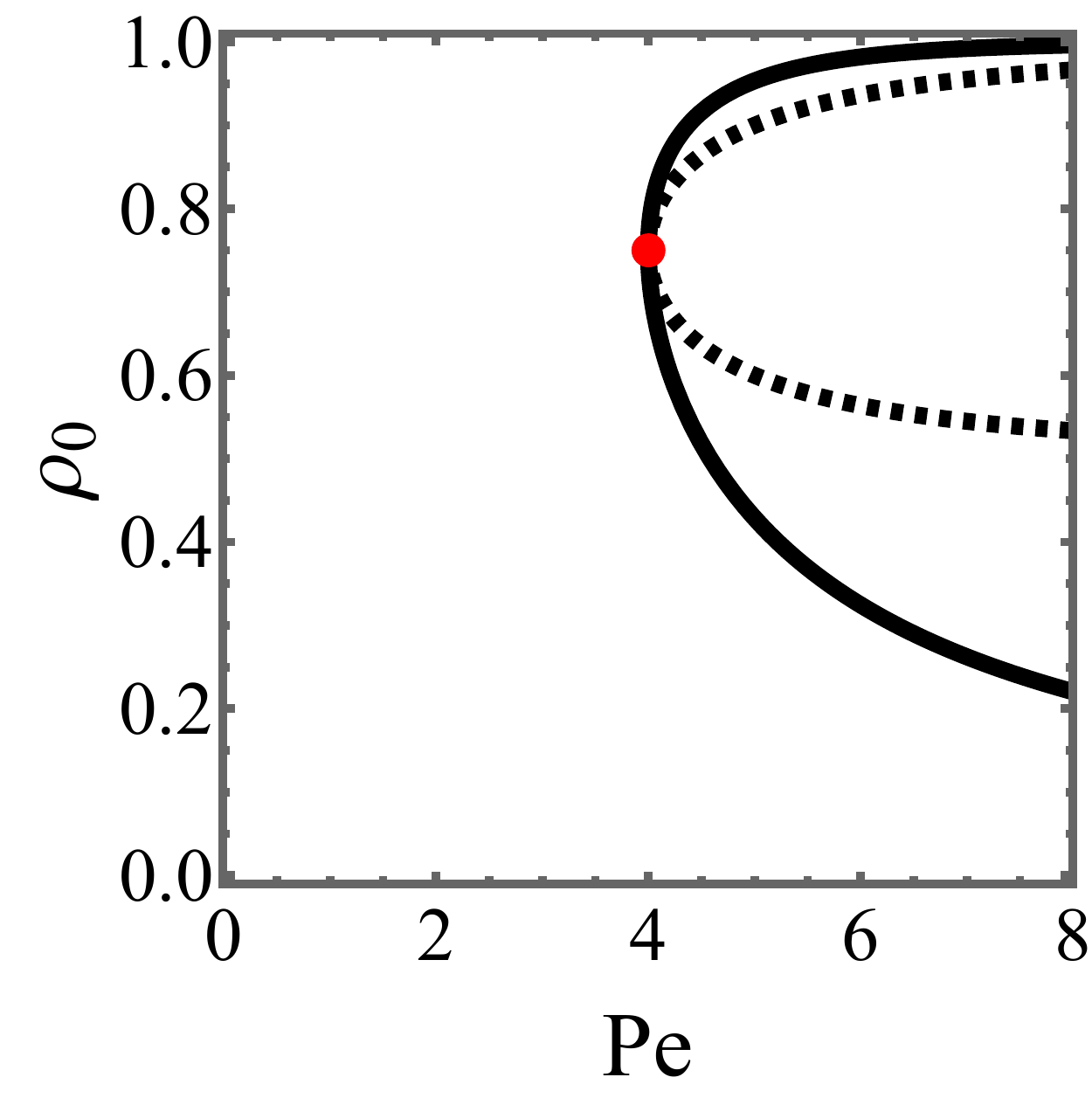}
	\caption{The phase diagram for MIPS in the active lattice gas model. The thick black curve is the binodal while the black dashed curve is the spinodal. These meet at the critical point given by the red dot.}
	\label{mips}
\end{figure}

The binodal and spinodal curves for MIPS in this model have been calculated (for $\ell\ll1$) in~\cite{kourbane-houssene_exact_2018-1}, and are shown in Fig.~\ref{mips} for completeness. The critical point of the model 
$(\rho_c,\text{Pe}_c)=(3/4,4)$
is
located where the binodal and spinodal curves meet.

\subsection{Hydrodynamic fluctuations}

We now describe the fluctuating hydrodynamics of this model, as derived in~\cite{agranov_exact_2021} and \cite{agranov_macroscopic_2022}.
For passive lattice gasses, such a description was widely employed in the context of the Macroscopic Fluctuation Theory (MFT), see \cite{bertini_macroscopic_2015-1} for a review. At the macroscopic level ($L\gg 1$), the system is overwhelmingly likely to follow the deterministic hydrodynamic equations given above, but significant deviations are still possible: they can be analysed by large deviation theory. To obtain an explicit formula for the probability of these rare events, we consider a trajectory (or history) of the system where the hydrodynamic fields are measured, as are the corresponding fluxes: denote such a trajectory as
\begin{equation}
{\cal X} = \left\{\rho(x,t),m(x,t),J_{\rho,m}(x,t),K(x,t)\right\}_{x\in[0,1],t\in[0,T]}
\label{pathX}
\end{equation}
where $T$ is the trajectory duration (measured on the hydrodynamic scale).
For any trajectory, the measured fluxes and tumbling rates $J,K$ always obey (\ref{eq:rho},\ref{eq:m}), but they differ in general from the deterministic values given in \eqref{mean1}.
The probability of such a trajectory is then
\begin{equation}
P({\cal X}) \simeq {\rm e}^{-L{\cal A}({\cal X})} 
\label{equ:path-LDP}
\end{equation}
where the path action\footnote{%
Our notation for the action here is different from previous works \cite{agranov_macroscopic_2022} so as to avoid confusion with the sign for the IEP in the following \eqref{entro0}.}
is an integral of a Lagrangian density that contains two terms:
\begin{equation}
\mathcal{A}= \int_0^T dt\int_0^1 \!dx\, (\mathcal L_J+\mathcal L_K).\label{act}
\end{equation}
The first contribution controls fluxes and is given by the Gaussian quadratic form
\begin{equation}\label{fluxrate}
\mathcal L_J=\frac{1}{2}
\begin{bmatrix}
J_{\rho}-\bar{J}_{\rho}\\
J_m-\bar{J}_m
\end{bmatrix}^\dag
\mathbb{C}(\rho,m)^{-1}	
\begin{bmatrix}
J_{\rho}-\bar{J}_{\rho}\\
J_m-\bar{J}_m
\end{bmatrix},
\end{equation}
with a symmetric correlation matrix
\begin{equation}\label{cmat}
\mathbb{C}(\rho,m)=D
\begin{bmatrix}
\sigma_{\rho}(\rho) &\sigma_{\rho,m}(\rho,m) \\
\sigma_{\rho,m}(\rho,m) &\sigma_{m}(\rho,m)
\end{bmatrix}
\end{equation}
where $\sigma_{\rho}(\rho)=2\rho(1-\rho)$, $\sigma_{m}(\rho,m)=2(\rho-m^2)$, and $\sigma_{\rho,m}(\rho,m)=2m(1-\rho)$.
The contribution to the large deviation functional controlling tumbling, $\mathcal L_K$, is non-Gaussian:
\begin{eqnarray}
\mathcal L_K=\gamma\left\{ \rho-\sqrt{K^2+(\rho^2-m^2)}+	K\ln \left[\frac{\sqrt{K^2+(\rho^2-m^2)}+K}{(\rho+m)}\right]\right\}.\label{lk}
\end{eqnarray} 
This form arises from the difference of the large deviation functionals for the two Poisson processes ($\pm\to\mp$) that together determine the net rate $K$; see \cite{agranov_macroscopic_2022} for details.

We have neglected contributions to \eqref{act} from the initial condition of the trajectory, these will be negligible in the following because we always consider very long times $T$.

\section{entropy production functional} \label{secentro}

This work concentrates on fluctuations of the informatic entropy production in the active lattice gas.  For a given trajectory ${\cal X}$, the informatic entropy production is defined~\cite{fodor_irreversibility_2022,obyrne_time_2022} by comparing its probability with that of a corresponding time-reversed path ${\cal X}^R$, as\footnote{It has recently been emphasized that the proper evaluation of informatic entropy production using a field theory requires a careful definition of time discretization, say It\=o or Stratonovich \cite{cates_stochastic_2022}. The scaling of fluctuation magnitude with $L^{-1/2}$ implies that this choice is not important here.}
\begin{equation}
\Sigma({\cal X})=\ln [P({\cal X})/P({\cal X}^R)].
\end{equation}
 When defining ${\cal X}^R$, one has to specify the parity under time reversal of the observable fields: different choices can reveal different aspects of breaking of time reversal symmetry~\cite{fodor_irreversibility_2022}.
In the present context, there is only a single choice that results in a nontrivial expression for the entropy production: we must treat the fluxes and the tumbling rate $K$ as odd under time reversal (this is natural since these are velocity-like quantities). That is,
\begin{equation}
{\cal X}^R = \left\{\rho(x,T-t),m(x,T-t),-J_{\rho,m}(x,T-t),-K(x,T-t) \right\}_{x\in[0,1],t\in[0,T]} \label{negexttr}
\end{equation}
which may be compared with~\eqref{pathX}. Notice that both $J$ and $K$ must be odd under time reversal, to ensure consistency with the dynamical equations \eqref{eq:rho} and \eqref{eq:m}.

In non-equilibrium steady states, the entropy production $\Sigma(\cal X)$ is proportional to the trajectory length $T$ so it is natural to define a time-averaged IEPR as
$\mathbb S_T({\cal X})=\Sigma({\cal X})/T$.  Writing this in terms of the action ${\cal A}$ and using the definition of ${\cal X}^R$ yields
\begin{equation}
\mathbb S_T=\frac{L}{T}\int_0^T dt\int_0^1 dx
\left\{2\begin{bmatrix}
J_{\rho}\\
J_m
\end{bmatrix}^\dag
\mathbb{C}(\rho,m)^{-1}	
\begin{bmatrix}
\bar{J}_{\rho}\\
\bar{J}_m 
\end{bmatrix}+\gamma K\ln\left(\frac{\rho+m}{\rho-m}\right)\right\}.\label{entro0}
\end{equation}
This expression can be significantly simplified by singling out a `non-equilibrium' contribution to \eqref{entro0}. Indeed, notice that at vanishing bias $\lambda=0$ the system becomes equilibrium. Correspondingly, we decompose the expected currents \eqref{mean1} into their `equilibrium' piece $-D\partial_x\rho$ and $-D\partial_xm$, and their non-equilibrium piece $\lambda m(1-\rho)$ and $\lambda\rho(1-\rho)$ for the density and magnetization fluxes respectively. For the equilibrium piece one finds that
	\begin{equation}
2\mathbb{C}(\rho,m)^{-1}	
\begin{bmatrix}
-D\partial_x\rho\\
-D\partial_xm 
\end{bmatrix}=\begin{bmatrix}
-\partial_x\frac{\delta F}{\delta\rho}\\
-\partial_x\frac{\delta F}{\delta m}
\end{bmatrix}\label{thermo}
	\end{equation}
with $F=L\int_0^1dx\mathcal F(\rho,m)$ being the free energy functional of the equilibrium ($\lambda=0$) system with density
\begin{equation}
\mathcal F(\rho,m)=\frac{\rho+m}{2}\ln\left(\frac{\rho+m}{2}\right)+\frac{\rho-m}{2}\ln\left(\frac{\rho-m}{2}\right)+(1-\rho)\ln\left(1-\rho\right).
\end{equation}	
The expression Eq.\eqref{thermo} defines a vector of thermodynamic forces that drives particles flow. The thermodynamic force that is conjugate to particles flipping is derived from the same free energy
\begin{equation}
\frac{1}{2}\ln\left(\frac{\rho+m}{\rho-m}\right)=\frac{\delta F}{\delta m}.
\end{equation}
	Plugging these expressions into \eqref{entro0} we have
\begin{equation}
	\mathbb S_T
	=
		\frac{L}{T}\int_0^T dt\int_0^1dx\left\{\begin{bmatrix}
	J_{\rho}\\
	J_m
	\end{bmatrix}^\dag\begin{bmatrix}
-\partial_x\frac{\delta F}{\delta\rho}\\
-\partial_x\frac{\delta F}{\delta m}
	\end{bmatrix}
	+2\gamma K\frac{\delta F}{\delta m}\right\}
+
\frac{L\lambda}{T}\int_0^T dt\int_0^1 dx
	\left\{2\begin{bmatrix}
	J_{\rho}\\
	J_m
	\end{bmatrix}^\dag
	\mathbb{C}(\rho,m)^{-1}	
	\begin{bmatrix}
	m(1-\rho)\\
	\rho(1-\rho) 
	\end{bmatrix}\right\},
	\label{entro1}
	\end{equation}
 Now using integration by parts and the dynamics \eqref{eq:rho} and \eqref{eq:m} allows to distinguish two contributions to the IEPR
\begin{equation}
	\mathbb S_T=-\frac{\Delta F}{T}+\frac{L\lambda}{T}\int_0^T dt\int_0^1 dx
	\left\{2\begin{bmatrix}
	J_{\rho}\\
	J_m
	\end{bmatrix}^\dag
	\mathbb{C}(\rho,m)^{-1}	
	\begin{bmatrix}
	m(1-\rho)\\
	\rho(1-\rho) 
	\end{bmatrix}\right\},\label{entro15}
	\end{equation}
with $\Delta F=F(t=T)-F(t=0)$ the free energy difference between the final and initial states of the system. As this difference is sub-extensive in time, then at large times its contribution to \eqref{entro15} becomes negligible (for initial and final conditions of bounded free energy $F$). From now on we will omit this term.
Finally, using the expression for $\mathbb C$ \eqref{cmat} one arrives at the following simple expression for the IEPR
\begin{equation}
\mathbb S_T=\frac{L\gamma\text{Pe}^2}{T}\int_0^T dt\int_0^1 dx
\alpha(x,t) ,\label{entro2}
\end{equation}
where we defined the local rate of entropy production field
\begin{equation}
\aalph(x,t) = \frac{1}{\lambda}\left[ (1-\rho(x,t))J_m(x,t)+ m(x,t)J_\rho(x,t) \right]. \label{locent}
\end{equation}

The only contribution to this IEPR stems from the flux probability cost $\mathcal{L}_J$, which alone encodes the irreversible, biased dynamics. There is no explicit contribution from tumbling, but of course this reversible process does influence the trajectories and hence the value of the IEPR.

It is interesting to compare this expression with the EPR for the underlying microscopic model. A detailed analysis is given in Appendix \ref{mic}, which shows that the macroscopic IEPR \eqref{locent} coincides with the microscopic one. It is unusual for the full microscopic EPR to be preserved under coarse-graining, but this is a feature of the current model (which was, after all, constructed to make the coarse-grained hydrodynamic equations more exact than usual). This coincidence means that we can identify the coarse-grained IEPR in \eqref{entro2} with the rate at which active work is expended at the particle level \cite{seifert_stochastic_2012,pietzonka_entropy_2018}.
\subsection{The average IEPR}
\label{sec:ave-epr}

At macroscopic level, the mean IEPR can be computed from the most likely trajectory.  In homogeneous steady states, any such trajectory has $\rho=\rho_0$, $J_m=\lambda\rho_0(1-\rho_0)$ and $m=J_{\rho}=0$. The
mean IEPR is then
\begin{equation}
\langle {\mathbb S}_T \rangle =  L\gamma\text{Pe}^2\bar{a}(\rho_0) 
\end{equation}
where
\begin{equation}
\bar{a}(\rho)=\rho (1-\rho)^2\label{ent2}
\end{equation}
is a local rate of entropy production in a system at density $\rho$. To interpret the non-trivial density dependence in \eqref{ent2}, it is useful to consider the microscopic dynamics.

 In this model, entropy production arises from particles hopping to adjacent vacant sites, as these are the only irreversible dynamical moves. The rate at which the system performs such moves is related to the total current of particles in their assigned ($\pm$) direction of motion, which is proportional to  $\rho_0(1-\rho_0)$. Yet the total hopping rate into \textit{vacant} sites is only a fraction of the total current, as the latter also includes a contribution from swaps between two occupied sites. With our chosen dynamical rules, the latter are reversible and so do not contribute to the IEPR. As we show in Appendix \ref{mic}, the relevant fraction of the total current is exactly $(1-\rho_0)$ which leads to the expression \eqref{ent2}.

It will be important in the following that the function $\bar{a}(\rho_0)$ loses convexity for  $\rho_0<2/3$.  This property will have far-reaching consequences including a phase diagram of considerable complexity (arising, as happens in equilibrium thermodynamics, from global minimization of a concave thermodynamic potential). To the best of our knowledge, previous works, which studied macroscopic fluctuations in related passive lattice gases, did not consider observables with such a property, see e.g. \cite{jack_hyperuniformity_2015,dolezal_large_2019,lecomte_inactive_2012}. In a forthcoming publication \cite{agranov_notitle_nodate} we will further explore the phenomena related to fluctuations of such observables in more general settings.

 \section{IEPR deviations: summary of main results}\label{over}
 
 \subsection{Large deviations of the IEPR}
 \label{sec:dev-iepr}
 
Above we have defined the model and its IEPR.  The central part of this work is an analysis of large deviations of the IEPR.
This section defines the problem and gives an overview of the results.  We work at the hydrodynamic level (which is $L\gg 1$, with $L$ the number of lattice sites) and we further assume that the trajectories are very long, $T\gg \gamma^{-1}$.  Since $T$ is defined on the hydrodynamic scale, this means that trajectories are long enough for typical particles to tumble many times, and for a typical particle to explore the whole system via diffusive hops.
 
 For such long times, one expects a large-deviation form for the time-averaged IEPR: that is, the the probability distribution for $\mathbb S_T$ behaves as 
 \begin{equation}
-\frac{1}{\gamma LT } \ln {\rm Prob}\left(\frac{\mathbb S_T}{L\gamma\text{Pe}^2} \approx a \right)\simeq I(a)\label{f}
\end{equation} 
where $I$ is the rate function.  In this formula, the notation $\approx$ indicates that $\frac{\mathbb S_T}{L\gamma\text{Pe}^2}$ should lie in some small interval around $a$, while $\simeq$ indicates asymptotic equality in the joint limit of large $L$ and large $T$, see for example~\cite{touchette_large_2009-1}.
This formula is valid when both $L$ and $T$ are large.
For large $L$, Eq.~\eqref{equ:path-LDP} already gives the probability of any individual hydrodynamic path, so it is sufficient to minimise the action ${\cal A}$ over all paths that achieve the desired IEPR, that is 
\begin{equation}
I(a)=\inf_{{\cal X} \colon \mathbb S_T({\cal X})=a L\gamma\text{Pe}^2 }\frac{ \mathcal{A}({\cal X}) }{\gamma T} \; ,\label{min}
\end{equation}
where the right hand side should be evaluated as $T\to\infty$. This means that initial conditions can be neglected in the path probability formula \eqref{equ:path-LDP}. From the definition of the IEPR it can be seen that if ${\cal X}$ solves the minimisation problem for some $a>0$ then ${\cal X}^R$ solves the equivalent problem for $a<0$, leading to a fluctuation theorem
\begin{equation}
I(-a)=I(a)+\text{Pe}^2\,a\label{negext}
\end{equation}
see {\em e.g.} \cite{seifert_stochastic_2012,jack_ergodicity_2020-1}.

\begin{figure*}
	\begin{tabular}{ll}
		\includegraphics[scale=0.25]{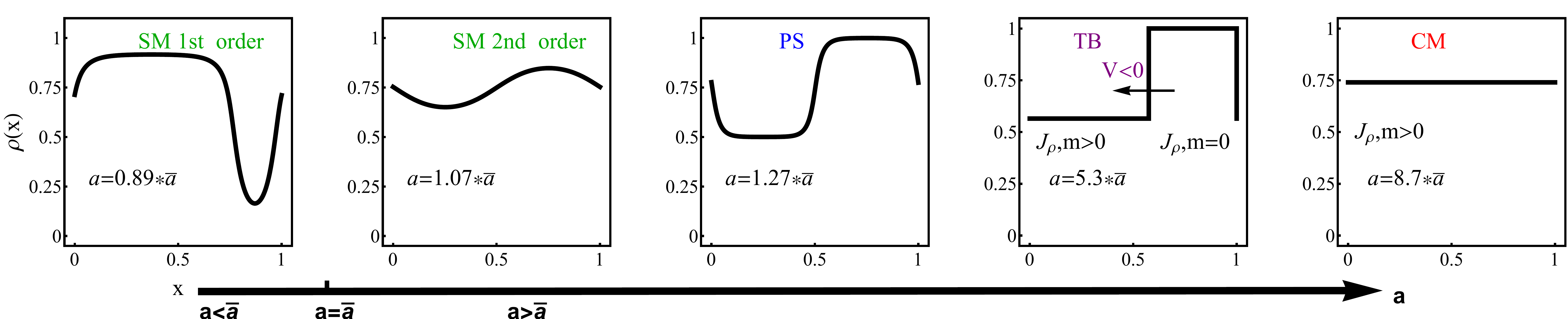}
	\end{tabular}
	\caption{States of the system at $\rho_0=0.75$ with increasing values of $a$, the scaled IEPR. Here $\text{Pe}=2.5$ and $\ell=0.02$ in the panels of SM and PS. The discontinuous profile in the panel TB corresponds to vanishing $\ell=0^+$ where at finite $\ell$ the discontinuities are smoothed over a scale $\ell$ as exemplified in the panel PS. The profile in the panel CM is $\ell$-independent. The distinction between the two SM states is discussed in Sec.~\ref{sm}. The homogeneous phase that is found in the vicinity of $a=\bar{a}$ is omitted from the sequence. 	}
	\label{profiles}	
\end{figure*}

Solving this minimization problem fully describes the large deviations of $\mathbb S_T$. The solutions will be computed in the following sections. They depend on the mean density $\rho_0$ and on the parameters Pe and $\ell$.  As discussed in Sec.~\ref{sec:hydro-eq}, we shall focus on the behaviour for very small $\ell$, for which we discuss the behaviour of $I(a)$ as a function of $a,\text{Pe}$.  As these variables are varied, the optimal trajectories can change in character, leading to singularities in $I(a)$, which are interpreted as dynamical phase transitions. We find five different dynamical phases in this model, whose main characteristics are summarized below and in Table \ref{table:1} and Fig.~\ref{profiles}.
	\begin{enumerate}
		\item Homogeneous (H) phase. All fields are constant both in space and in time. Moreover, the system is nonpolar with zero `magnetization', $m=0$,  and also has vanishing mass flux $J_{\rho}=0$. 	
		\item Collective motion (CM) phase. All fields are still constant but the system breaks reflection symmetry by developing a nonvanishing magnetization $m\neq0$ and mass flux $J_{\rho}\neq0$.
		\item Phase separated (PS) phase. The system breaks translational invariance with the formation of sharply phase separate profiles. Reflection symmetry is unbroken and there is no magnetization or current: $m,J_{\rho}=0$.
			\item Traveling band (TB) phase. Here a phase separated solution involves a magnetized state $m,J_{\rho}\neq0$ resulting in a sharply phase separated traveling band solution. This state breaks translational and temporal invariance and also reflection symmetry.
		\item Smoothly modulated (SM) phase. Here the system only breaks translational invariance, but this time with a smoothly varying spatial profile.
	\end{enumerate}

	\begin{table}
	\centering
	\begin{tabular}{||c c c c||} 
		\hline
		Phase & Net current $J_{\rho}$ & Spatial structure & Temporal structure \\ [0.5ex] 
		\hline\hline
		Homogeneous, H & No &Homogeneous & Constant \\ 
		Collective motion, CM & Yes & Homogeneous & Constant \\
		
		Phase separated, PS & No & Sharp interface & Constant \\
		Traveling band, TB & Yes & Sharp interface & Traveling wave \\ 
		Smoothly modulated, SM & No & Smooth modulation & Constant \\[1ex] 
		\hline
	\end{tabular}
\caption{The main characteristics of the different phases }
	\label{table:1}
\end{table}

Note that these phases (including those with non-zero magnetization) arise in our model without any aligning interactions, for which typical steady-state trajectories are homogeneous with zero mass flux. (The model also supports MIPS steady states but we do not consider those parameter values in this work: see Sec.~\ref{conc} for a discussion of that case.) 

\begin{figure}
	\includegraphics[scale=0.45]{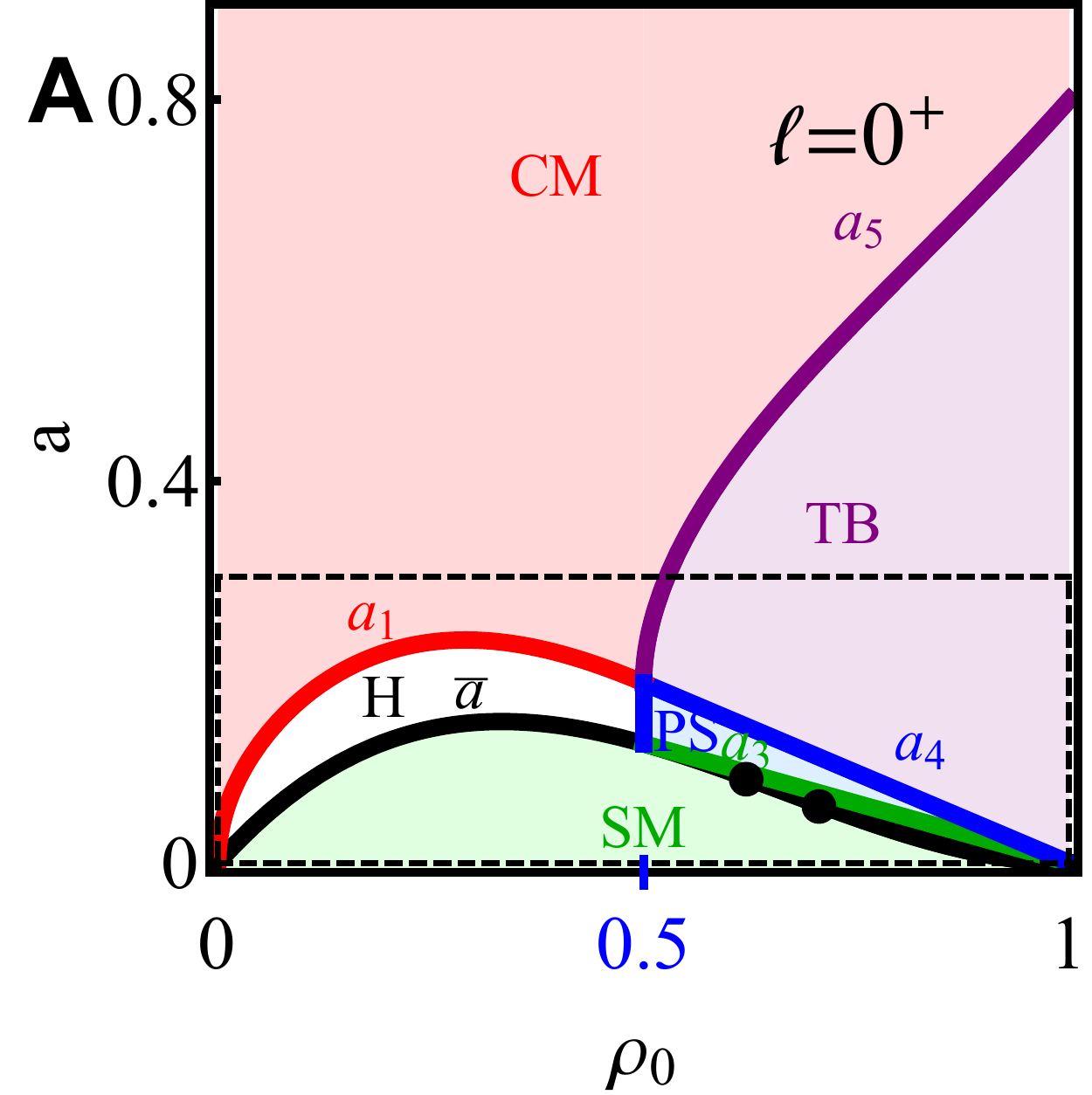}\includegraphics[scale=0.45]{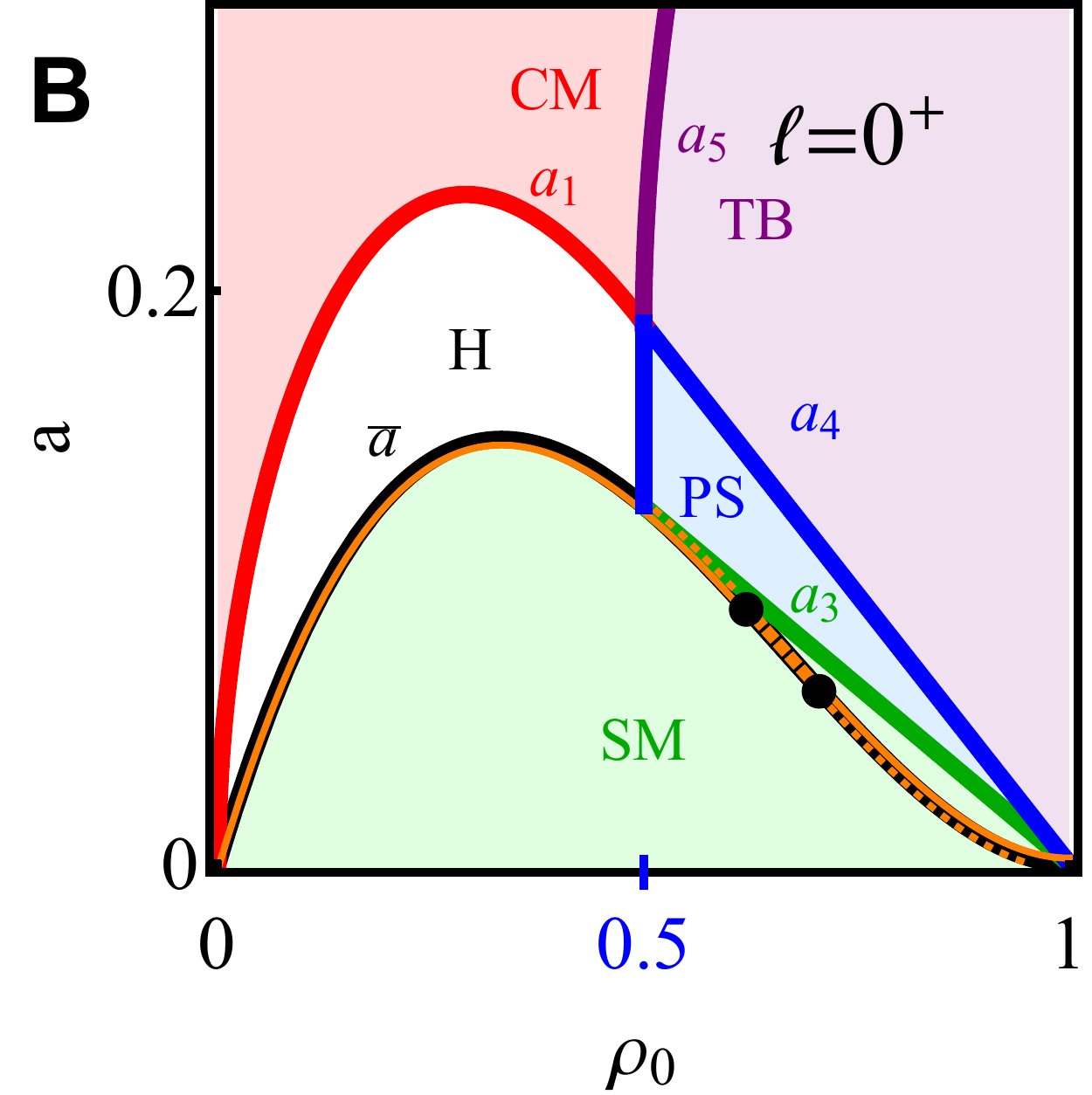}\\\includegraphics[scale=0.45]{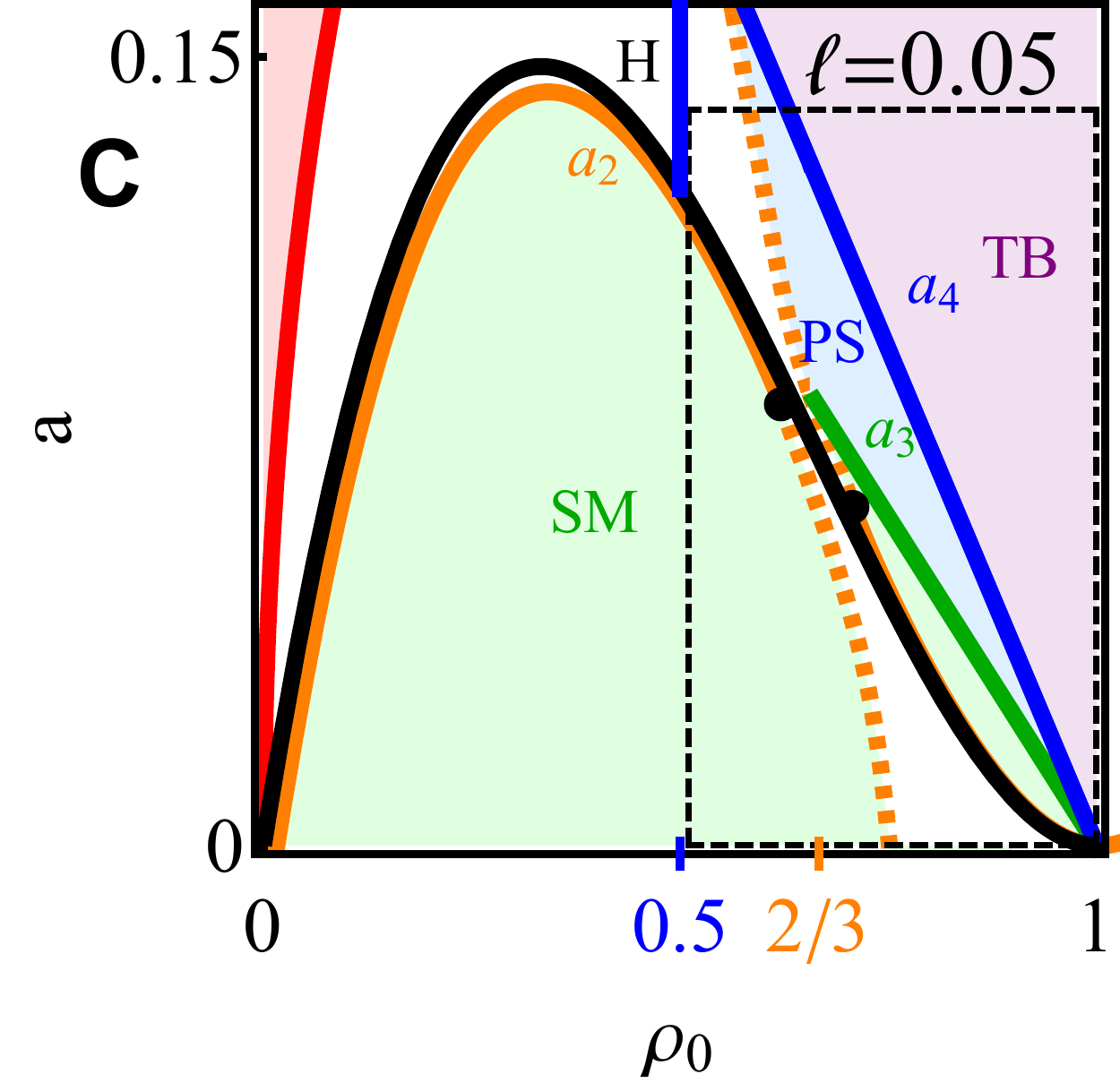}
	\includegraphics[scale=0.45]{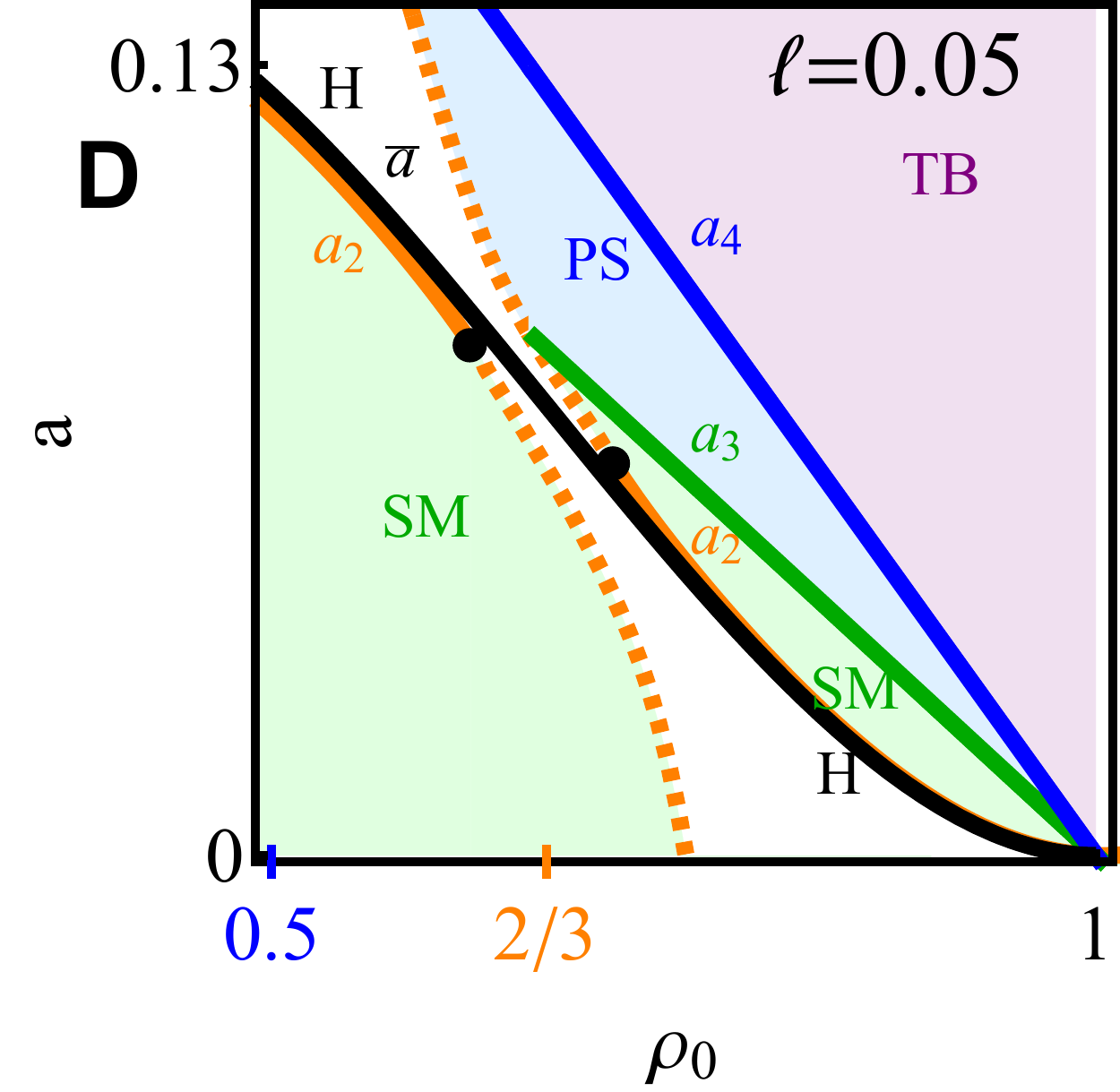}	
	\caption{The nonequilibrium phase diagram for entropy production deviations given by the minimization problem \eqref{min} at $\text{Pe}=2.5$ and vanishing $\ell=0^+$ (A,B) and finite $\ell=0.05$ (C,D). (Panels B and D are enlargements of the regions within the black dotted line in A and B respectively.) The black solid line is the expected entropy production curve \eqref{ent2}. The black dots are tricritical points where a second order phase transition line (solid orange) and first order (dashed orange) meet. The miscibility gap that corresponds to the first order transition, not shown here, is discussed in Sec.~\ref{sm}, see also Fig~ \ref{phasefig2}. The critical lines in red, orange, green, blue and purple are given by $a_{1-5}$ in Eqs.~(\ref{ac1}, \ref{ac2}, \ref{ac3}, \ref{ac4}) and \eqref{ac5} respectively. The vertical blue segment is at $\rho_0=1/2$. The re-entrant H phase arising for negative $a-\bar{a}$ at large density $\rho_0$ in (C,D) is stabilized by interfacial contributions  and disappears as $\ell \to 0^+$ (A,B). Note that the first order (dashed orange) lines are associated with miscibility gaps with tie-lines vertical (signifying time-like phase separation). These gaps are too narrow to be shown here; see Fig. \ref{phasefig2} }
	\label{phasefig}
\end{figure}

Fig.~\ref{profiles} shows the density profiles within the phases. The complete phase diagram, as derived and explained in subsequent sections, is depicted in Fig.~\ref{phasefig}. The lines denote transitions between the different phases; these transitions are accompanied by singularities of the rate function $I(a)$ as detailed later in Fig.~\ref{singfig}.

\subsection{Finding the phases: overview }

Since the derivation of these new phases and of the overall phase diagram is a somewhat involved task, we give an overview of the subsequent calculations here. 
The black line $a=\bar{a}(\rho)$ in the phase diagram (Fig.~\ref{phasefig}) corresponds to the typical behaviour of the system; all other points correspond to unusual values of the IEPR, and the corresponding optimal trajectories are obtained by solving \eqref{min}.  Note that dynamical phase diagrams are often plotted as a function of a biasing field, which parameterises the distance of $a$ from $\bar{a}$~\cite{fodor_irreversibility_2022,jack_ergodicity_2020-1}: in this case the typical behaviour of the system would correspond to the zero-bias axis of the phase diagram, instead of a non-trivial curve.  Here, we plot the phase diagram in terms of the IEPR itself. In the analogy with thermodynamics, introducing the biasing field corresponds to a change of ensemble~\cite{touchette_large_2009-1}.  In such phase diagrams, first-order phase transitions lead to regions of phase coexistence in the phase diagram.  In this work, we refer to such regions as miscibility gaps, by analogy with the thermodynamics of fluid mixtures.  These miscibility gaps are very narrow in the system studied here so they are not shown in Fig. \ref{phasefig} but they are included in Fig. \ref{phasefig2}, see Sec.~\ref{hsm-sm2ps}.

Our derivation of the phase diagram starts with the H and CM phases in Sec.~\ref{hum}. In these cases all fields are constant both in space and time which means that the minimization problem \eqref{min} becomes algebraic. We denote by $I_H(\rho_0,a)$ the rate function that corresponds to such homogeneous solutions. In the H phase, the optimal trajectory in \eqref{min} has $m=0$ and vanishing mass flux $J_{\rho}=0$. The resulting rate function is quadratic, see Fig.~\ref{singfig} below. This quadratic form provides an upper bound for the rate function everywhere in parameter space. The same upper bound was employed in a series of recent works \cite{pietzonka_universal_2016-1,horowitz_proof_2017} that established a non-equilibrium thermodynamic uncertainty relation for current fluctuations. In this model, the bound is saturated only in the H phase. 
In the other four phases, the minimisation problem is solved by different kinds of trajectory, so $I(a) < I_H(\rho_0,a)$, and dynamical phase transitions occur.  

The CM phase occurs when $a$ is larger than its typical value: here the system spontaneously breaks the reflection symmetry, leading to a non-zero magnetization $m\neq0$ and mass flux $J_{\rho}\neq0$.  In Sec.~\ref{hum} we identify the mechanism behind this transition by looking at the microscopic dynamics. A similar phase transition was reported in previous numerical studies of scalar active systems \cite{nemoto_optimizing_2019-1,fodor_dissipation_2020,grandpre_entropy_2021,keta_collective_2021}. Our analysis confirms that, within the active lattice gas, this transition is described by a mean-field Landau theory.

Inhomogeneous phases are discussed in Sec.~\ref{exp}. Analytical progress is possible here by extending an analogy with equilibrium statistical mechanics that was recently established in \cite{agranov_macroscopic_2022} for the study of current fluctuations in this model. Within this analogy, $I_H(\rho,a)$ takes the role of the bulk term in a free energy functional of two conserved order parameters. These are the spatially varying local entropy production rate $a(x)$ and density $\rho(x)$ fields. As we show, the small length scale $\ell\ll1$ controls the analogue of the interfacial energy. Consequently, the state of the system can be described by an analogy with the classical equilibrium theory for phase separation where the rate function $I$ \eqref{f} is given by a convex hull (common tangent plane) construction over $I_H$. This construction is valid when the inhomogeneity involves sharp interfaces whose width is controlled by $\ell$ and which contribute subdominantly relative to the `bulk' term $I_H$.

This leads to sharply phase-separated (PS) solutions that are found via a common tangent construction over $I_H$, as discussed in  Sec.~\ref{pstb}.  In this phase the sharply separated profile does not break reflection symmetry ($m,J_{\rho}=0$) and is stationary. This phase was previously reported for related systems when biased towards suppressed EPR values \cite{grandpre_entropy_2021,yan_learning_2022,fodor_dissipation_2020,chiarantoni_work_2020,nemoto_optimizing_2019-1,cagnetta_large_2017}. We show here that this transition can also occur at \textit{enhanced} IEPR biasing. We identify the source of this phenomenon as the non-convexity of $I_H$ at enhanced IEPR deviations for this model. We also report a second sharply phase-separated state, for which the common tangent construction brings in a magnetized state with $m,J_{\rho}\neq0$, resulting in phase-separated profile that moves with a constant velocity. This traveling band (TB) phase was not reported in previous studies of other models, yet we expect it to be generic in the sense that it it should appear whenever non-convexity of $I_H$ occurs alongside reflection symmetry breaking ($m\neq0$). Indeed, this phase was reported in a recent study of the same model under bias for a different dynamical quantity \cite{agranov_macroscopic_2022}. 

Lastly we analyze the SM phase in Sec.~\ref{sm}. Here the bulk term $I_H$ fails to account for the state of the system due to a degeneracy in the common tangent plane construction.  This occurs because $I_H$ has the same value for all unbiased states. (This could conceivably occur in the analagous problem of a free energy function for a equilibrium system with two conserved order parameters, but we have never seen an instance of it.) As a result, the state of the biased system is determined by the sub-leading surface-tension-like term in the rate functional. Solving the spatial minimization problem for this functional, we find that the optimal solution is smoothly varying in space rather than comprising bulk phases separated by sharp interfaces as in the PS and TB states.

The SM phase was anticipated in a similar active context in \cite{nemoto_optimizing_2019-1} based on generic arguments. Here, by establishing its existence in the active lattice gas model, we provide the first conclusive evidence for its existence. Moreover, we find that the transition to this phase can be either first or second order, depending on the value of the averaged density $\rho_0$, see Fig.~\ref{phasefig}. Similar phase transitions were extensively studied in the context of passive lattice gases \cite{bodineau_current_2010-1,lecomte_inactive_2012,jack_hyperuniformity_2015,baek_dynamical_2017,baek_dynamical_2018,dolezal_large_2019}, but for each given model were found to be of fixed order (first or second). Here we find that the H$\to$SM transition occurs along both a second order and a first order transition line, with these meeting at a tricritical point. We trace back this phenomenon to the change of convexity in \eqref{ent2}. 

In the next four sections we derive all the above-mentioned results. The fluctuation theorem \eqref{negext} means that it is sufficient to consider only $a\geq 0$ since the rate function with $a<0$ can be then be deduced, see Sec.~\ref{negative}.

\section{Homogeneous phases}\label{hum}

We first consider solutions to the minimization problem \eqref{min} that are homogeneous in space and time:
\begin{equation}
\rho(x,t)=\rho_0\quad,\quad m(x,t)=m\quad,\quad J_\rho(x,t)=J_\rho\quad,\quad J_m(x,t)=J_m.\label{homassume}
\end{equation}
These will serve as a baseline for the following analysis of non-homogeneous phases.  Under these conditions, the minimization problem \eqref{min} is algebraic. In the context of passive lattice gases the stationarity hypothesis is known as the ``additivity principle''~\cite{bodineau_current_2004} and was shown to hold in a variety of scenarios \cite{bodineau_current_2004,hurtado_test_2009,hurtado_large_2010,hurtado_symmetries_2011,prados_large_2011,meerson_survival_2014,meerson_full_2015,agranov_survival_2016,agranov_fluctuations_2017,agranov_narrow_2018,agranov_occupation-time_2019}.
Its breaking signals a dynamical phase transition, which has attracted significant interest in the literature, see e.g. \cite{bertini_current_2005,bodineau_distribution_2005,bertini_non_2006,shpielberg_chatelier_2016,zarfaty_statistics_2016}. As we find below, however, a phase transition is possible in the active lattice gas model even among simple time- and space-independent solutions.

We now turn to minimize \eqref{min} subject to \eqref{homassume}. Since $\rho_0$ is a parameter, the quantities to be optimized are $m,J_\rho,J_m,K$. The dynamical equation \eqref{eq:m} requires the net tumbling rate $K=0$.  Second, constraining the IEPR in \eqref{entro2} to obey $\mathbb S_T = a L\gamma\text{Pe}^2$ requires
\begin{equation}
(1-\rho_0)J_m+J_\rho m=a\lambda. \label{j1}
\end{equation}
Plugging these expressions into \eqref{act} and minimizing with respect to $J_\rho$ provides an additional algebraic relation between the fluxes, and one finally arrives at
\begin{equation}
J_\rho=\frac{ma\lambda}{\rho_0(1-\rho_0)+m^2}\quad,\quad J_m=\frac{\rho_0a\lambda}{\rho_0(1-\rho_0)+m^2}.\label{flux}
\end{equation}

With these relations, the minimization problem \eqref{min} reduces to a single minimization over the magnetization $m$: 
\begin{equation}
I_H(\rho_0,a;\text{Pe})=\min_{m}\tilde{I}(m),\label{ih}
\end{equation}
 with
\begin{equation}
\tilde{I}(m)=\rho_0-\sqrt{\rho_0^2-m^2}
+\text{Pe}^2\,\frac{\left\{a-(1-\rho_0)\left[\rho_0(1-\rho_0)+m^2\right]\right\}^2}{4(1-\rho_0)\left[\rho_0(1-\rho_0)+m^2\right]}
\end{equation}
Anticipating a second-order phase transition to the CM phase, we expand in powers of $m$ to obtain 
\begin{equation}
\tilde{I}(m) =\text{Pe}^2 \frac{(a-\bar{a})^2}{4\bar{a}} 
-\text{Pe}^2\,(1-\rho_0)\frac{a^2-a_{1}(\rho_0,\text{Pe})^2}{4\bar{a}^2}m^2
+\frac{\bar{a}^2+2\text{Pe}^2\,\rho_0^2 a^2 }{8\rho_0^3\bar{a}^2}m^4+\mathcal O(m^6)\label{tildf}
\end{equation}
where $\bar{a}=\bar{a}(\rho_0)$ is given by \eqref{ent2} and 
\begin{equation}
a_{1}(\rho_0,\text{Pe})= \bar{a}(\rho_0) \sqrt{\frac{2+\text{Pe}^2\,\rho_0(1-\rho_0)}{\text{Pe}^2\,\rho_0(1-\rho_0)}}.
\label{ac1}
\end{equation}
Interpreting Eq.~\eqref{tildf} as a Landau free energy, one sees that $m$ acquires a non-zero value whenever $a$ exceeds the critical value $a_{1}(\rho_0,\text{Pe})$.  
Clearly $a_{1}(\rho_0,\text{Pe}) > \bar{a}(\rho_0)$, so this symmetry breaking transition occurs when biasing the system towards increased IEPR. This is a second order phase transition: the optimal value $m$ (and the corresponding mass flux $J_{\rho}$ \eqref{flux}) grows continuously from zero on increasing $a$ through $a_1$. It follows that near the critical point
\begin{equation}
m,J_\rho\sim \left[ a-a_{1}(\rho_0,\text{Pe}) \right]^{1/2}
\end{equation}
which is the usual critical exponent for such transitions.

Inserting these results into
\eqref{ih}, one obtains 
\begin{numcases}
{I_H(\rho_0,a)=}
\frac{\text{Pe}^2}{4}\frac{\left[a-\bar{a}(\rho_0)\right]^2}{\bar{a}(\rho_0)}, & $a\leq a_{1}(\rho_0,\text{Pe})$, \label{lows}\nonumber\\
\rho_0-\sqrt{\rho_0^2-m}
+\frac{\text{Pe}^2}{4}\frac{\left[a-\bar{a}(\rho_0) -(1-\rho_0)m^2\right]^2}{\bar{a}(\rho_0) +(1-\rho_0)m^2},& $a> a_{1}(\rho_0,\text{Pe})$, \label{highs}\label{sto}
\end{numcases}
where the relation $m=m(a;\rho_0,\text{Pe})$ is given implicitly by
\begin{equation}
a=\left[\bar{a}(\rho_0)+(1-\rho_0)m^2\right]\sqrt{1+\frac{2}{\text{Pe}^2}\frac{1}{(1-\rho_0)\sqrt{\rho_0^2-m^2}}}.\label{mag}
\end{equation}
This $I_H$ is the variational estimate for the rate function, under the assumption that the deviations in the IEPR occur via homogeneous trajectories.

As usual in the Landau-like scenario, the expression \eqref{sto} for the minimized rate function (analogous to a free energy) has a discontinuous second derivative at $a=a_1$. Note that the quadratic form in \eqref{sto}, valid for $a<a_{1}$, coincides with a universal bound on current fluctuations  \cite{pietzonka_universal_2016-1,horowitz_proof_2017}. This bound is saturated in the H phase, but not in the CM (collective motion) phase, which is stationary and homogeneous, but has nonzero magnetization $m$.

The mechanism behind this transition can be illuminated by looking at the microscopic picture. As mentioned above in Sec.~\ref{secentro} and detailed in Appendix \ref{mic}, the entropy production arises from the total particle current into adjacent vacant sites. This current is only a fraction of the total current which, acccording to the dynamical rules in the chosen model, includes a contribution also from reversible swaps between adjacent sites (Fig.~\ref{schem}). By increasing the imbalance $m$ between right-moving ($+$) and left-moving ($-$) particles, the part of the particle current $J_\rho$ carried by reversible swaps is reduced. At the extreme $|m|=\rho_0$, where all particles are of the same type, reversible swaps contribute nothing to the current, maximizing the rate of entropy production alongside the current itself. 

This mechanism is closely related to the one behind a similar transition that was reported in a system of active Brownian particles with repulsive pairwise interactions \cite{nemoto_optimizing_2019-1}. That model undergoes a transition to a collective motion phase, when considering sufficiently large values of the active work (which is effectively the microscopic EPR). In this CM state, particles align their orientation and move collectively in a common direction. This alignment decreases the rate of collisions between the particles, thereby enhancing the rate at which particles perform irreversible motion (through self-propulsion), and increasing the EPR. In the model we study here, particle collisions are less effective, since particles can always swap positions by diffusive dynamical moves (Fig.~\ref{schem}). Nevertheless, encounters between oppositely oriented particles still decrease the rate of irreversible moves, similar to the active Brownian particles. Thus, by aligning the particles' orientation and reducing the rate of those encounters, the IEPR is enhanced.

\section{Concavities and inhomogeneous phases}\label{exp}

\subsection{Squared gradient approximation (small $\ell$)}

The expression (\ref{sto}) in the previous Section was obtained by optimizing (\ref{min}) under the assumption of homogeneous trajectories.  As such, it is a valid bound on the rate function $I$, but there are cases where the global minimiser of \eqref{min} is an inhomogeneous trajectory. Studying the 
expression \eqref{sto} over the entire $(\rho_0,a)$ parameter space will allow us to understand why the homogeneous trajectory becomes sub-optimal in some regions, and expose the inhomogeneous phases that take over. 

We first assume that the system is stationary with zero mass flux $J_{\rho}=0$, so the fields vary only in space, not time (this is the additivity principle \cite{bodineau_current_2004}). 
One can make progress in the ensuing minimization problem by exploiting the small length scale $\ell$. This length scale controls the gradient terms appearing in the action functional \eqref{act}. We show in Appendix \ref{convex} that the magnetization $m$ can be optimized analytically, yielding a two-field minimization problem over the density $\rho(x)$ and the flux $J_m(x)$.  It is convenient to reparameterise $J_m$ in terms of the local rate of entropy production \eqref{locent}
$\aalph(x) = J_m(x)(1-\rho(x))/\lambda$.

At leading order in a gradient expansion, the minimization problem \eqref{min} becomes
\begin{equation}
I(a)\simeq\min_{\aalph(x),\rho(x)}\int_0^1dx\left\{I_H (\rho(x),\aalph(x))+\ell^2\begin{bmatrix}
\partial_x\rho\\
\partial_x \aalph
\end{bmatrix}^\dag\mathbb{A}
\begin{bmatrix}
\partial_x\rho\\
\partial_x \aalph
\end{bmatrix}\right\},\label{Imin0}
\end{equation}
subject to the constraints
\begin{equation}
a=\int_0^1\aalph(x)dx\quad,\quad \rho_0=\int_0^1\rho(x)dx.\label{cons4}
\end{equation}
with the matrix $\mathbb{A}(\rho,\aalph)$ given in Eq.\eqref{fina} of Appendix \ref{convex}.

The bulk term $I_H$ could have been anticipated by the following argument. Neglecting the gradient terms in the action functional altogether, we have a minimization problem that is locally algebraic and coincides with that of the previous section \ref{hum}. This means that the local contribution to the action functional should match that for homogeneous profiles \eqref{sto}, yet now these profiles can vary in space. Meanwhile the derivation of the sub-leading term in \eqref{Imin0} necessitates retaining the gradient terms in the functional \eqref{act} and is derived in Appendix \ref{convex}.

\subsection{Optimization in states with sharp interfaces: analogy with thermodynamics }\label{opt}

To perform the minimization in \eqref{Imin0}, we again take inspiration from equilibrium statistical mechanics: $I_H(\rho,\aalph)$ can be interpreted as an analogue of a bulk free energy density as a function of two order parameters $\aalph$ and $\rho$ that are spatially conserved due to \eqref{cons4}. This is analogous to globally minimizing the free energy of a ternary fluid mixture, which is a well-studied problem in chemical physics (see, e.g., \cite{andelman_structure_1987}). So long as the interfacial energy term in \eqref{Imin0} is negligible to leading order, then whenever $I_H$ becomes non-convex, the homogeneous solution becomes sub-optimal across a region that encloses the non-convexity (but generally extends beyond it into locally convex regions). The optimal solution then comprises a sharply phase-separated state whose coexisting phases have $(\rho_0,a)$ values found by a common tangent construction over $I_H$. The resulting sum of `bulk free energies' represents the lower convex envelope of $I_H$.  (In standard thermodynamic terminology, this envelope is referred to as the convex hull of $I_H$, we adopt this convention throughout the following.)

For two conserved densities, as here, there can generically be either two phases or three phases in coexistence, but the latter does not arise in our model. Within any region of two-phase coexistence, each globally unstable density point $(\rho_0,a)$ lies on a unique `tie-line', which is one of a family of straight lines each connecting two stable points on the perimeter of that region. Upon global minimization, the uniform state at $(\rho_0,a)$ phase-splits into the two phases that are end-points of its tie line, with phase volumes determined by \eqref{cons4}. The interfacial energy in such a phase-separated state scales as  $\ell$ and is negligible for $\ell\ll1$. Notice however, that for such a sharply phase separated solution, the gradient expansion \eqref{Imin0} breaks down in the region of the sharp interface where higher order gradient terms contribute. Nonetheless, the interfacial contribution is sub-extensive so that the minimized rate function $I$ consists, to leading order in $\ell$, of the convex hull of $I_H$. Wherever $I_H$ does not coincide with its convex hull, inhomogeneous trajectories are preferred to homogeneous ones in \eqref{min}.

\begin{figure}[t]
	\includegraphics[scale=0.4]{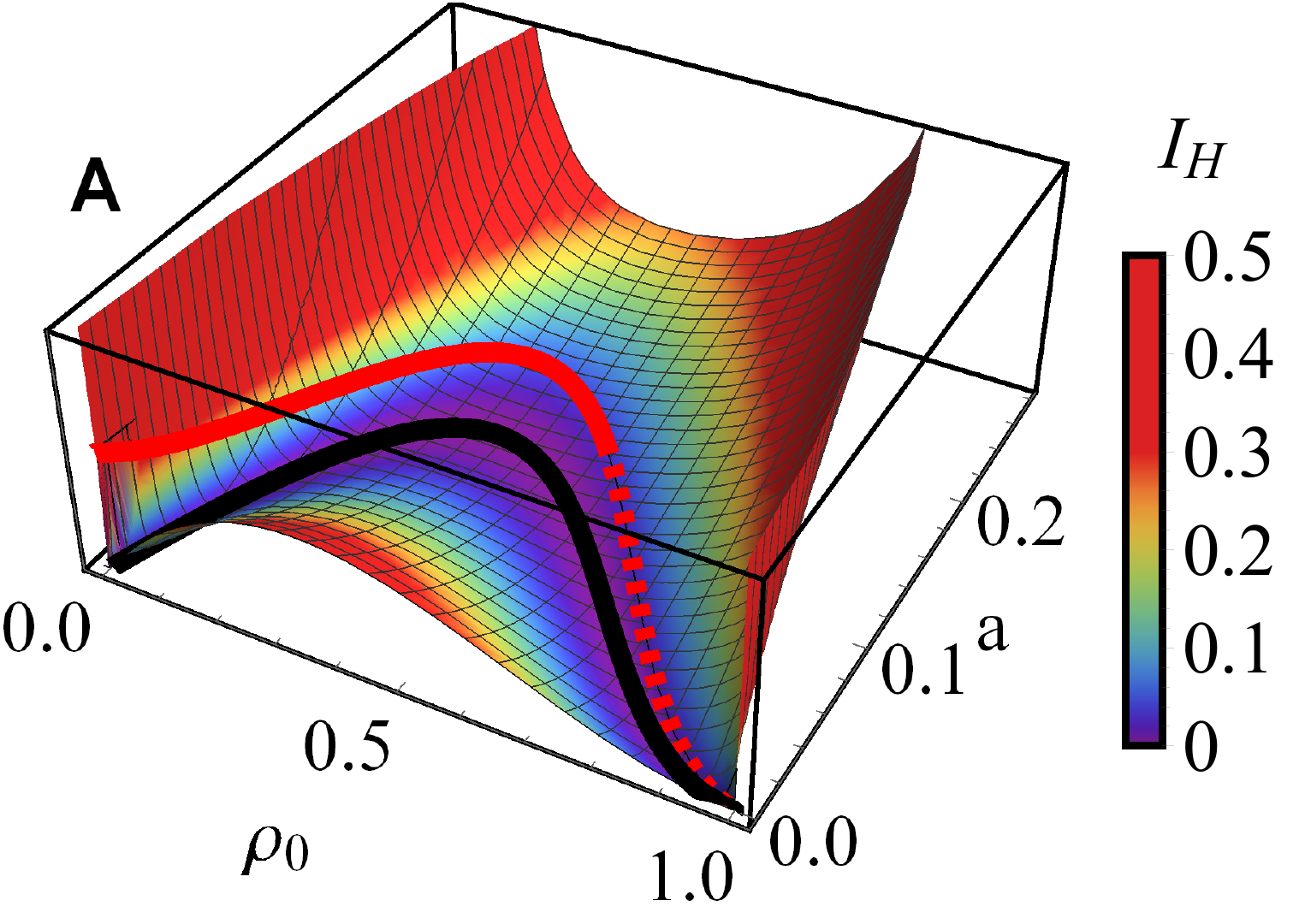}	\includegraphics[scale=0.35]{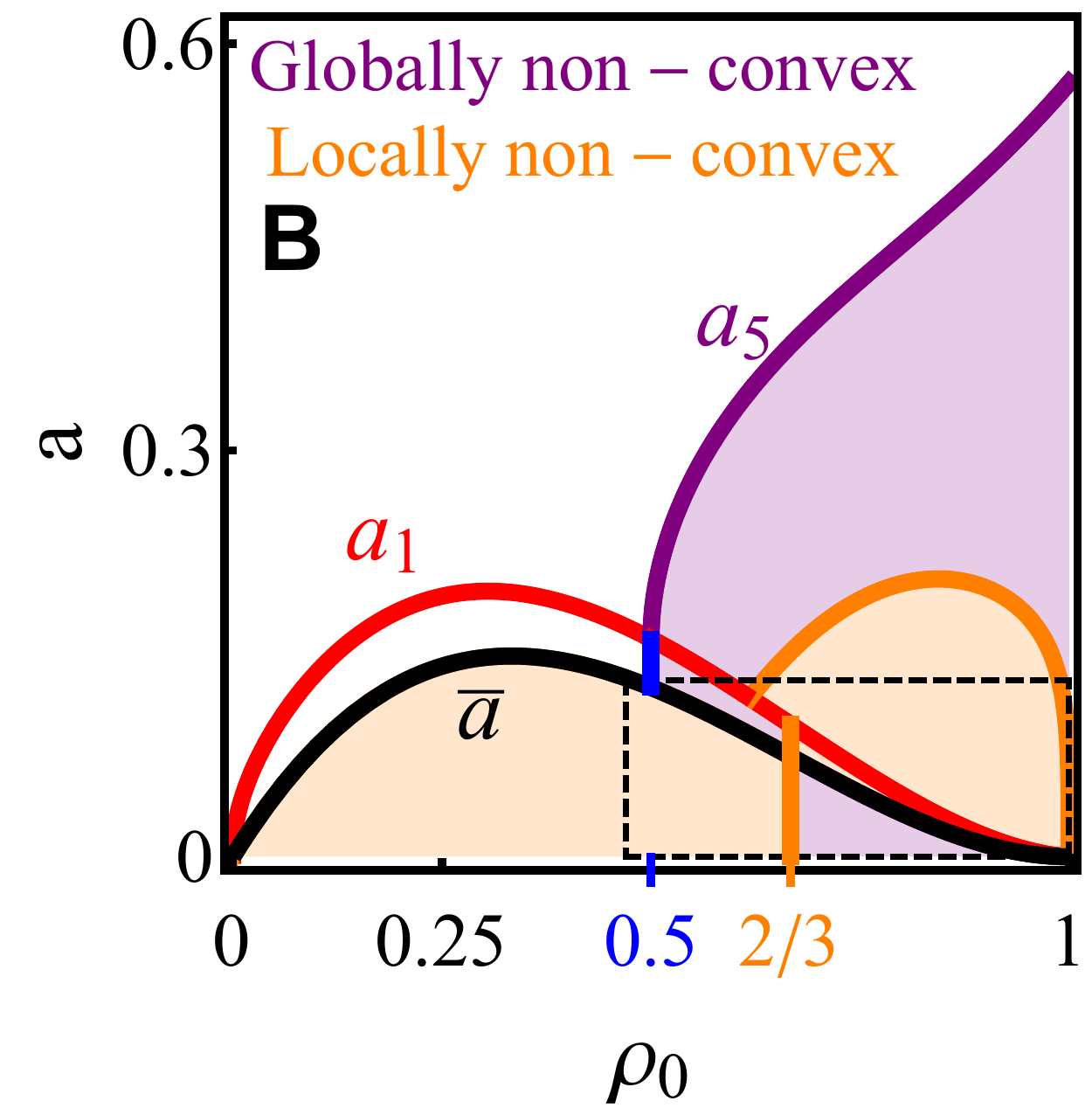}
	\includegraphics[scale=0.35]{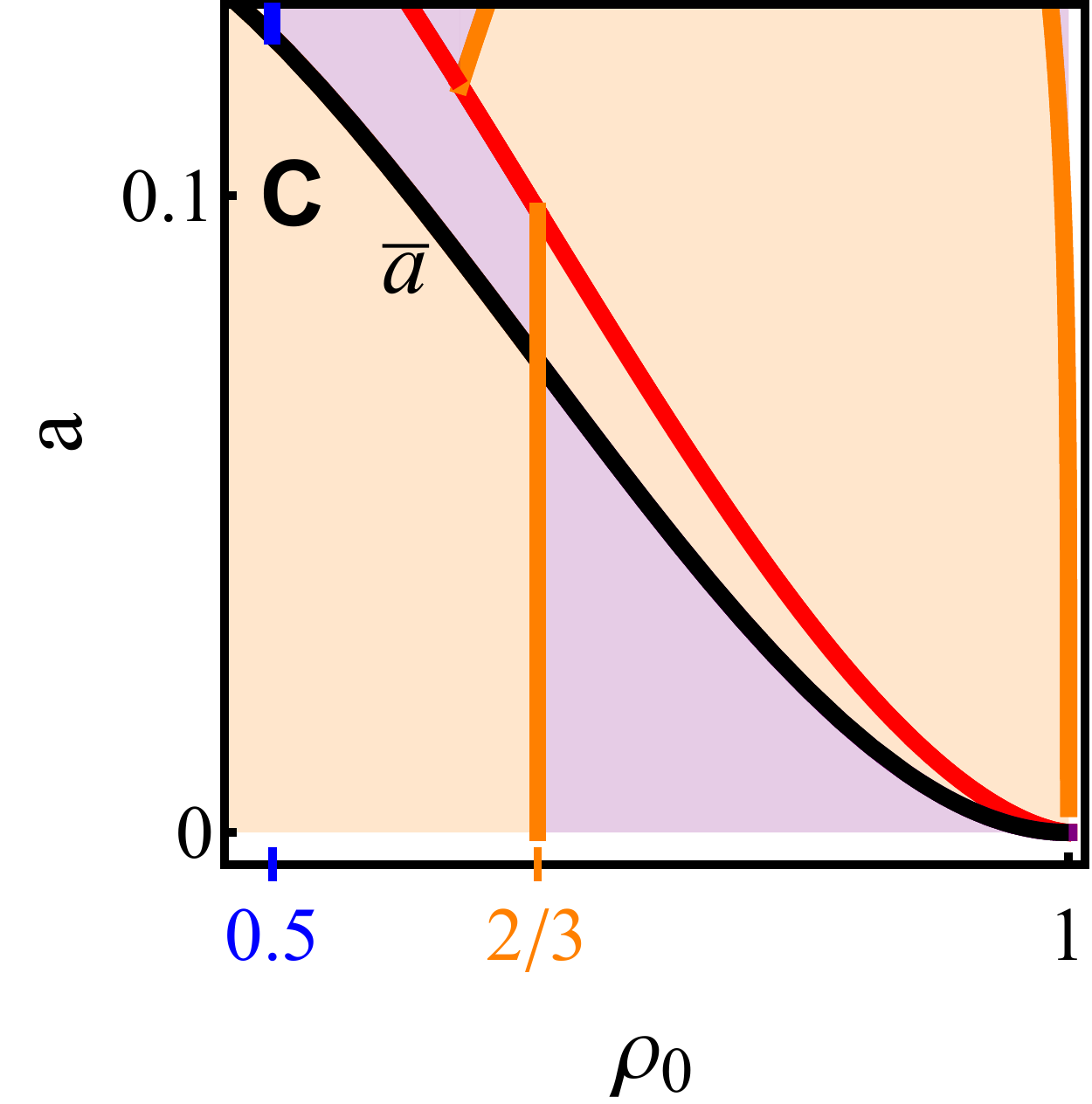}	
	\caption{(A) The function $I_H$ \eqref{sto} at $\text{Pe}=3.5$. The black curve marks the minimum $I_H=0$ obtained for $a=\bar{a}(\rho_0)=\rho_0(1-\rho_0)^2$. This is the locus of unbiased states each of which is homogeneous. The red curve marks the critical line \eqref{ac1} between H and CM phases. It is denoted by a dashed line within the region where $I_H$ becomes non-convex, and the optimal solution is no longer homogeneous. (B) and the blow up (C) show the regions of non-convexity of $I_H$, see \eqref{sto}. Local non-convexity $\text{Det}\left[\text{Hess}(I_H)\right]<0$ in orange, and global non-convexity in purple. See the Appendix \ref{convexrig} for explicit expressions and details of their derivation. The purple curve is given by \eqref{ac5} and denotes the boundary of the TB phase.}
	\label{nonzeromag}	
\end{figure}

The problem is thereby reduced to finding the required convex hull. Notably, $I_H(\rho_0,a)$, is everywhere convex when viewed as a function of the single variable $a$. However, viewed as a two dimensional function over the parameter space $(\rho_0,a)$ it is not everywhere convex. Fig. \ref{nonzeromag}(A) shows a plot of $I_H$ \eqref{sto}.  To analyse the convexity of this function, we separate two ways in which it can be non-convex.  For some values of $(\rho_0,a)$, the function $I_H$ is locally non-convex (its Hessian has at least one negative eigenvalue).  For other values, the Hessian may be positive definite but $I_H$ does not coincide with its convex hull: we refer to such regions as globally non-convex.   All these regions can be found analytically, see Appendix \ref{convexrig}; they are plotted in Fig.~\ref{nonzeromag}(B,C).

Although these arguments assumed stationarity, they can also be easily adapted to the travelling band (TB) phase in which $J_{\rho}\neq0$ and a phase-separated state translates at fixed speed through the system. As a result, the leading order term that controls the rate function $I$ is controlled solely by $I_H$ in the TB phase as well, with the same construction of the convex hull via the common tangent plane. In this case, the tangent construction then describes the two phases which comprise the traveling band state. This is proven in Appendix \ref{convex}. Thus both the PS and TB regions are accessible by this quasi-thermodynamic approach to the minimization in \eqref{Imin0}, essentially because in each case the minimiser constists of two phases separated by a narrow domain wall whose contribution to $I_H$ is sub-leading at small $\ell$. The quantitative results of this procedure are detailed in Section~\ref{pstb} below.

\subsection{The problem of SM states}
\label{sec:sm-problem}

The analysis so far has considered inhomogeneous states with sharp interfaces.  However, the system studied here also supports 
SM (smoothly modulated) phases, where this approach fails. This is attributable to a degeneracy that is generic in the large deviation context, although it would be quite exceptional in a thermodynamic setting.  Namely, the function $I_H(\rho_0,a)$ in \eqref{sto} attains its minimum $I_H=0$ everywhere along the curve $a=\bar{a}(\rho_0)$, see Fig.~\ref{nonzeromag}.  (In the thermodynamic context, one expects instead to find minima of the bulk free energy at one or more isolated points.)  The curve where $I_H=0$ represents the set of unbiased states, one for each $\rho_0$, all of which have zero rate function. As a result, at any point in the two-dimensional parameter space $(\rho_0,a)$ that is engulfed by the convex hull of the curve $a=\bar{a}(\rho_0)$, the rate function drops onto the plane $I=0$ (modulo interfacial contributions). 
The convex hull of the curve $a=\bar{a}(\rho_0)$, refers to the convex hull of the two-dimensional domain $0<a<\bar{a}(\rho_0)$, in the $(\rho_0,a)$ plane: this is the green shaded region in Fig.~\ref{phasefig}(A,B). It is  given by
\begin{equation}
a < \begin{cases} \bar{a}(\rho_0), & \rho_0 \leq (1/2) \\
a_3(\rho_0) &  \rho_0 > (1/2) \end{cases}
\label{equ:not-convex}
\end{equation}
with the line
\begin{equation}
a_{3}(\rho_0)=\frac{(1-\rho_0)}{4}\label{ac3}
\end{equation}
which is tangent to the curve $\bar{a}$ at the point of contact at $\rho_0=1/2$.
 This makes the tangent construction degenerate, as any point on that plane is now shared by infinitely many possible tie lines, each representing a different pair of bulk phases in coexistence. More importantly, in all of these phases the bulk contribution $I_H$ vanishes, so that the interfacial tension terms now dominate. 

To find the optimal solution in the presence of the degeneracy we therefore retain the interfacial tension term in \eqref{Imin0}. We show in Section~\ref{sm} that the solution to this minimization problem is given by a smoothly  modulated solution that is not characterized by a sharp interface, making the minimized rate function scale as $\mathcal O(\ell^2)$.

\section{Results for states with sharp interfaces: PS and TB}\label {pstb}

Having set out the general principles for solving the minimization problem \eqref{min} in the presence of inhomogeneous states, we now analyse
the PS and the TB states in detail.  (Recall that these states contain sharp interfaces between bulk phases.
 They differ qualitatively  in that the TB phase has a nonvanishing mass flux and is nonstationary.)  Both states are found by the common tangent construction over $I_H$. 
 Note that both these states are restricted to densities above half filling, $\rho_0\geq1/2$.  The resulting high-density phase has $\rho(x)=\rho_h$ and the low-density one has $\rho(x)=\rho_l$; the corresponding local EPRs are $a_h,a_l$.  
 
\begin{figure}[t]
	\includegraphics[scale=0.4]{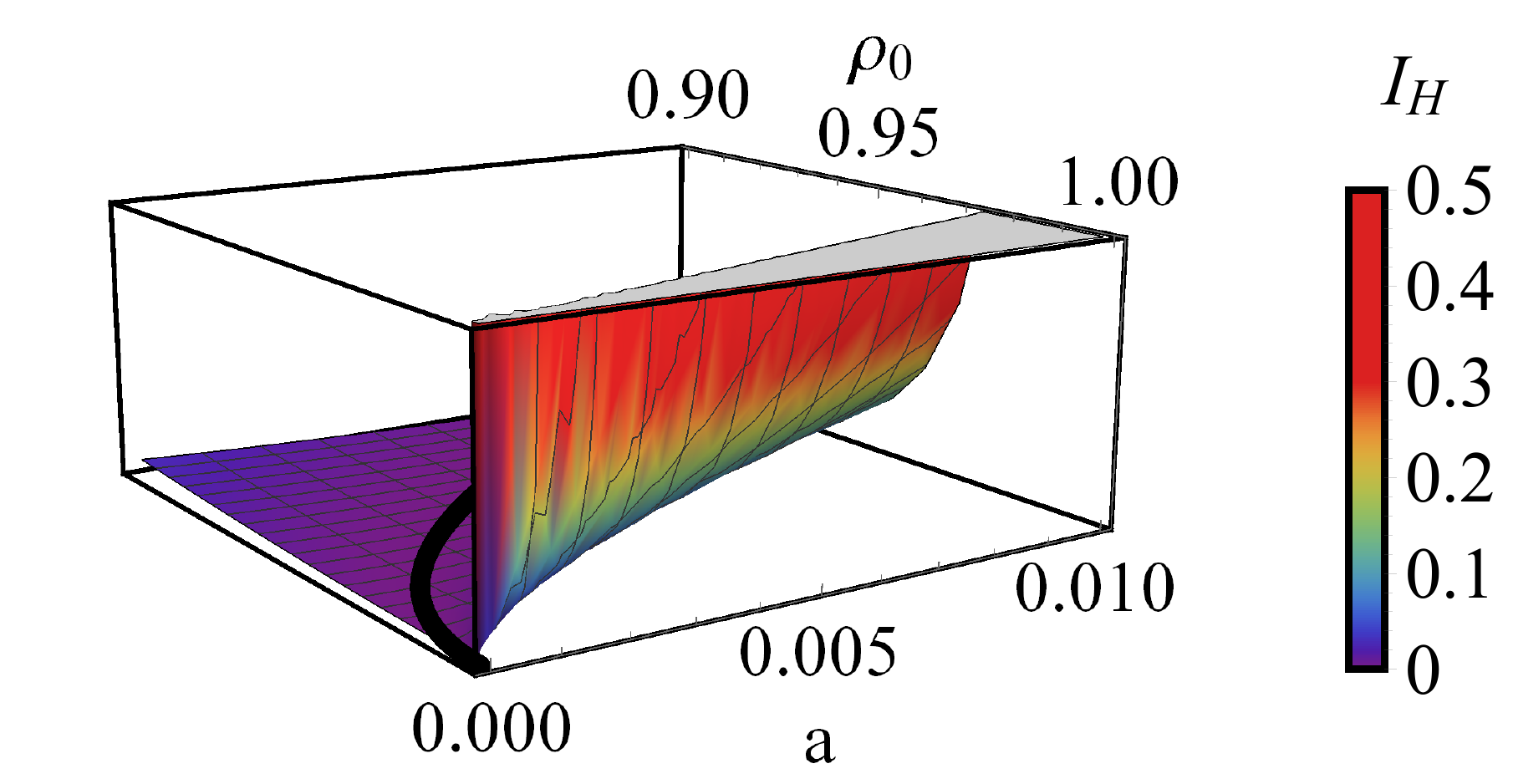}
	\caption{A blow up of the `pinch point' singularity at $(\rho_0=1,a=0)$ of the homogeneous rate function \eqref{sto}}
	\label{pinch}
\end{figure}

It is explained in Appendix \ref{convexrig} that the common tangent construction is dominated by the point $(\rho_0=1,a=0)$ where the function $I_H$ has a `pinch point' singularity: $I_H\sim a^2/(1-\rho_0)^2$, see Fig. \ref{pinch}. Indeed, at close packing $\rho_0=1$, the rate function diverges for any non vanishing IEPR $a\neq0$, as all accessible dynamical moves are now reversible. At the same time, the rate function vanishes at $a=0$ since this is the expected value $\bar a$ when $\rho_0=1$.
This point forms part of the convex hull of $I_H$ and in fact represents the high-density phase for {\em all} coexisting states. (Note that the usual requirement for local convexity at a point of common tangency is bypassed because this point lies at the very edge of the allowed $(\rho_0,a)$ domain; `tangency' here does not constrain the derivatives.) Thus the high-density phase in both the PS and TB phases is always the state of close packing which has vanishing IEPR $(\rho_h=1,a_h=0)$.

 On the other hand, the low-density phase $(\rho_l,a_l)$ is given by different expressions in the PS and TB cases, as follows.
For the phase separated (PS) phase, the tangent construction involves the parabolic branch of the rate function $I_H$ \eqref{sto}. The resulting low density phase is given by half filling ($\rho_l = 1/2$) but with an IEPR that varies with the global density: $a_l=a/2(1-\rho_0)$. 

\begin{figure}[t]
	\includegraphics[scale=0.5]{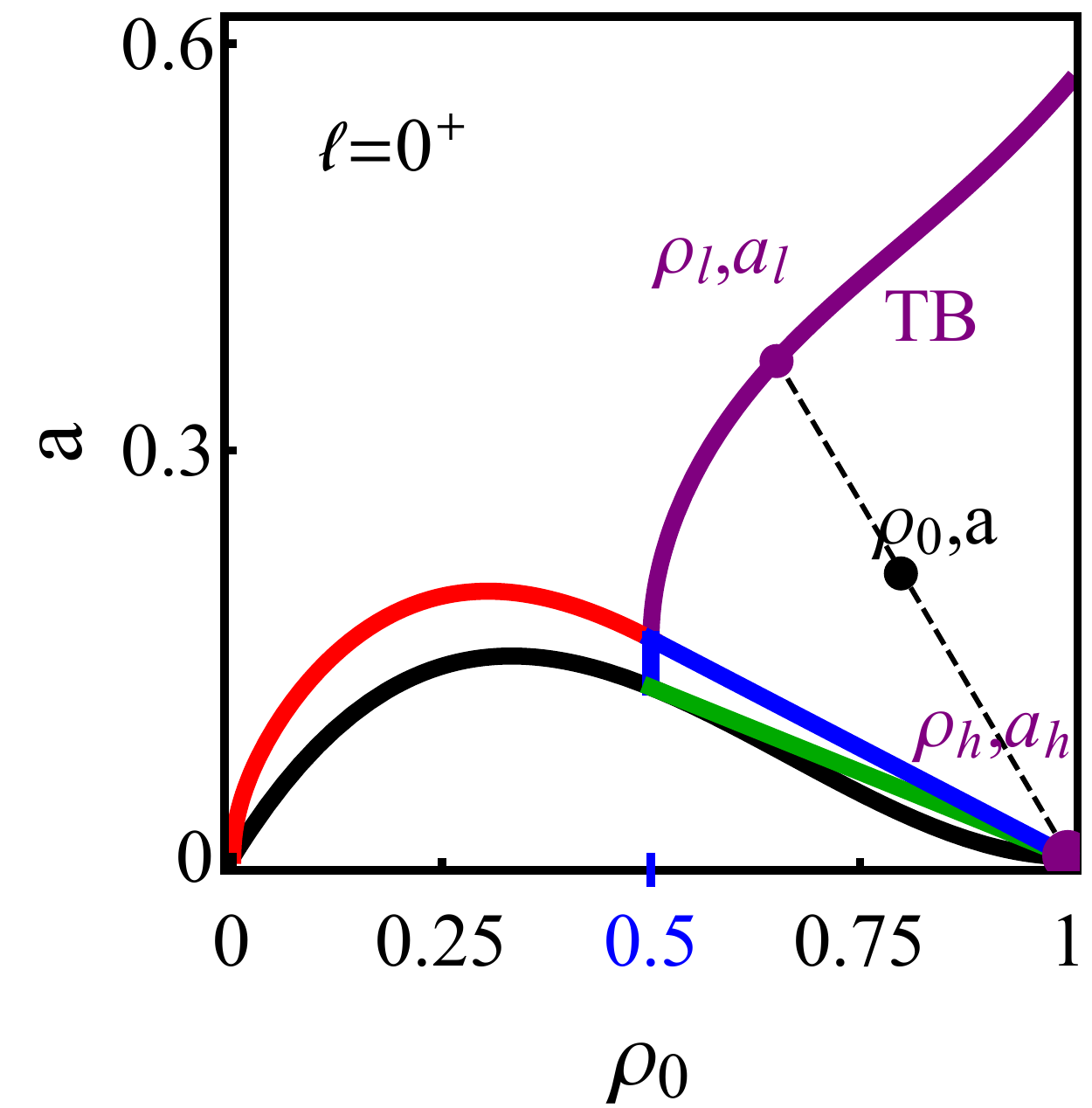}	
	\caption{Geometrical construction for the high and low density phases in the TB region. Here $\text{Pe}=3.5$. }
	\label{tc}
\end{figure}

Finding the low-density phase in the travelling band (TB) phase is more involved. The main result (Appendix \ref{convexrig}) is that the low-density state develops a spontaneous magnetisation when $a>a_4(\rho_0,\text{Pe})$ with 
\begin{equation}
a_{4}(\rho_0,\text{Pe})=\frac{1}{4}(1-\rho_0)\sqrt{1+\frac{8}{\text{Pe}^2}}\label{ac4}
\end{equation}
The resulting symmetry-broken state is a TB: it corresponds to a low-density CM region, while the high-density region remains symmetric ($m=0$) with $\rho_h=1$. The properties of the low-density phase are $(\rho_l,m_l,J_{\rho,l},J_{m,l})$; they can be computed by constructing the relevant tie-line on the $(\rho_0,a)$ plane as we detail in Appendix \ref{globconc}. Since the high density phase is invariant, this is simpler than in most equilibrium problems.  We first construct a line connecting the high density phase at $(\rho_h=1,a_h=0)$ to the point of interest $(\rho_0,a)$ set by the chosen global mean values for the density and IEPR. Now extend this line until it intersects the critical curve
\begin{equation}
a_{5}(\rho_0,\text{Pe})=\frac{\rho_0+\sqrt{2\rho_0-1}}{2}\sqrt{(1-\rho_0)^2+\frac{2\sqrt{2}(1-\rho_0)}{\text{Pe}^2\,\sqrt{\rho_0-\sqrt{2\rho_0-1}}}},\label{ac5}
\end{equation}
which marks the boundary between the TB and CM phases. This curve is derived in Appendix \ref{globconc}. The intersection point marks the low-density phase, $(\rho_l,a_l)$, whose values then determine $m_l$ via \eqref{mag}, and the fluxes via \eqref{flux} (see Fig.~\ref{tc}).
As in the PS state (or any equilibrium-like phase coexistence), the relative amounts of the two coexisting phases is set by global conservation such that as the point of interest approaches one end of the tie-line, the amount of the other phase vanishes linearly. Upon crossing the curve $a_5(\rho_0)$ set by \eqref{ac5} this causes a jump in the second derivative of the rate function \eqref{f}.

Until now we did not discuss the dynamical equations \eqref{eq:rho} and \eqref{eq:m} that must be satisfied by any phase-separated solution. These are trivially met in the bulk of the phases, and also at the interfaces in the PS state where there is no mass flux in either coexisting phase. However, in the TB state, a stationary solution would violate the conservative equation \eqref{eq:rho} since the high density phase has zero mass flux, while the low density phase has a nonzero mass flux. To resolve this issue, the entire phase separated solution must propagate with velocity
\begin{equation}
V=-\frac{J_{\rho,l}}{1-\rho_l},
\end{equation}
as indicated in Fig.~\ref{profiles}. Meanwhile the dynamical constraint \eqref{eq:m} can always be accommodated with the help of the additional tumbling rate degree of freedom $K$  which is nonvanishing across the sharp interface. This results in a negligible $\mathcal O(\ell)$ contribution to the rate function that is negligible here.

\section{Results for the smoothly modulated (SM) phase }\label{sm}

The last region of the phase diagram to be analysed is the SM state, which occurs in the green shaded regions of Fig.~\ref{phasefig}(A,B).  As discussed in Sec.~\ref{sec:sm-problem}, this state occurs in regions where the common tangent construction is degenerate, which means that the behaviour as $\ell\to0$ cannot be deduced by considering bulk phases separated by sharp interfaces.   At the level of the minimisation problem \eqref{Imin0}, this can be seen by considering profiles such that the local entropy production is (almost) typical for the local density 
 \begin{equation}
\aalph(x)=\bar{a}(\rho(x))+o(\ell) \; . \label{choise}
\end{equation}
Inserting any such profile in \eqref{Imin0} yields $I(a)=O(\ell^2)$, where $a=\int_0^1 \aalph(x) dx$.  That is, the ``interfacial'' (gradient) terms in \eqref{Imin0} determine which density profile minimises the action.

We therefore assume \eqref{choise}, so the minimization problem \eqref{Imin0} reduces to a single field minimization
\begin{equation}
I(a)\simeq \ell^2 \inf_{\rho(x)}\int_0^1 \!dx\, M(\rho)(\partial_x \rho)^2,\label{Imin3}
\end{equation}
where $M(\rho)$ is derived in Appendix \ref{convex}, as
\begin{equation}
M(\rho)=\frac{[2+\text{Pe}^2\,(1-\rho)(1-2\rho)]^2}{8\rho(1-\rho)[2+\text{Pe}^2(1-\rho)]} \; ,\label{m}
\end{equation}
and the constraints \eqref{cons4} become
\begin{equation}
\rho_0 = \int_0^1 dx\, \rho(x) \qquad , \qquad a=\int_0^1dx\, \bar{a}(\rho(x)).
\label{equ:con-aalph}
\end{equation}  
It can be verified that $M(\rho)\geq 0$. Notice that the spatial averages \eqref{equ:con-aalph} over any density profile $\rho_0(x)$ are bounded within the convex hull of the curve $\bar{a}(\rho_0)$ which comprises the SM region \eqref{equ:not-convex}. 

The next step is to analyse the single-field minimisation problem \eqref{Imin3}.  This can be performed numerically (indeed, that method was used to generate the SM profiles in Fig.~\ref{profiles}).  However, we will show that analytic progress is also possible, and allows phase transitions to be characterised.
The integral constraint $a = \int \aalph(x) dx$ makes the analysis more challenging so we employ a standard tactic in large deviation theory \cite{touchette_large_2009-1} which corresponds to a change of ensemble.  This is equivalent in this case to using a Lagrange multiplier to enforce the constraint.

\subsection{Scaled cumulant generating function}\label{canonical}

To deal with the constraint, we consider a scaled cumulant generating function (CGF)~\cite{jack_ergodicity_2020-1,touchette_large_2009-1}
  for a suitably rescaled version of the IEPR $\mathbb S_T$,
\begin{equation}
\psi(\Lambda) = \frac{1}{\gamma LT}\ln\left\langle e^{T\Lambda \mathbb S_T/(\gamma\text{Pe}^2) }\right\rangle,\label{cgf}
\end{equation}
where the average is over the system dynamics.  Since individual paths obey the large deviation principle \eqref{equ:path-LDP}, the average can be computed by the saddle point method, which gives
\begin{equation}
-\psi(\Lambda) =  \frac{1}{\gamma T}\inf_{\cal X} \left[ {\mathcal A}(\mathcal X) - \Lambda \mathbb{S}(\mathcal{X}) T/(\gamma\text{Pe}^2 L) \right]
\label{equ:psi-min}
\end{equation}
Comparing with \eqref{min}, we have exchanged a constrained minimisation over ${\cal X}$ for a ``biased minimisation'' that includes the Lagrange multiplier $\Lambda$.   If one now optimises over trajectories ${\cal X}$ with $\mathbb S_T({\cal X})=a L\gamma\text{Pe}^2$, one sees from \eqref{min} that 
\begin{equation}
\psi(\Lambda)=\sup_a\left[\Lambda a-I(a)\right] \; ,\label{leg}
\end{equation}
which we recognise as a Legendre-Fenchel transform.  (In the context of large deviation theory, this is Varadhan's lemma \cite{touchette_large_2009-1}, which relates the CGF of $a$ to its rate function.)
In the homogeneous state where $a<a_1$ one has the Legendre transform of $I_H$
	\begin{equation}
	\psi_H(\Lambda)=\left(\frac{\Lambda^2}{\text{Pe}^2}+\Lambda\right)\bar{a}(\rho_0) \; .
	\label{equ:psi-H}
	\end{equation}
Computation of $\psi(\Lambda)$ also allows reconstruction of the rate function by inverse Legendre transform as
\begin{equation}
I(a)=\sup_\Lambda\left[\Lambda a-\psi(\Lambda)\right] \; .\label{GE}
\end{equation}

If the derivative $\psi'(\Lambda)$ has discontinuities, some care is required with this transformation~\cite{touchette_large_2009-1}.  Such situations correspond to first-order dynamical phase transitions.  In the system considered here, \eqref{GE} does hold even in the presence of such phase transitions; this leads to a common-tangent construction for $I(a)$, associated with time-like phase separation as we find in the subsequent Sec.\ref{bls}.

\subsection{Stability of homogeneous state : Landau-like theory}

We now consider the minimisation problem \eqref{equ:psi-min} and we analyse the stability of the homogeneous state.  The aim is to characterize the system state in the SM phase and the associated phase transition. In contrast to previous Sections where these were computed in the limit $\ell\to0$, here these are determined by the sub-leading  $\mathcal O(\ell^2)$ corrections. As a side result we will derive also the boundaries of the (green) SM regions in Fig.~\ref{phasefig} up to $\mathcal O(\ell^2)$.

To this end, we construct a variational ansatz for \eqref{equ:psi-min}.
It is sufficient to consider trajectories $\cal X$ that are homogeneous in time (consistent with the additivity principle).  
For parameters in the non-convex region given by \eqref{equ:not-convex}, we also take the local entropy production rate to satisfy \eqref{choise}.  Then the action ${\cal A}$ in (\ref{equ:psi-min}) can be approximated as in (\ref{Imin3}) by 
${\cal A}(\mathcal X)=\ell^2 \int dx M(\rho) (\partial_x \rho)^2$.  Since the action is $O(\ell^2)$ it is convenient to define 
$\tilde{\Lambda}=\Lambda/\ell^2$ and $\tilde{\psi}(\tilde\Lambda)=\psi(\ell^2\tilde\Lambda)/\ell^{2}$.  Eq.~(\ref{equ:psi-min}) finally becomes
\begin{eqnarray}
-\tilde{\psi}(\tilde\Lambda)=\inf_{\rho(x)} \int_0^1dx\left[M(\rho)(\partial_x \rho)^2-\tilde{\Lambda}\bar{a}(\rho(x))\right] . \label{square}
\end{eqnarray}
where the minimisation is subject (as usual) to $\rho_0 = \int_0^1 dx \rho(x)$.
Now consider a variational ansatz for the density profile, as:
\begin{equation}
\rho(x) = \rho_0 + A\cos(2\pi x) + B\cos(4\pi x) \; .
\label{lin}
\end{equation}
 This ansatz is similar to that of \cite{bodineau_cumulants_2007,lecomte_inactive_2012,baek_dynamical_2018,dolezal_large_2019}.
It is sufficient in this context because the instabilities of the homogeneous state occur in the largest wavelength term (with coefficient $A$), but it will be necessary to consider also the second harmonic (with coefficient $B$), to accurately describe the behaviour of $A$ beyond leading order.

Injecting \eqref{lin} into \eqref{square}, the functional minimisation over $\rho$ reduces to a minimisation over $A$ and $B$.  Details of the computation are given in appendix \ref{land}.  For homogeneous profiles ($A=B=0$), one reproduces the linear term in the homogeneous CGF $\psi_H$ \eqref{equ:psi-H}, as the quadratic part is sub-leading $\mathcal O(\ell^4)$ in the scaling regime $\Lambda\sim\ell^2$.
Assuming now that $A$ and $B$ are small, one finds that $B=O(A^2)$ and this parameter may be minimised analytically. 
Collecting terms up to quartic order in $A$, \eqref{square} becomes 
\begin{equation}
\tilde\psi(\tilde\Lambda) 
= \tilde\psi_H(\tilde\Lambda_c) - \min_A \Psi(A) \,.\label{landau-full}
\end{equation}
where
\begin{equation}
\Psi(A) = 
(\tilde\Lambda_c-\tilde\Lambda) \bar{a}''(\rho_0) \frac{A^2}{4}+\frac{\beta(\rho_0,\text{Pe})}{2} A^4+\mathcal O(A^6)
\label{landau-Psi}
\end{equation}
is a Landau-like free energy with $\tilde{\Lambda}_c=8\pi^2M(\rho_0)/\bar{a}''(\rho_0)$. A formula for $\beta$ is given in Eq.\eqref{theo} of Appendix \ref{land}.

The boundaries of the SM state are obtained by computing values of $(\rho_0,\text{Pe},\tilde\Lambda)$ for which $\Psi$ is minimised by a value $A\neq0$, corresponding to an inhomgeneous density. Since $\Psi$ is known to $\mathcal O(A^4)$ we can perform this calculation for small $A$.
The behaviour of this Landau-like theory depends on the signs of $\beta(\rho_0,\text{Pe})$ and $\bar{a}''(\rho_0)$, as well as that of $(\tilde\Lambda-\tilde\Lambda_c)$. We separate two cases according to the sign of $\beta$.

\subsection{Linear instabilities and second-order phase transitions ($\beta>0$)}

\begin{figure}[t]
	\includegraphics[scale=0.4]{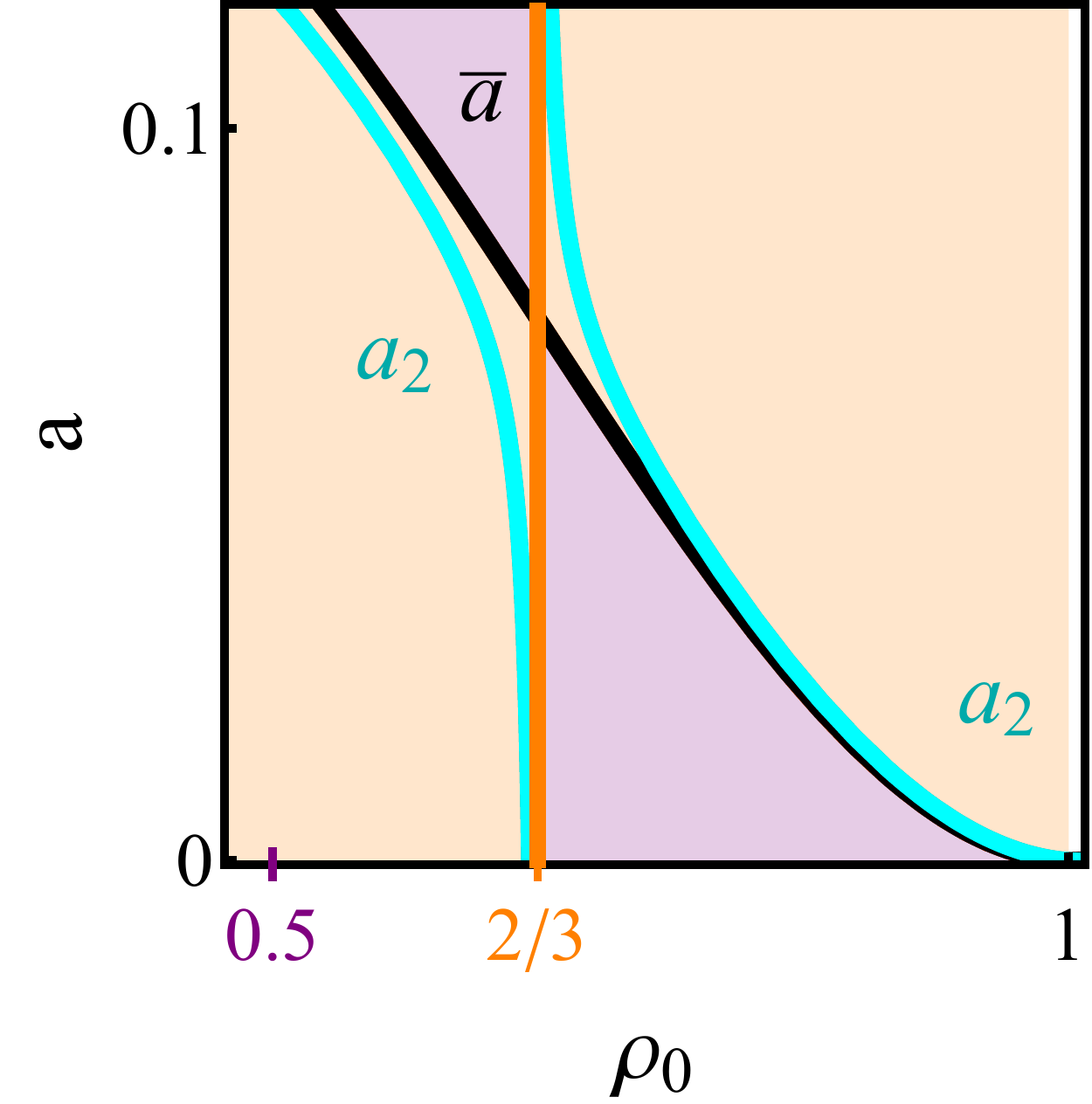}	
	\caption{Illustration of the boundaries of linear instability dented by cyan line. They are given by the curve $a_{2}(\rho_0)$ \eqref{ac2}. As $\ell\to0$ these tend to the boundaries of local non-convexity of $I_H$ denoted by orange shading. The purple shading denote regions where $I_H$ is globally non-convex yet locally convex, see also Fig. \ref{nonzeromag} (B) and (C). The black curve is the expected IEPR curve \eqref{ent2}. Here $\text{Pe}=2.5$ and $\ell=0.1$.}
	\label{phasefig3}	
\end{figure}

Assume first that $\beta>0$.  In this case, analysing the stability of the homogeneous state only requires that we consider \eqref{landau-full} up to quadratic order in $A$.
From \eqref{landau-full}, the system will be inhomogeneous ($A\neq0$ and $\tilde\psi > \tilde\psi_H$) if
\begin{equation}
(\tilde\Lambda-\tilde\Lambda_c) \bar{a}''(\rho_0) < 0. 
\label{equ:inhom}
\end{equation}
From \eqref{ent2} then
\begin{equation}
\bar{a}''(\rho_0) = 6\rho_0 - 4
\end{equation}
which changes sign at $\rho=2/3$.
Note also that $\tilde\Lambda_c$ has the same sign as $\bar{a}''(\rho_0)$.  Using these facts with \eqref{equ:psi-H}, the result is that the homogeneous phase has two regions of linear instability, which must correspond to $A\neq0$, see also Appendix \ref{secondvar}.  The first occurs for $\rho_0<2/3$, it is $a<a_2(\rho_0,\text{Pe},\ell)<\bar{a}(\rho_0)$ with 
 \begin{equation}
a_{2}(\rho_0,\text{Pe},\ell)=\left[1+\ell^2\frac{8\pi^2M(\rho_0)}{\text{Pe}^2\,(3\rho_0-2)}\right]\bar{a}(\rho_0).\label{ac2}
\end{equation}
The second linearly unstable region occurs for $\rho_0>2/3$ and corresponds to $a>a_2(\rho_0,\text{Pe},\ell)>\bar{a}(\rho_0)$, where $a_2$ is given by the same formula (but note this function has two branches).
See Fig.~\ref{phasefig3} and Appendix \ref{secondvar} for further details. 
In both cases, the minimum of the Landau-like free energy occurs at
 $A= \bar{A}$ with
 \begin{equation}
\bar{A} =  \sqrt{\frac{(\tilde\Lambda_c-\tilde\Lambda)\bar{a}''(\rho_0)}{4\beta(\rho_0,\text{Pe})}} .
\label{equ:bar-A}
\end{equation}
This is a continuous transition, with $\bar{A}\sim |\tilde\Lambda_c-\tilde\Lambda|^{1/2}$.

To put these results in context, we note from \cite{lecomte_inactive_2012,baek_dynamical_2018,dolezal_large_2019,jack_hyperuniformity_2015} that for functional minimisations of the form \eqref{square}, one expects inhomogeneous states to form for either positive or negative values of the biasing parameter $\Lambda$, according to the sign of $a''$.  In contrast to previous studies, the function $a''$ considered here changes sign, leading to two separate regions of instability.
We will see in the following that the situation is in fact more complex than this linear analysis can reveal.

As anticipated above, one also sees that for $\ell\to 0$, the critical curve \eqref{ac2} tends to the boundary of the region where $I_H$ is locally convex, which is given by $(2/3-\rho_0)(a-\bar{a}(\rho_0))<0$, see Fig.~\ref{phasefig3}. This arises from \eqref{ac2} by virtue of the fact that the $\ell^2$ piece has coefficient $\pm\infty$ for $\rho_0 \to 2/3$, shown as a vertical orange line in Fig.~\ref{phasefig3}, so that as $\ell\to 0$, the stability boundary is the union of this vertical line with the unbiased IEPR curve $\bar{a}(\rho_0)$. 
In consequence, at $\rho_0<2/3$ the solution is unstable for vanishingly small {\em suppressions} of the entropy production,  $a<a_{2}=\bar{a}-\mathcal O(\ell^2)$, while at $\rho_0 >2/3$ it is unstable for vanishingly small {\em enhancements}, $a>a_{2}=\bar{a}+\mathcal O(\ell^2)$.

\subsection{Beyond linear stability ($\beta<0$): First-order phase transition and  timelike phase coexistence}\label{bls}
 
To go beyond linear stability analysis, we return to the Landau-like free energy \eqref{landau-Psi}.  If the coefficient $\beta<0$ [but assuming that the theory is ultimately stabilised by terms at $O(A^6)$ or higher], this minimisation problem can support first-order phase transitions.  That is, while the homogeneous state may be a local minimum of $\Psi$, there may be additional minima corresponding to SM states.  On varying the parameters, the global minimum can then change discontinuously from homogeneous to SM, in direct analogy with first-order thermodynamic phase transitions. 
Such transitions occur while the homogeneous state is locally stable: they cause the SM states to appear for {\em even smaller} IEPR deviations than \eqref{ac2} would predict, see also \cite{bodineau_cumulants_2007}.

\begin{figure}[ht]
	\includegraphics[scale=0.26]{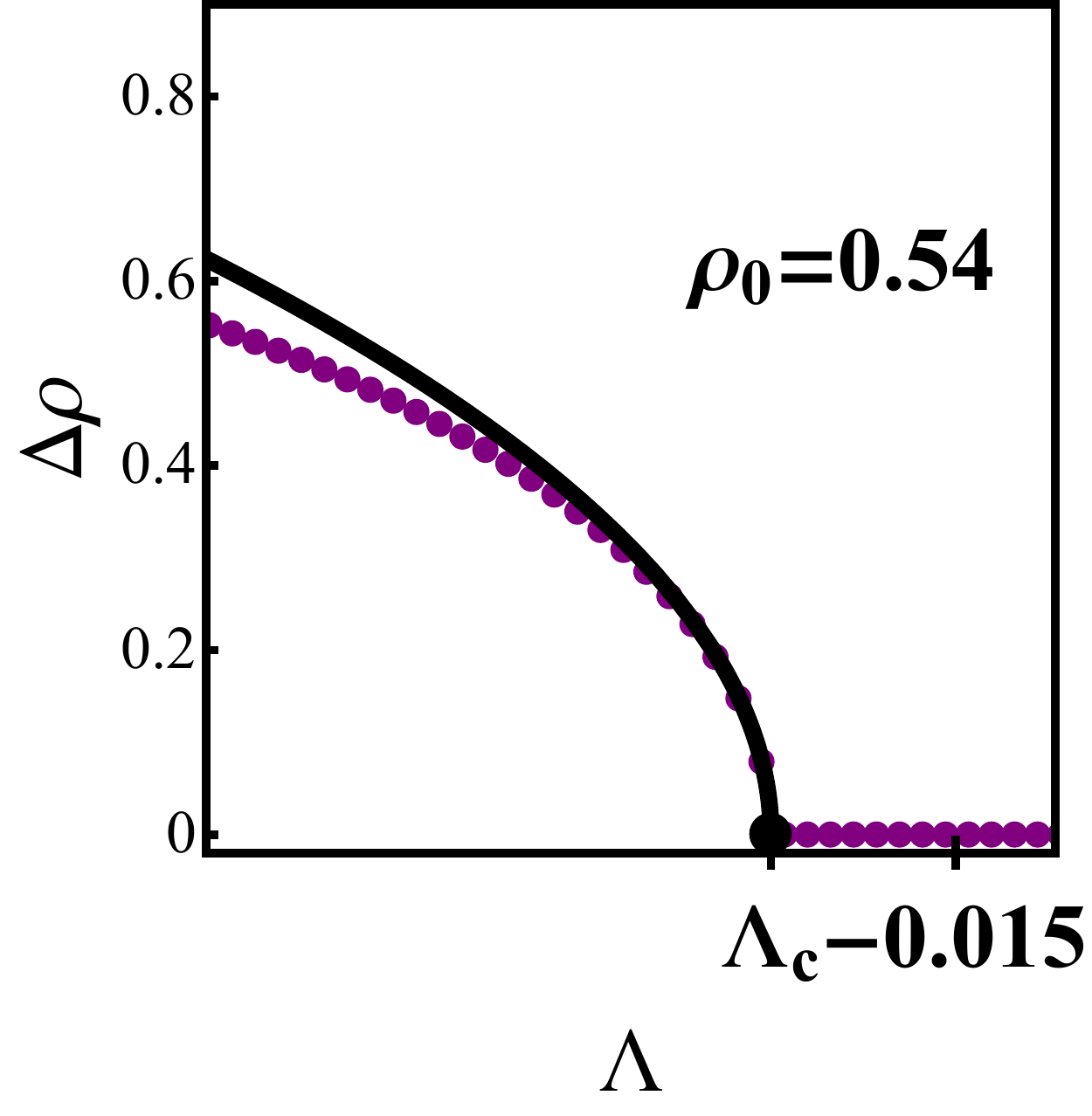}
	\includegraphics[scale=0.27]{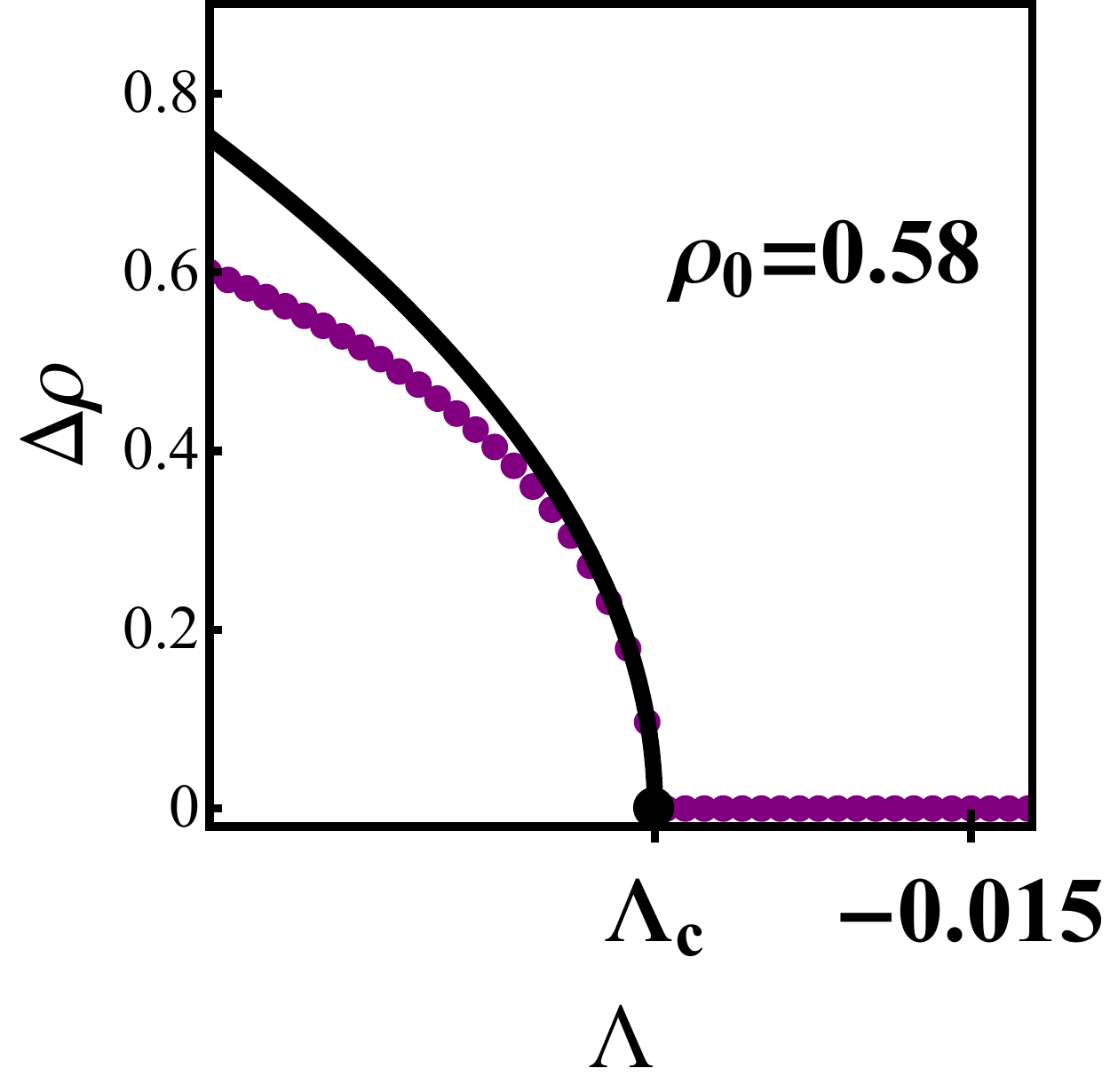}
	\includegraphics[scale=0.27]{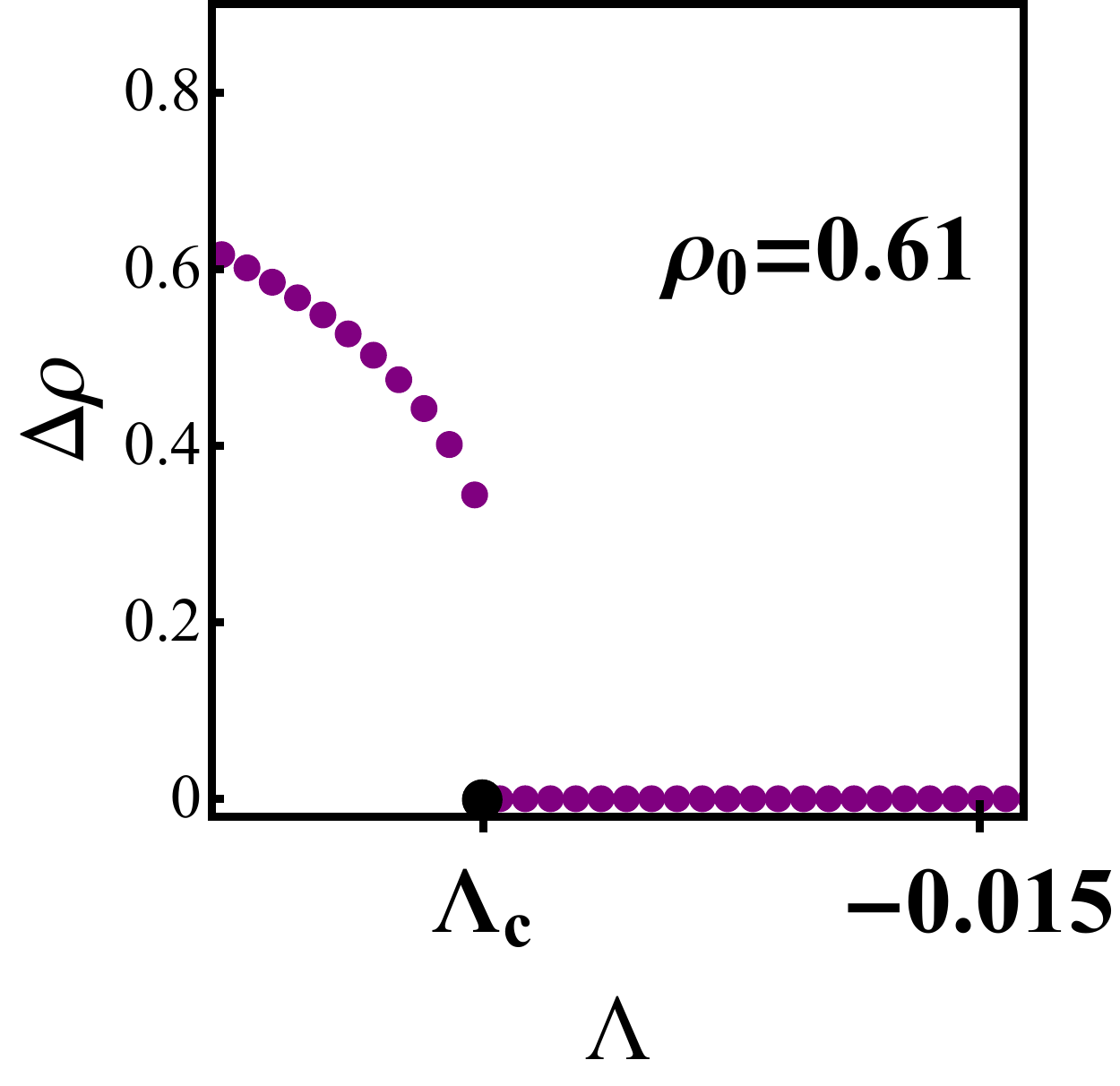}
	\includegraphics[scale=0.27]{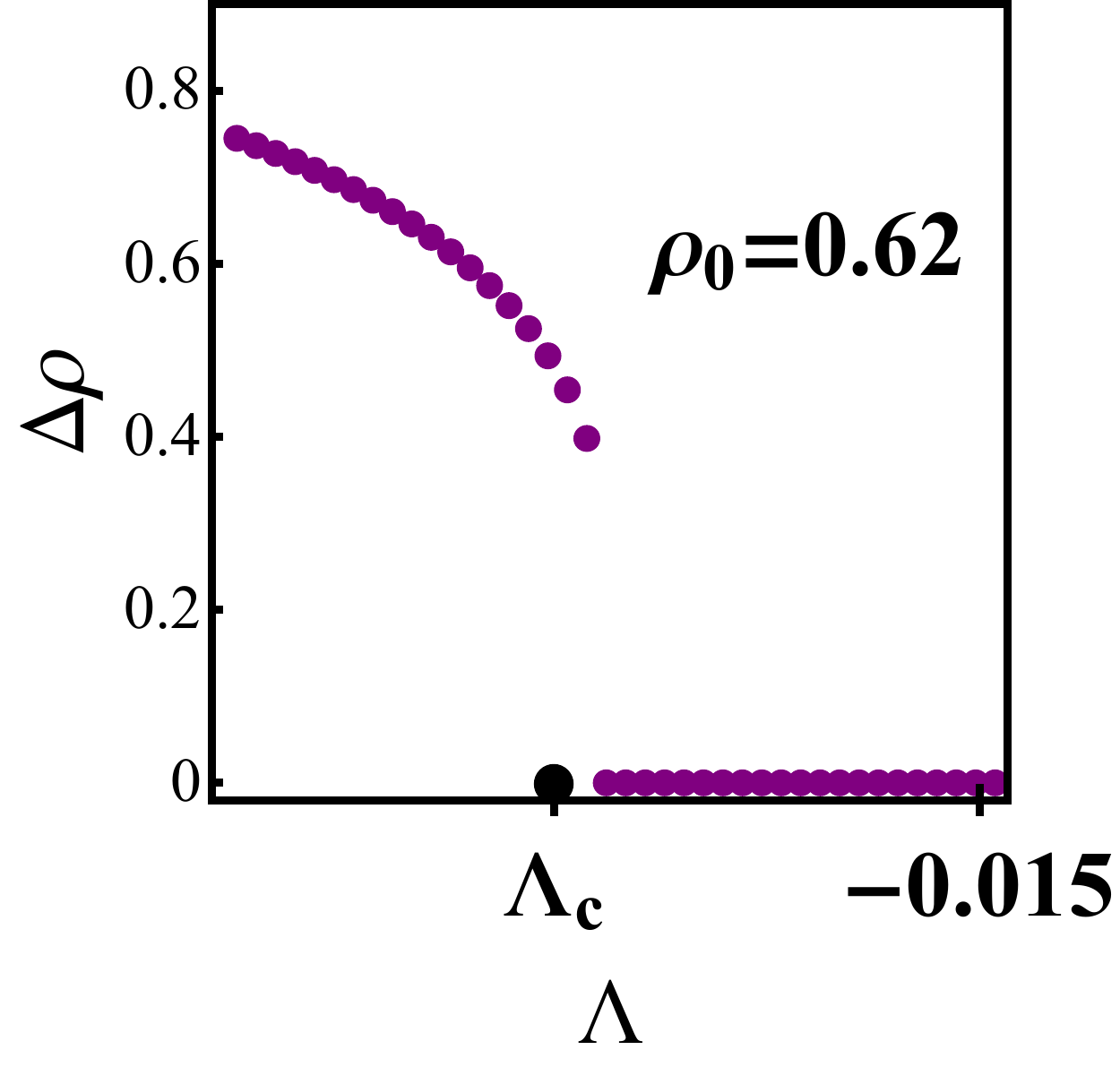}
	\includegraphics[scale=0.27]{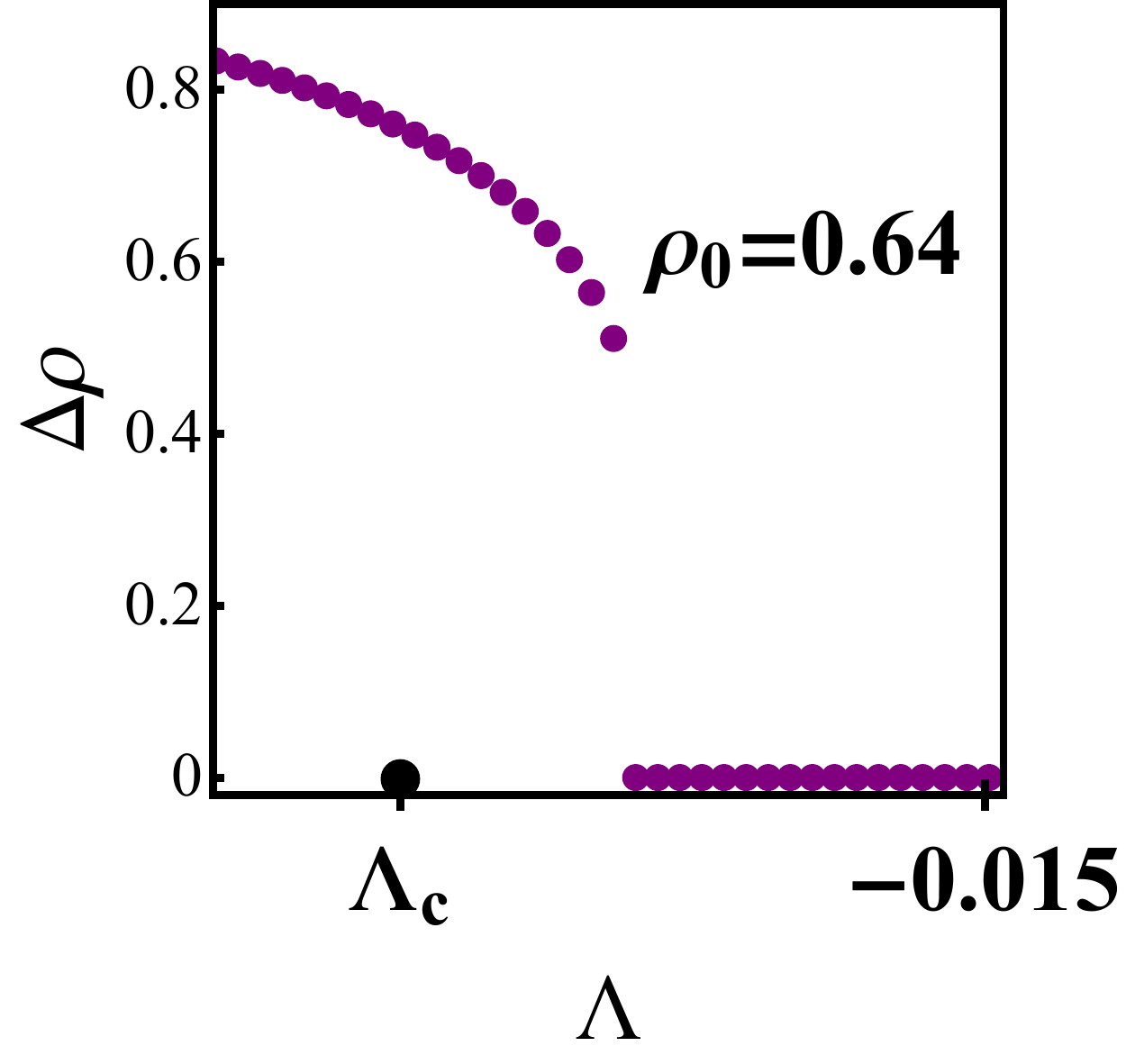}
	
	\caption{The spatial modulation amplitude $\Delta\rho=(\rho_{\text{max}}-\rho_{\text{min}})$, at different values of averaged density $\rho_0$ in purple points. these are obtained from numerical solution of \eqref{square}. The black dot marks the critical value for linear instability. The black curves are the theoretical prediction close to criticality $\Delta\rho=2A$ with $A$ given by \eqref{equ:bar-A}. The point $\rho_0=0.58$ is just below the tricritical value $\rho_{c1}=0.600\dots$, while $\rho_0=0.61$ is just above it. For densities $\rho_0<\rho_{c1}$ the transition is continues and second order. For densities $\rho_0>\rho_{c1}$ the transition is first order and occurs before the critical value for linear instability $|\Lambda_1|<|\Lambda_c|$. Here $\text{Pe}=2$ and $\ell=0.02$.  }
	\label{delta3}	
\end{figure}

 To see this, recall (\ref{equ:bar-A}).   For $\beta>0$ we have a continuous phase transition, as noted above.
 For $\beta<0$, higher-order terms in $\Psi$ must be considered, in order that the minimum can be found.  
 However, the order of the phase transition can be deduced without computing these terms.  For convenience, consider the case $\rho_0>2/3$  so that $\bar a''(\rho_0)>0$.  Suppose that $\beta<0$ and $\tilde\Lambda$ is increased from zero towards $\tilde\Lambda_c$.  For $\tilde\Lambda=0$ the free energy $\Psi$ has its global minimum at zero.  At $\tilde\Lambda=\tilde\Lambda_c$, this minimum becomes unstable: the negativity of $\beta$ shows that the global minimum of the Landau-like free energy must be already at $A\neq 0$.  That is, the global minimum of $\Psi$ must become non-zero for some $\tilde\Lambda_1<\tilde\Lambda_c$, at which point the local minimum at zero is still locally stable. This corresponds to a first-order (discontinuous) phase transition with a jump discontinuity in the magnitude $A$. 
  For
 $\rho_0<2/3$, a similar analysis holds, except that $\tilde\Lambda_c<0$ so one must consider the behaviour on reducing $\tilde\Lambda$ towards $\tilde\Lambda_c$. Within the Landau theory, this argument shows that points with $(\Lambda_c,\beta)=(\tilde\Lambda_c,0)$ are tricritical.\footnote{%
 Technically, they might be critical points of even higher order, since we have not checked the positivity of the $A^6$ terms in \eqref{landau-Psi}, but this does not affect whether the transition is continuous or discontinuous, which is our main concern here.}  We find that for $\text{Pe}<\text{Pe}^*\simeq2.641$ the Landau theory supports two such points $\rho_{c1,2}$, one at $\rho_{c1}<2/3$ and one at $\rho_{c2}>2/3$. For $\text{Pe}>\text{Pe}^*$ additional points appear on both sides of $\rho_0=2/3$, see also Sec.\ref{mipsrel}. Fig. \ref{delta3}, shows the jump discontinuity in the spatial modulation amplitude $\Delta\rho=\rho_{\text{max}}-\rho_{\text{min}}$
which grows smoothly when advancing along the first order transition line away from the tricritical point $\rho_{c1}$.  

 \begin{figure}[t]
	\includegraphics[scale=0.5]{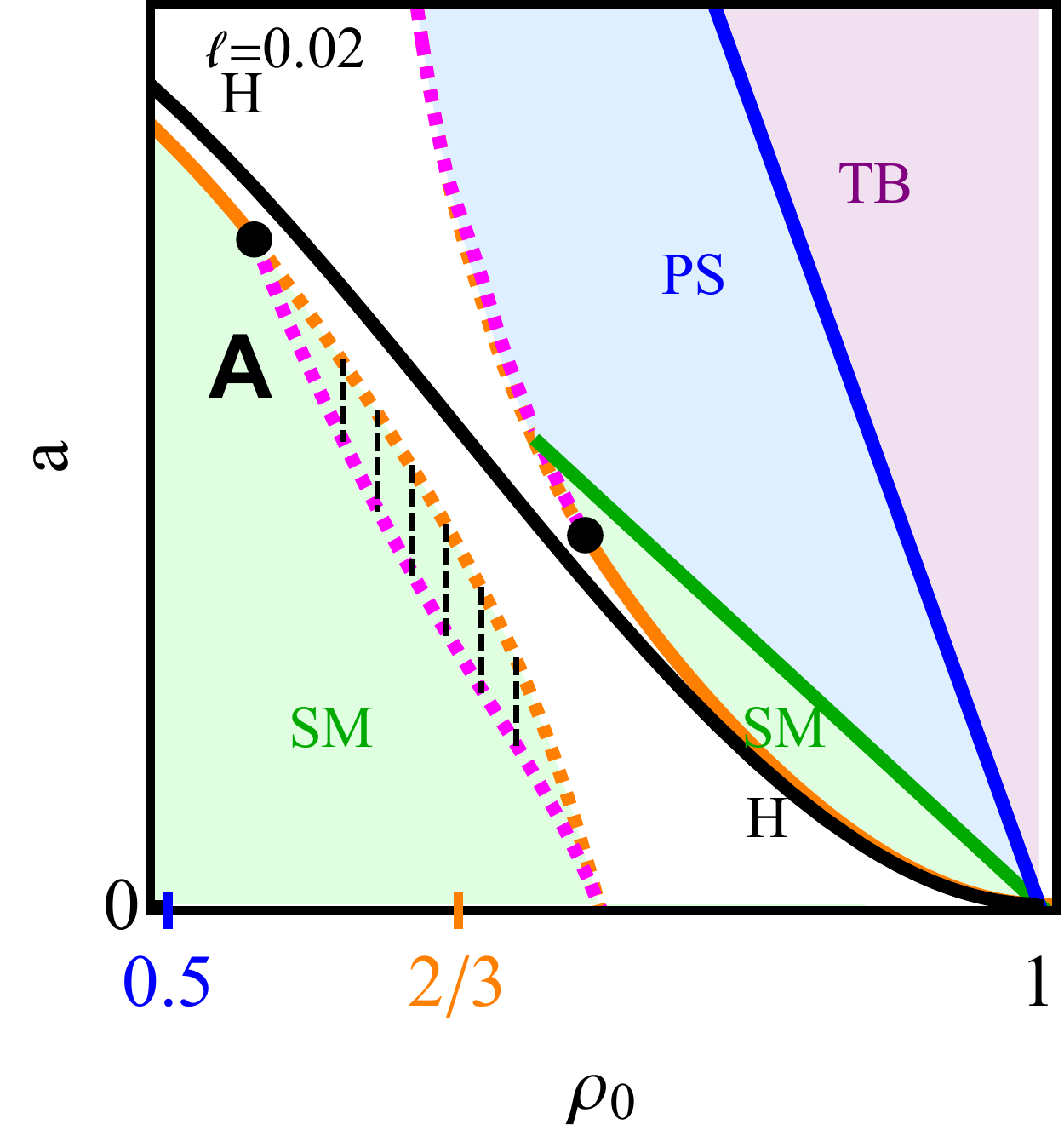}	
	\includegraphics[scale=0.5]{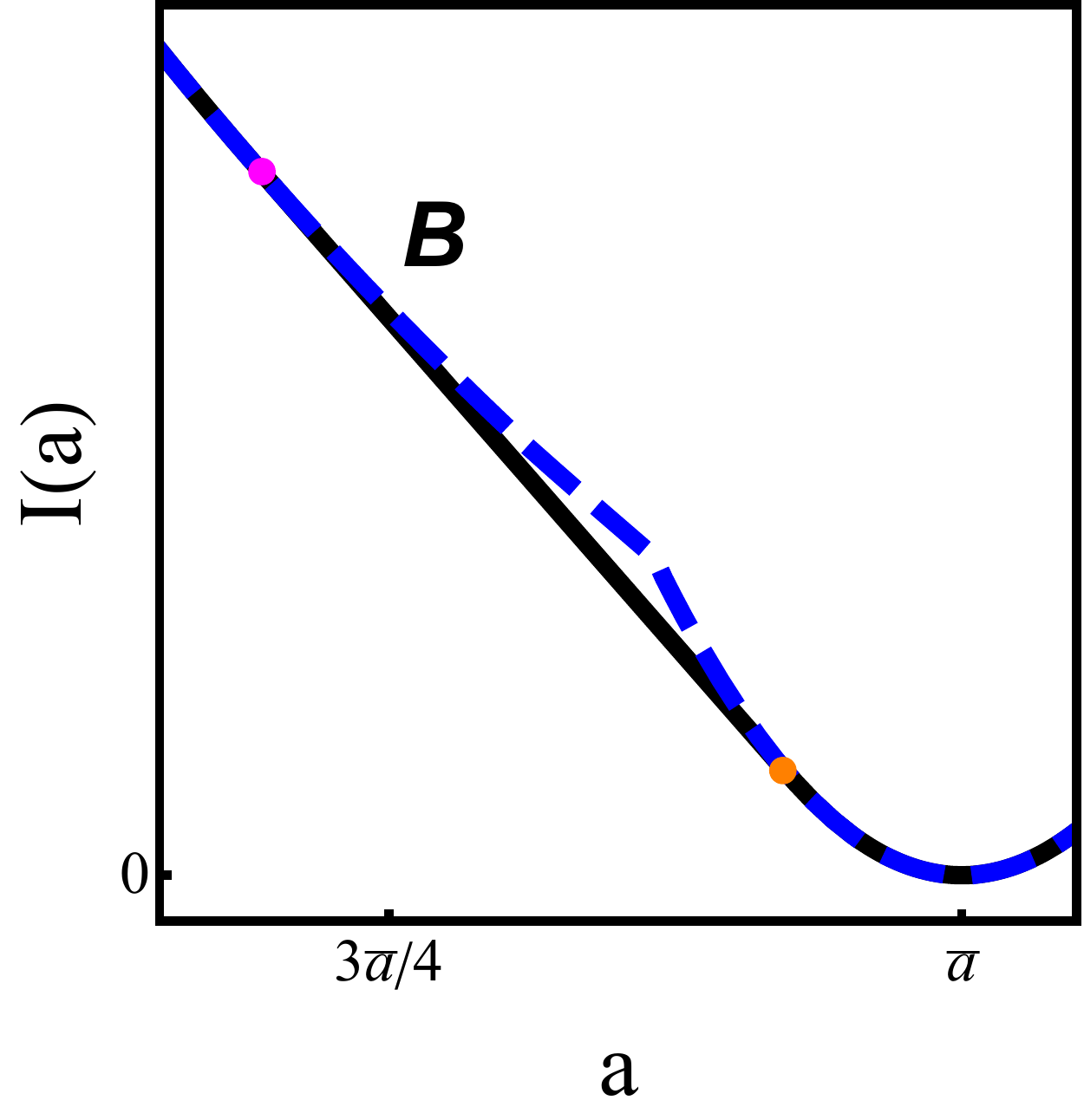}
	\caption{(A) Dynamical phase diagram for representative parameters  $\text{Pe}=1$ and $\ell=0.02$.  Phases are labelled as in Fig.~\ref{phasefig}.  Orange solid lines: second order transitions between H and SM, given by \eqref{ac2}.  Dashed orange and magenta lines: boundaries of the miscibility gap associated with discontinuous transitions between H and SM.  Black circles: tricritical points that separate first-order and second-order transitions.  Black dashed lines: tie-lines connecting points at equal density, within the miscibility gap.
	(B) Illustration of the miscibility gap for $\text{Pe}=2$, $\rho_0=0.74$ and $\ell =0.02$. Solid black line: rate function $I(a)$. Dashed blue line: the solution of the minimisation problem \eqref{Imin3}, which assumes trajectories are homogeneous in time, which shows a kink at the point where the minimiser of that problem changes from H to SM.  The true rate function is obtained by \eqref{GE} as the convex hull of the dashed blue line: the dominant trajectories in this case exhibit time-like phase separation.  Orange and magenta dots mark the edges of the miscibility gap.  As $\ell\to0$, the rate function is $O(\ell^2)$ throughout this range, the orange dot approaches $\bar{a}$, while the magenta dot approaches a finite value of $a$, so the miscibility gap stays open.
	}
		\label{phasefig2}
\end{figure}

The resulting dynamical phase diagram is shown in Fig.~\ref{phasefig2}(A).  
We focus on the behaviour for $a<\bar{a}$ (below the black line in that Figure), which includes the tricritical point at $\rho_0<2/3$.   The orange line indicates the linear instability of the H state, which is one boundary of the SM phase.  This line terminates at the tricritical point at $\rho_{c1}\simeq0.55$ (black dot), where $\beta=0$.  Beyond this point, the transition between H and SM states is first-order, corresponding to a jump discontinuity in $a$ as $\tilde\Lambda$ is reduced from zero.  The two ends of the jump are indicated by orange and magenta dashed lines.  For values of $a$ below the magenta dashed line, the system is SM. 

	The first order transition means that the rate function defined by \eqref{Imin3} is not convex when viewed as a function of the \textit{single} variable $a$ but instead has a kink singularity along the $a$ axis, see Fig.~\ref{phasefig2} (B). Similarly, the CGF defined by \eqref{square} has a kink singularity along the $\tilde{\Lambda}$ axis. Correspondingly, the relation \eqref{leg} is not invertible \cite{jack_ergodicity_2020-1,touchette_large_2009-1}. Acting with a Legendre-Fenchel transform on $\tilde{\psi}$ we will now arrive at the convex envelop of \eqref{Imin3} obtained by a common tangent construction over the `miscibility gap' in $a$. This results in the final expression for the true rate function \eqref{f}, which does not equal the expression \eqref{Imin3}, see Fig.~\ref{phasefig2} (B).	
This construction corresponds to a timelike phase coexistence between the magenta and orange dashed lines:
 the optimal path in \eqref{min}  spends part of its time period in the H state and the remainder in the SM state~\cite{jack_ergodicity_2020-1}.  These time periods are separated by instantons (domain walls in time) whose contribution to the action is negligible~\cite{bertini_non_2006,baek_dynamical_2017,baek_dynamical_2018} in the limit $T\to\infty$.
In the thermodynamic context, the domain of phase coexistence in the phase diagram is called a miscibility gap.

All these phenomena -- tricritical point, miscibility gap, and time-like phase coexistence -- are duplicated for $a>\bar{a}$, although the fine detail is not graphically resolved in Fig.~\ref{phasefig2}(A).  In this regime, it is 
evident from Fig.~\ref{phasefig2}(A) that the transition from homogeneous to inhomogeneous profiles may proceed directly to PS state, skipping the SM state altogether (possibly subject to a small miscibility gap). However, when $\ell$ becomes small, 
the orange lines (both solid and dashed) approach the black curve, $a = \bar{a}(\rho_0)$, and the first instability of the H state is always to an SM state.
On the other hand, the miscibility gap remains finite even as $\ell\to0$.  Indeed, its endpoints correspond to the minima of $\Psi$ \eqref{landau-Psi} which are $\ell$ independent, as is the gap between these endpoints. For further discussion see Appendix \ref{misc}.

Finally, we note that timelike phase coexistence is only possible when $I(\rho_0,a)$ is concave in $a$ at fixed $\rho_0$, so that the two phases that correspond to the ends of the miscibility gap can share the same $\rho_0$. This is why we did not consider such solutions when discussing the convex hull of $I_H$.

\subsection{Character of the H $\to$ SM and SM $\to$ PS transitions}\label{hsm-sm2ps}

The preceding arguments based on \eqref{square} and \eqref{landau-full} are sufficient to derive the boundaries of the SM region in Fig.~\ref{phasefig2}(A), as advertised above.  We give a brief summary of the underlying physics.

 For $a=\bar{a}$ (typical values of the IEPR) then the system is homogeneous (H).  For any $a<\bar{a}$, the convex hull of $I_H$ is zero, so one expects inhomogeneous states to appear as $\ell\to0$, but the structure of these states cannot be predicted by the quasi-thermodynamic (common tangent) approach used in Sec.~\ref{pstb}.  Instead, the physics in this regime is entirely under the control of the gradient terms in \eqref{Imin0}. 
The result is that the H state survives in a narrow region with $a$ close to $\bar{a}$, beyond which the system becomes SM.  This may take place either by a linear instability (continuous transition) or by a finite jump in the density profile, from a constant density to one with a finite modulation amplitude.  In this latter case, the phase diagram includes a miscibility gap, where SM and H phases coexist within a single trajectory (in different time periods).  The boundary between the continuous and discontinous transitions is a tricritical point.  

The distance of the H$\to$SM transition from the line $a=\bar{a}$ is determined by the balance of two costs in the minimisation problem~\eqref{Imin0}.  The H state has a cost from the reduction in IEPR at fixed $\rho$, coming from the bulk term $I_H$, while the SM state has an interfacial cost proportional to $\ell^2$.  At moderate densities, the costs balance for $\bar{a}-a = O(\ell^2)$, leading to a narrow band of the H phase between $a=\bar{a}(\rho)$ and the SM state in Fig.~\ref{phasefig2}. However, if the density is  large enough that $\bar{a}(\rho_0)=O(\ell)$, this band of the H phase can extend all the way to $a=0$. This is a non-perturbative effect of $\ell$ on the phase boundary; the range of densities over which it can occur must vanish as $\ell\to0$. [In that limit, Fig.~\ref{phasefig} already shows that the SM state extends throughout the range $0<a<\bar{a}(\rho_0)$.]

For $a>\bar{a}$ and $\rho>1/2$, one observes again an H$\to$SM phase boundary on which a tricritical point separates regions with first-order and second-order phase transitions.  In this case, the H state occurs when $a-\bar{a}=O(\ell^2)$. In addition, the system supports PS states for $a>a_3(\rho_0)$, as defined in \eqref{ac3}.
On approaching this line, the SM solution develops increasingly sharp gradients in order to accommodate the constraints \eqref{cons4} and the assumption \eqref{choise} simultaneously: the SM profiles start to resemble coexisting bulk phases whose densities are approximately   $\rho_{l,h} = 1/2,1$.  At the SM to PS transition, the interfaces sharpen (their widths become $O(\ell)$), while the coexisting densities are (exactly) $\rho_l=1/2$ and $\rho_h=1$, consistent with common tangent construction of Sec.~\ref{pstb}.  Note that the SM and PS states are distinguished by whether their interfacial widths are sharp $O(\ell)$ or smooth $O(1)$: this distinction is only meaningful in the limit $\ell\to0$.  For finite $\ell$, there is a smooth crossover between these states.

In fact, a similar SM$\to$PS transition also occurs upon approaching the $a=0$ line from the the SM phase with $a>0$. Here too, the inhomogeneous SM solution must develop increasingly sharp gradients until becoming sharply phase separated with co-existing densities $\rho_{l,h}=0,1$. However, this PS state only occurs when $a=0$ exactly. In the next Section we address what lies beyond the $a = 0$ line, in the region at negative $a$.

\section{The rate function and its extension to the negative half plane} \label{negative}
The preceding sections provide a full description of the nonequilibrium phase diagram for positive values of the IEPR, $a>0$. This half plane contains regions of both  positive and negative deviation from the unbiased value $\bar{a}(\rho_0)$. Here we consider the resulting minimized rate function.  We also extend both the rate function and the phase diagram to $a<0$. 

\subsection{The small $\ell$ limit}\label{extsl}
\begin{figure*}
	\begin{tabular}{ll}
		\includegraphics[scale=0.43]{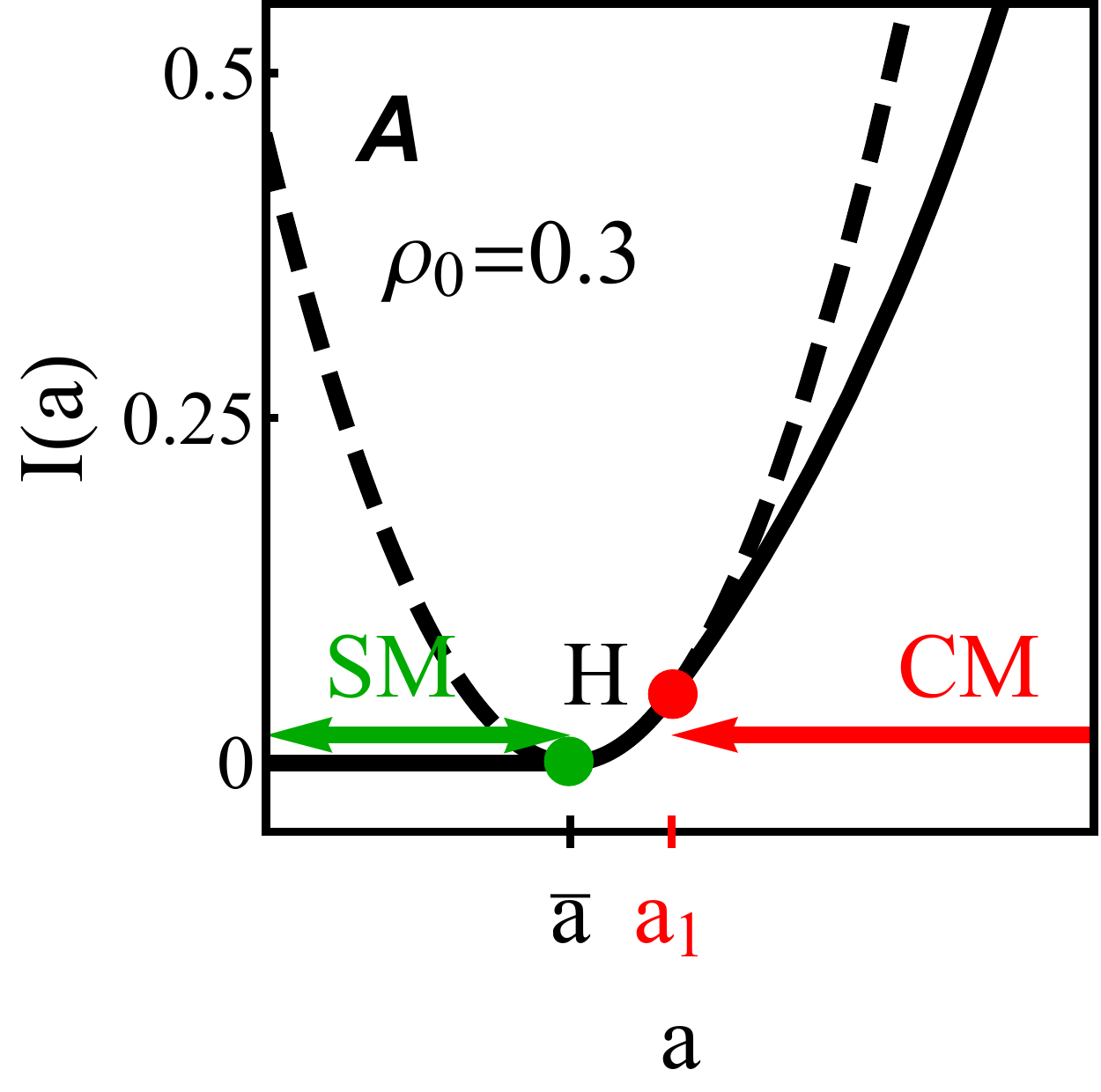}
		\includegraphics[scale=0.43]{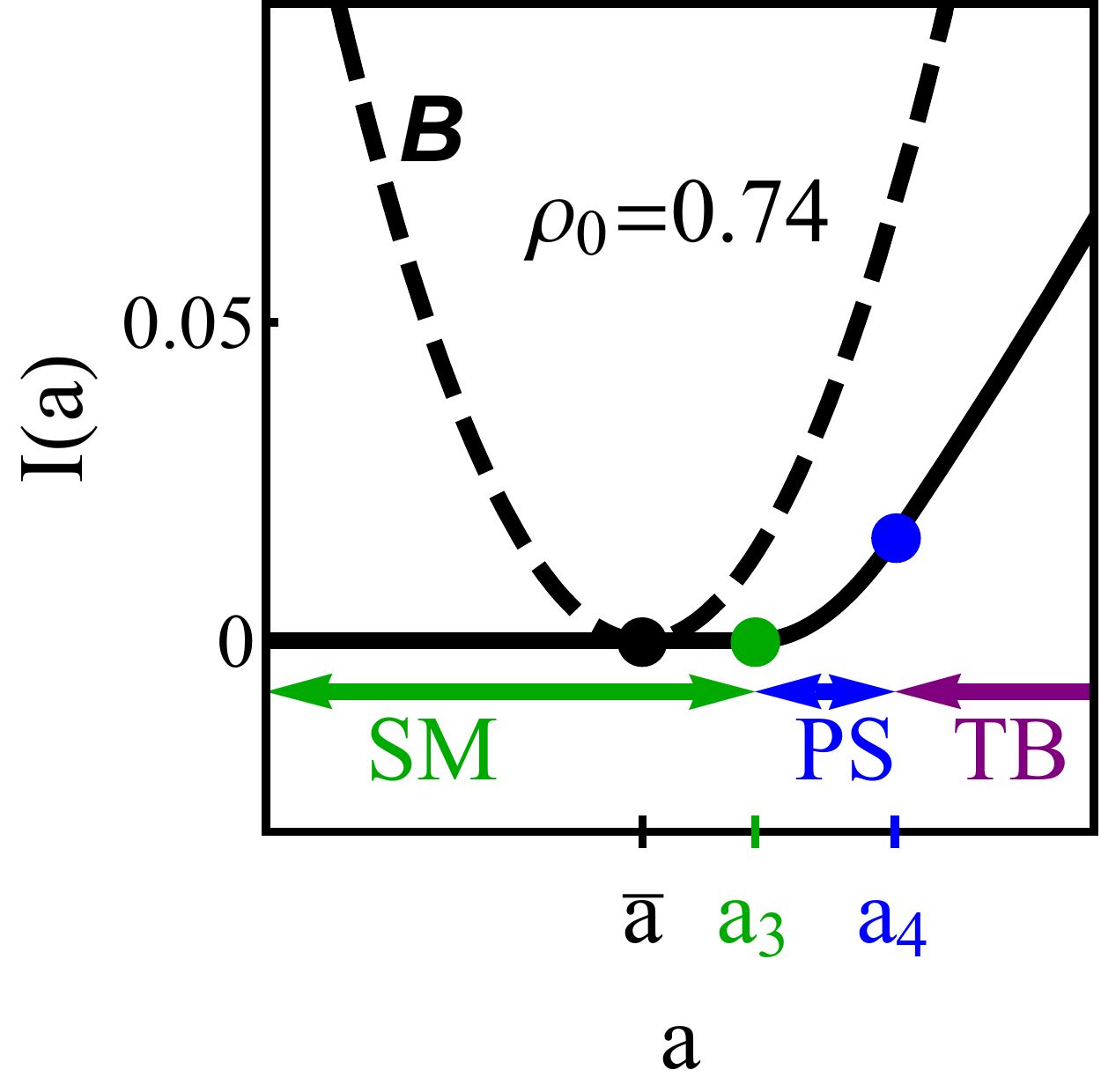}
		\includegraphics[scale=0.4]{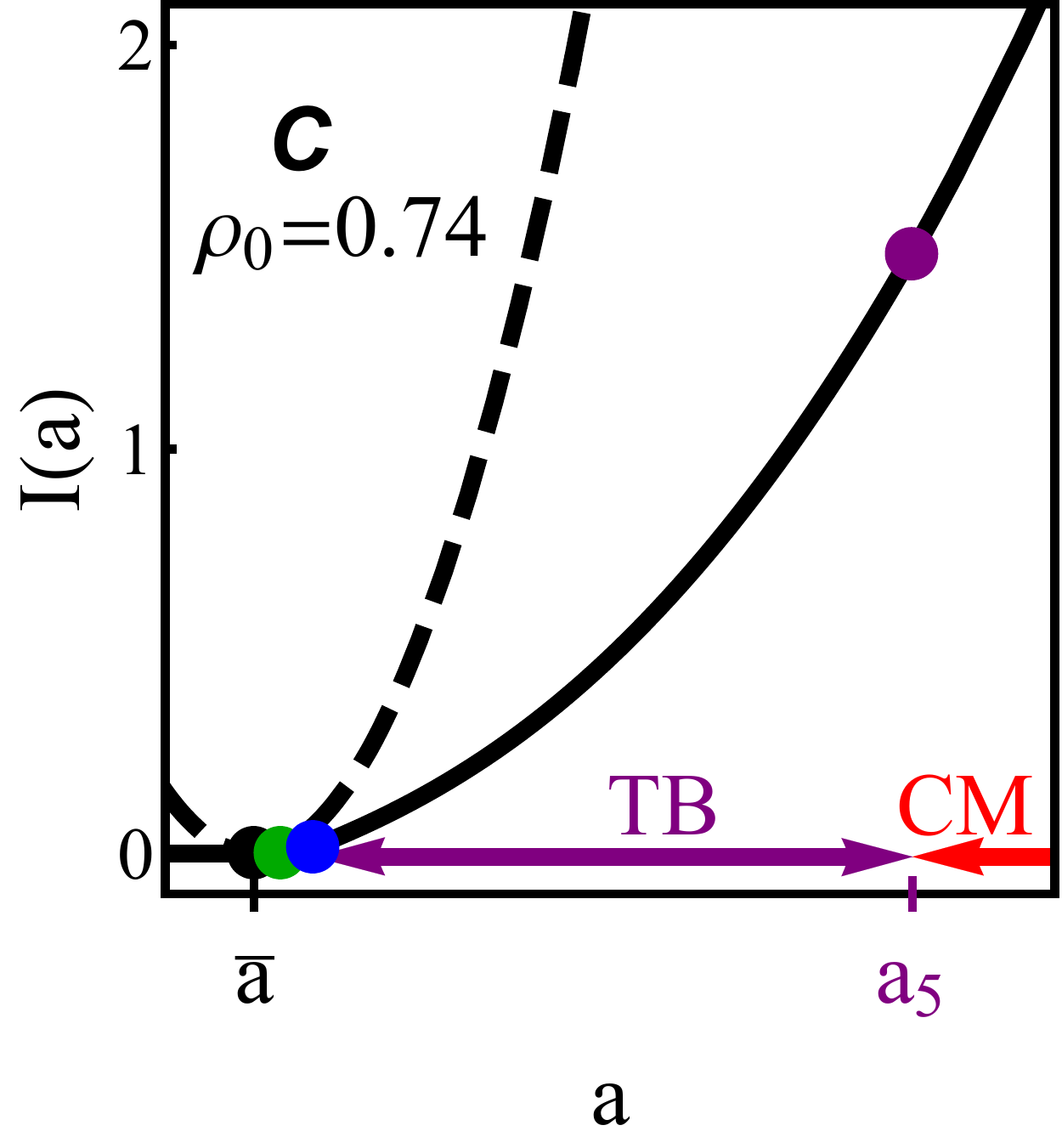}
		
	\end{tabular}
	\caption{The large deviation rate function $I(a)$ \eqref{f} in thick black line for $\rho_0=0.3$ (A) where it is given by \eqref{stot3}, and for $\rho_0=0.74$ (B) and (C) where it is given by \eqref{stot4}. The dashed black lines are the quadratic form in \eqref{highs} which serves as an upper bound on the rate function $I$ everywhere in parameter space. The colored points mark the critical values that correspond to the critical phase transition curves in Fig.~\ref{phasefig}. In all panels $\text{Pe}=3.5$. All the results shown are for $\ell = 0^+$. } 
	\label{singfig}	
\end{figure*}

Although the phase diagram for $a>0$ is complicated, even in the limit $\ell\to 0^+$ (Fig.~\ref{phasefig} (A,B)) the resulting rate function $I(\rho_0,a)$ becomes simple in this limit: it everywhere approaches the convex hull of $I_H$ obyeing \eqref{sto}. 
For $\rho_0<1/2$, this hull is given by 
\begin{numcases}
{I_{<}(a)=}
0, & $0\leq a\leq \bar{a}$,\nonumber\\
I_H(\rho_0,a), & $\bar{a}\leq a$, 
\label{stot3}
\end{numcases}
and for $\rho_0>1/2$:
\begin{numcases}
{I_{>}(a)=}
0, & $0\leq a\leq a_{3}$, \label{1}\nonumber\\
\frac{\text{Pe}^2}{(1-\rho_0)}(a-a_{3})^2, & $a_{3}\leq a\leq a_{4}$, \label{qc3}\nonumber\\
\frac{1-\rho_0}{1-\rho_l}I_H(\rho_l,a_l),& $a_{4}\leq a\leq a_{5}$, \label{qc4}\nonumber\\
I_H(\rho_0,a),& $a_{5}\leq a$. \label{qc5}
\label{stot4}
\end{numcases}
with $\rho_,a_l$ obtained by the geometrical construction that is shown in Fig. \ref{tc}, see Sec.\ref{pstb}. These expressions for $I_H$ are plotted in Fig.~\ref{singfig}.

These results can be extended to $a<0$ using the detailed fluctuation theorem~\eqref{negext}. The corresponding trajectories (which solve~\eqref{min}) are given by simple time reversal \eqref{negexttr} of the corresponding points with $a>0$.
As a result, all the critical lines $a_{1-5}$ and the various phases that they separate are simply reflected to the negative half plane. Note that the additional linear term in \eqref{negext} does not affect the construction of phase equilibria within this half plane (it has no effect on concavities there) and moreover, because its coefficient is positive, it cannot create additional concavities linking the two half planes either. The full minimization problem for the rate function is thus solved independently in each half plane, and the phase diagram for $a<0$ is simply the reflection, in the $a=0$ axis, of that found already for $a>0$ .

\begin{figure*}
	\begin{tabular}{ll}
		\includegraphics[scale=0.45]{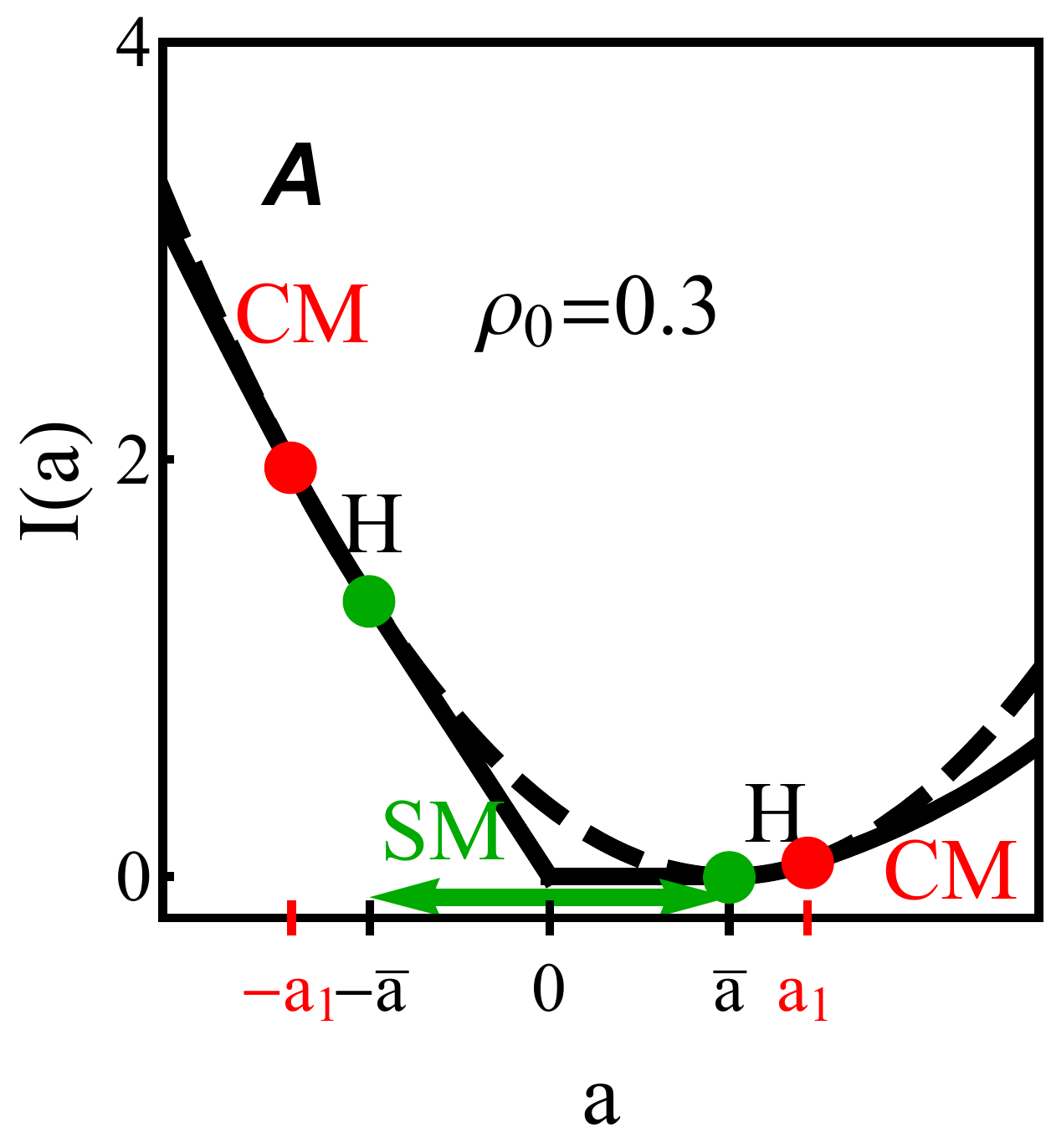}
		\includegraphics[scale=0.45]{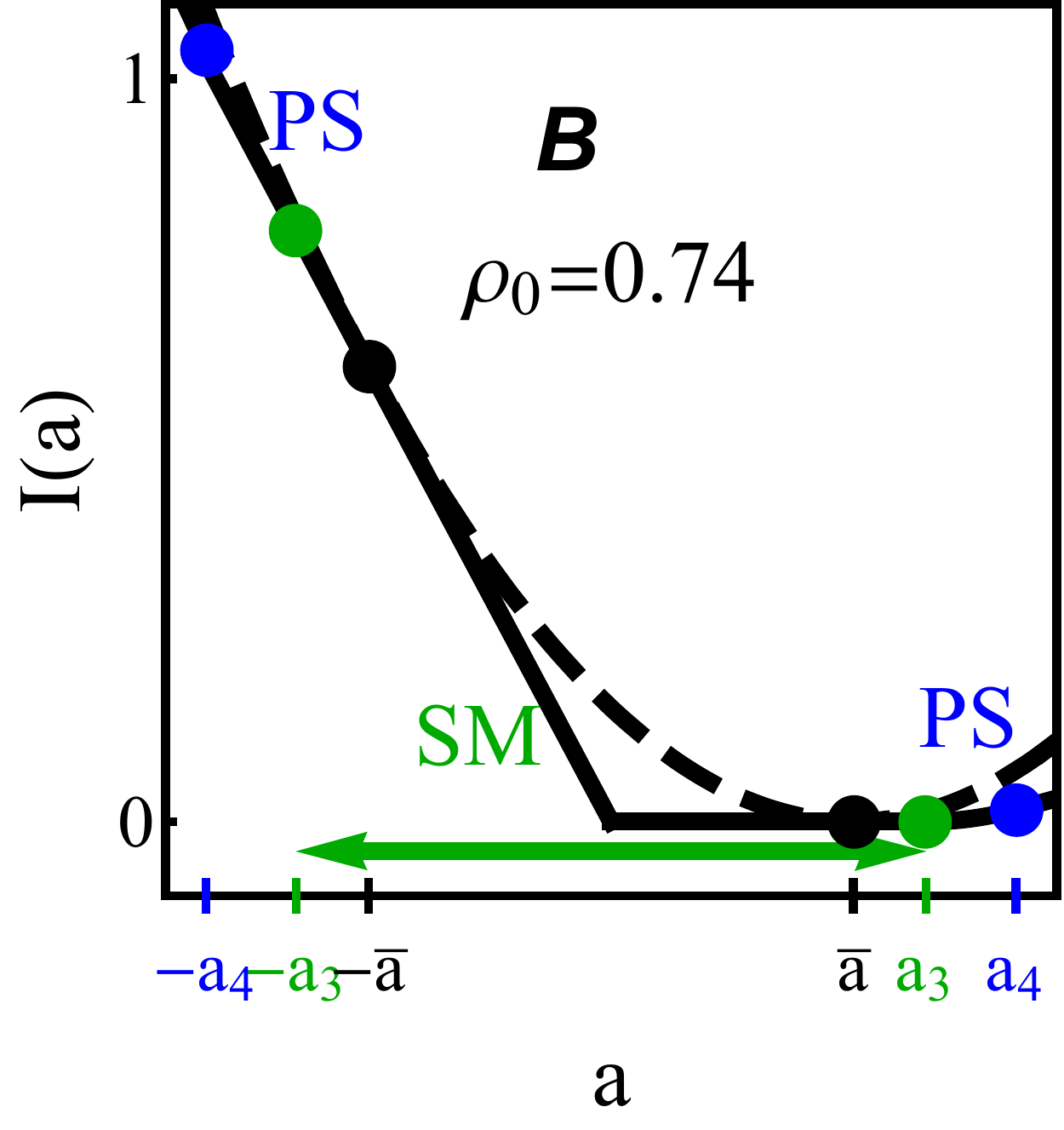}

	\end{tabular}
	\caption{The fluctuation theorem construction \eqref{negext} extending our results to the negative half plane $a<0$. All critical values are reflected accordingly. The large deviation function $I(a)$ \eqref{f} in thick black line for $\rho_0=0.3$ (A) and for $\rho_0=0.74$ (B). In both panels $\text{Pe}=3.5$. All the results shown are for $\ell = 0^+$. In panel (A) the rate function in solid line follows the dashed parabola (the quadratic from in \eqref{highs}) in the two H phases (between the red and green points), where in panel (B) there is no H phase and the rate function differs from the parabola everywhere. }
	\label{singfig2}	
\end{figure*}

Examples showing the resulting rate function (for $\ell\to0$) are presented in Fig.~\ref{singfig2}: these are composites of direct and mirrored segments for positive and negative $a$. This composite function exhibits a kink singularity at $a=0$. Indeed in the SM region with $\bar{a}(\rho_0)>a>0$ we have that $I=0$, while its reflection to the negative half plane has $I=\text{Pe}^2|a|$. The two surfaces meet along the line $a=0$, with discontinuous first derivative.

The fluctuation theorem \eqref{negext} applies for all $\ell$, so the dynamical phase diagrams shown in Fig.~\ref{phasefig} (C,D) are again mirrored in the lower half plane.  The identities of the phases are the same and the corresponding trajectories are related by time reversal.

\subsection{Finite $\ell$ corrections}\label{fl}

We have discussed the role of finite $\ell$ for these phases and transitions at various points in the preceding sections, but summarize them here for ease of reference. These effects are quite subtle and responsible for all the differences between the diagrams in panels (A,B) of Fig.~\ref{phasefig}  and those in (C,D) of the same figure, some of which are nonperturbative, such as the re-entrant H phase in the lower right corner. The reason for these relatively large effects is the degeneracy whereby the interfacial terms dominate the action everywhere within the two-dimensional convex hull of the $\bar{a}(\rho_0)$ curve. In most other regions of the phase diagram, the parameter $\ell$ has (at most) a perturbative effect on the phase boundaries.
For example, the transition between H and CM states occurs at $a=a_1(\rho_0,\text{Pe})$, as given in \eqref{ac1}.  Since these states both have homogeneous density, this phase boundary is not affected by $\ell$.

 On the other hand, the parameter $\ell$ does affect the phase boundaries of the SM state: as discussed in Sec.~\ref{bls} and Fig.~\ref{phasefig2}, increasing $\ell$ diminishes the SM regions of the phase diagram, making way for the re-entrant H phase as just described.  The SM phase is diminished by $\ell$ because its associated interfacial cost (proportional to $\ell^2$) may become greater than the cost of a homogeneous state with mildly reduced IEPR. The tricritical points on the boundary of the SM phase are present in the limit $\ell\to0$, while their positions depend perturbatively on $\ell$.   The widths of the miscibility gaps are finite as $\ell\to0$, and in general they depend non-perturbatively on $\ell$.

Turning to the SM $\to$ PS transition (which occurs for $a=a_3(\rho_0)$, as given by \eqref{ac3}): the distinction between the SM and PS states is whether there are sharp interfaces between bulk phases or a smooth modulation of the density.  This is a clear distinction only in the limit $\ell\to0$, since sharp interfaces are never observed for $\ell>0$.  Hence, the SM$\to$PS transition becomes a smooth crossover when $\ell$ is finite.  A similar rounding occurs of the kink that appears at $a=0$, when the rate functions for the upper and lower half plane get glued together; the discontinuous derivative described in Sec.~\ref{extsl} is smoothed across a boundary layer of width $\ell$.

Lastly, the domains of the PS and TB phases are diminished due to a finite $\mathcal O(\ell)$ cost in the action for sharply phase separated solutions. This means that the transitions at $a=a_4$ \eqref{ac4} and $a=a_5$ \eqref{ac5} are perturbatively distorted on increasing $\ell$ from zero, leaving the nature of the transition unchanged.

\section{Tricriticality near the MIPS critical point}\label{mipsrel}
All the work reported above addresses the large deviation structure for IEPR fluctuations in regimes of the active lattice gas model where the unbiased dynamics results in a homogeneous (H) phase. On the one hand, as found above, the biased H state loses linear stability beyond the critical IEPR level set by $a_2$ \eqref{ac2}. On the other hand, the same model has a MIPS critical point at $(\rho_0=0.75,\text{Pe}=4)$, where the {\em unbiased} homogeneous state loses linear stability towards an inhomogeneous (in this case PS) state. That is, the MIPS transition in the unbiased system and the transition from H to SM under bias share the same sort of criticality. Notice that this is not true for any of the other phase transitions shown in Fig.~\ref{phasefig}, which do not arise from the loss of linear stability. A similar property was reported in the study of current statistics of the same model \cite{agranov_macroscopic_2022}. However, the transition in that work was into a traveling band state which breaks reflection symmetry, in contrast to the SM state here which does not.

The MIPS criticality can then be interpreted as the point where the biased system is critical at exactly zero bias. 
Indeed, the difference $a_{2}-\bar{a}$ vanishes when approaching the MIPS critical point. This can be verified from the expression  \eqref{ac2} for $a_{2}$, because the mass term $M$ \eqref{m} entering this expression vanishes at the MIPS critical point. This behavior was anticipated without proof in a previous study \cite{nemoto_optimizing_2019-1}; the current work provides the first definitive substantiation of such a scenario.

\begin{figure}[t]
	\includegraphics[scale=0.4]{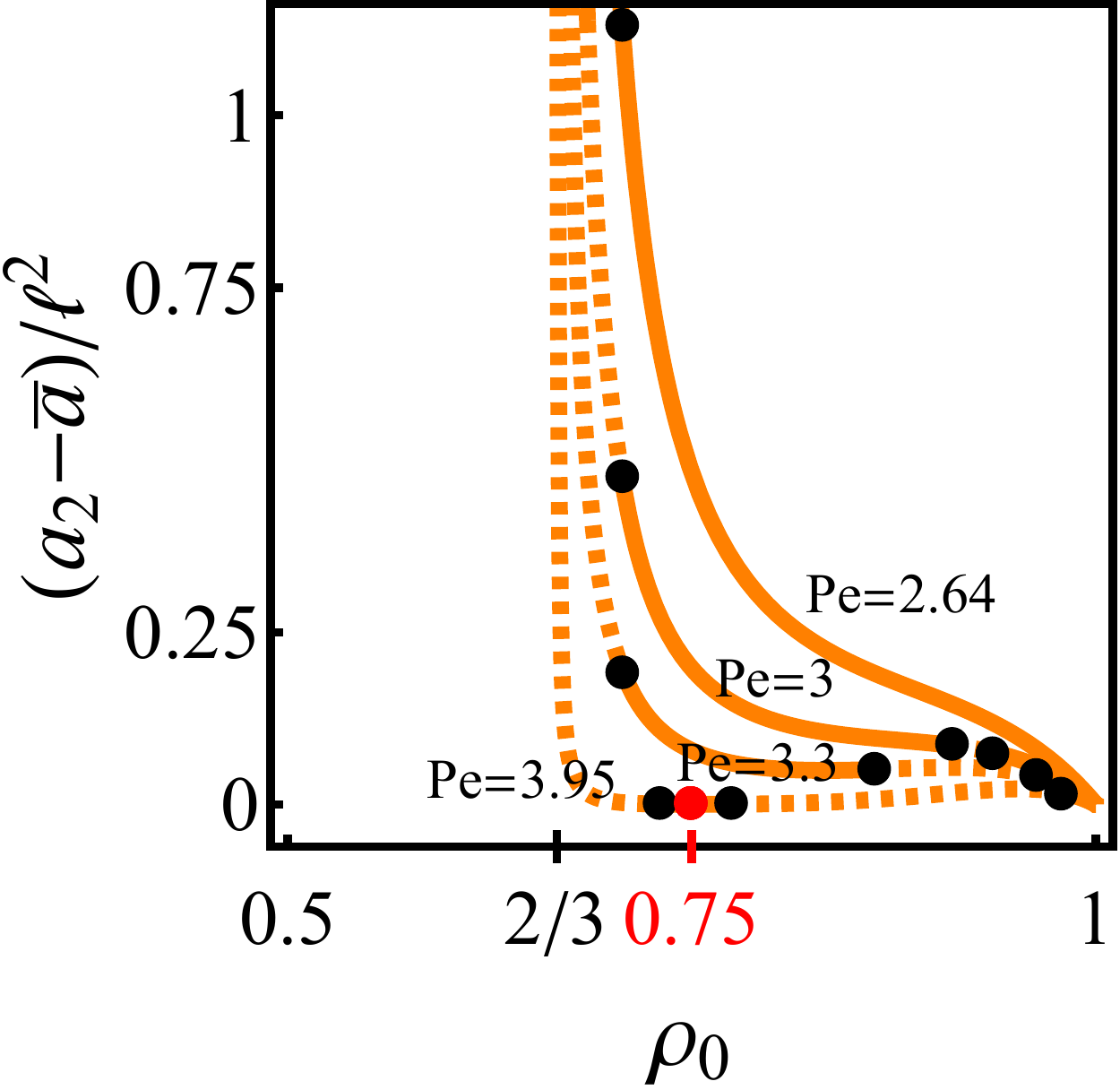}	
	\caption{The difference $(a_{2}-\bar{a})/\ell^2$ \eqref{ac2} tends to zero when approaching the MIPS criticality at $(\rho_0=0.75,\text{Pe}=4)$ denoted by the red dot. Here the solid line denotes the second order transition curve $a_{2}$ \eqref{ac2}, while the dashed line denotes its extensions beyond the tricritical points (the black dots) where the transition is first order and is not captured by the $a_{2}$ curve. Two tricritical points collide at the MIPS critical point of the unbiased model.
The leftmost of them is the one analyzed before in Sec.~\ref{bls} and shown in Fig.~\ref{phasefig} and Fig.~\ref{phasefig2}. Above $\text{Pe}>\text{Pe}^*\simeq2.64$ two additional tricritical points appear with the same phenomenology as discussed before. Of these two points, the one with a lower $\rho_0$ is the one of the two points that collide at the MIPS criticality.}
	\label{approachmips}	
\end{figure}

Intriguingly, we find that at the MIPS criticality condition two tricritical points collide as shown in Fig.~\ref{approachmips} and verified analytically in Appendix \ref{mipsver}. One of these points is the one we analyzed in Sec.~\ref{bls} above. The second one is an additional tricritical point that emerges at higher values of $\text{Pe}$ than we considered thus far, see Fig. \ref{approachmips}.
In terms of the biased phase diagrams, this collision occurs only at one specific Peclet number (Pe $=4$), the one for which the unbiased MIPS is critical. In this setting it is a higher order critical point of some kind.

Note however, that this does not imply that the MIPS critical point itself is any kind of tricritical or higher order critical point of the unbiased system.  Indeed, an ordinary critical point on an unbiased phase diagram might become first order at infinitesimal bias, but this does not make the unbiased model tricritical.

\section{Discussion : Comparison with numerical studies of off-lattice models}\label{conc}
We have studied the large deviation theory for fluctuations in the informatic entropy production rate (IEPR) of an active lattice gas model. Remarkably, the model admits an exact analytical derivation of the IEPR large deviation function \eqref{stot3} and \eqref{stot4}, and the corresponding phases, see Figs.~\ref{profiles} and \ref{phasefig}. 
It displays a rich phenomenology with an intricate phase diagram unlike any reported before in studies of related passive lattice gases, or other active systems. The IEPR fluctuations are characterized by five different phases that correspond to the breaking of different symmetries in the system. 
All these exact derivations are possible in the limit of small diffusive length scale $\ell\ll1$  which is where the system also admits MIPS. (Note however that we addressed only parameter values where the unbiased system is homogeneous rather than showing MIPS.) Yet finite $\ell$ corrections leave most aspects of the qualitative picture unchanged, albeit with some stronger effects in regions of the phase diagram where the rate function vanishes in the absence of these corrections. These can in turn be analysed in detail to leading order in $\ell$.

The nature of the dynamical phases and the associated phase transitions have been discussed extensively in previous sections, and we will not repeat them here.   However, it is useful in what follows to compare the results for this one-dimensional lattice-based model with off-lattice two-dimensional models for which large deviations of the IEPR have been estimated numerically~\cite{grandpre_entropy_2021,keta_collective_2021,nemoto_optimizing_2019-1,yan_learning_2022}.

\subsection{Fast vs slow magnetisation fields}

When comparing off-lattice models with the active lattice gas, it is important to recall that the tumbling rate in the lattice model is slower than particle hopping by a factor of $L^2$.  In terms of the hydrodynamic description  (\ref{eq:rho},\ref{eq:m}), this means that both the density and the magnetisation enter on an equal footing: the magnetisation is a slow field in the hydrodynamic sense.  By contrast, the off-lattice models are defined such that the time to move a diameter is of the same order (in system size) as the time for decorrelation of their direction of propulsion.  This means that the corresponding magnetisation fields for the off-lattice systems relax quickly; they are not hydrodynamic.

This difference is important for several reasons: it helps to make the lattice model tractable for analytical calculations, and it leads to large deviation probabilities with different dependences on the system size.  However, the main phenomenology of the models' steady states is similar, including MIPS.  To be precise about the scaling of the large deviation probabilities, it is useful to return to the microscopic time variable $\hat{t}$, which we recall is defined such that a time period $\hat{t}=1$ is long enough that a typical particle will diffuse by one lattice spacing.  Expressed in this way, the probability to observe IEPR $a$ scales in the off-lattice models as
\begin{equation}
-\ln P(a) \sim \hat{T} L^{d-2} I(a)
\label{equ:Pa-hydro}
\end{equation}
where the dimension $d=1$, we used (\ref{f}), and the trajectory length in microscopic time units is $\hat{T}=TL^2$.

In off-lattice systems, there are two possibilities for the scaling of large deviations: either (\ref{equ:Pa-hydro}), or alternatively
\begin{equation}
-\ln P(a) \sim \hat{T} L^{d} I_{m}(a) \,
\label{equ:Pa-micro}
\end{equation}
for some rate function $I_m$.  This latter form
corresponds to rare events with much smaller probability.   The scaling (\ref{equ:Pa-hydro}) occurs when the large deviation occurs by a hydrodynamic mechanism, which means that only the slow hydrodynamic fields behave in a non-typical way during the rare event.  For the active lattice gas model, all the large deviations that we have considered occur by hydrodynamic mechanisms and obey (\ref{equ:Pa-hydro}).  On the other hand, large deviations that require non-typical behaviour in fast (non-hydrodynamic) fields are expected to follow (\ref{equ:Pa-micro}).  In the off-lattice models of~\cite{keta_collective_2021,nemoto_optimizing_2019-1}, some large deviations do occur by such mechanisms, because the magnetisation is now a fast field.

When considering biased ensembles, it is important that the bias $\Lambda$ was chosen in \eqref{cgf} under the assumption that large deviations scale as in \eqref{equ:Pa-hydro}.  In contrast, the corresponding bias $s$ in~\cite{keta_collective_2021,nemoto_optimizing_2019-1} was defined under the assumption that \eqref{equ:Pa-micro} holds.  To access hydrodynamic scaling regimes one should define a corresponding quantity $\Lambda=s L^2$.  For a unified treatment of lattice and off-lattice models we could choose to consider large deviations of the IEPR by taking the limit $\hat{T}\to\infty$ at fixed $L$, and then consider the behaviour of the corresponding rate function as $L\to\infty$.  This  is a different limiting procedure to that described in Sec.~\ref{sec:dev-iepr}, but it leads to the same results.

\subsection{Dynamical phase transitions: inhomogeneous states and collective motion}

In off-lattice models, Ref.~\cite{nemoto_optimizing_2019-1} identified two dynamical phase transitions.  In the present context, those results correspond to inhomogeneous states for $a<\bar{a}$, and states with collective motion that occur for $a\geq a_c>\bar{a}$, where $a_c$ is a critical IEPR.  Fig.~\ref{phasefig} shows that this is similar (at least qualitatively) to the behaviour of the active lattice gas for low density, $\rho_0<1/2$.  This is the connection that we explore in the following.  (To our knowledge, the off-lattice models do not support complex phase behaviour like that of the active lattice gas for $\rho_0>1/2$, this point is discussed further below.)

Based on the discussion above, the expected scaling for large deviations in the off-lattice models is (\ref{equ:Pa-hydro}) for the inhomogeneous states (which occurs by a hydrodynamic mechanism), and (\ref{equ:Pa-micro}) for the collective motion state (which requires the participation of the fast magnetisation  field~\cite{keta_collective_2021}).  This is consistent with the results of~\cite{nemoto_optimizing_2019-1}: the inhomogeneous states were analysed in that work by assuming (\ref{equ:Pa-hydro}) with an $L$-dependent rate function $I_m$, which is one possible way to encapsulate both (\ref{equ:Pa-hydro}) and (\ref{equ:Pa-micro}) in a single framework.

Since the magnetisation of the active lattice gas is a slow field, the transition to CM behaviour is hydrodynamic in this model.  This is different from the off-lattice models.  However, it was already argued in Sec.~\ref{hum} that the microscopic origin of the transition is similar in both cases: the IEPR is suppressed by ``collisions'' between particles with different orientations, and CM states reduce the possibilities for such collisions.

\begin{figure*}
	\begin{tabular}{ll}
		\includegraphics[scale=0.45]{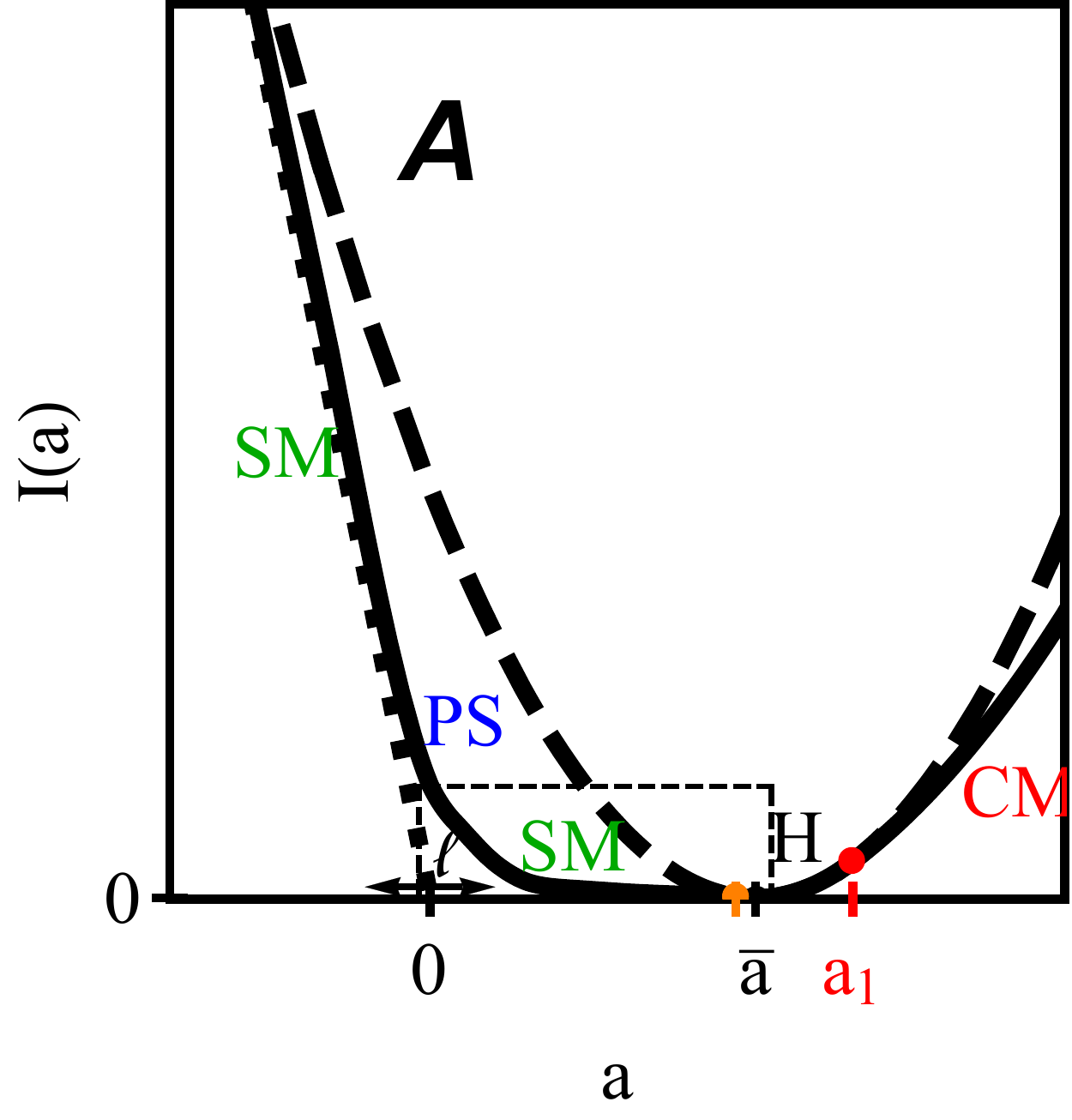}
		\includegraphics[scale=0.45]{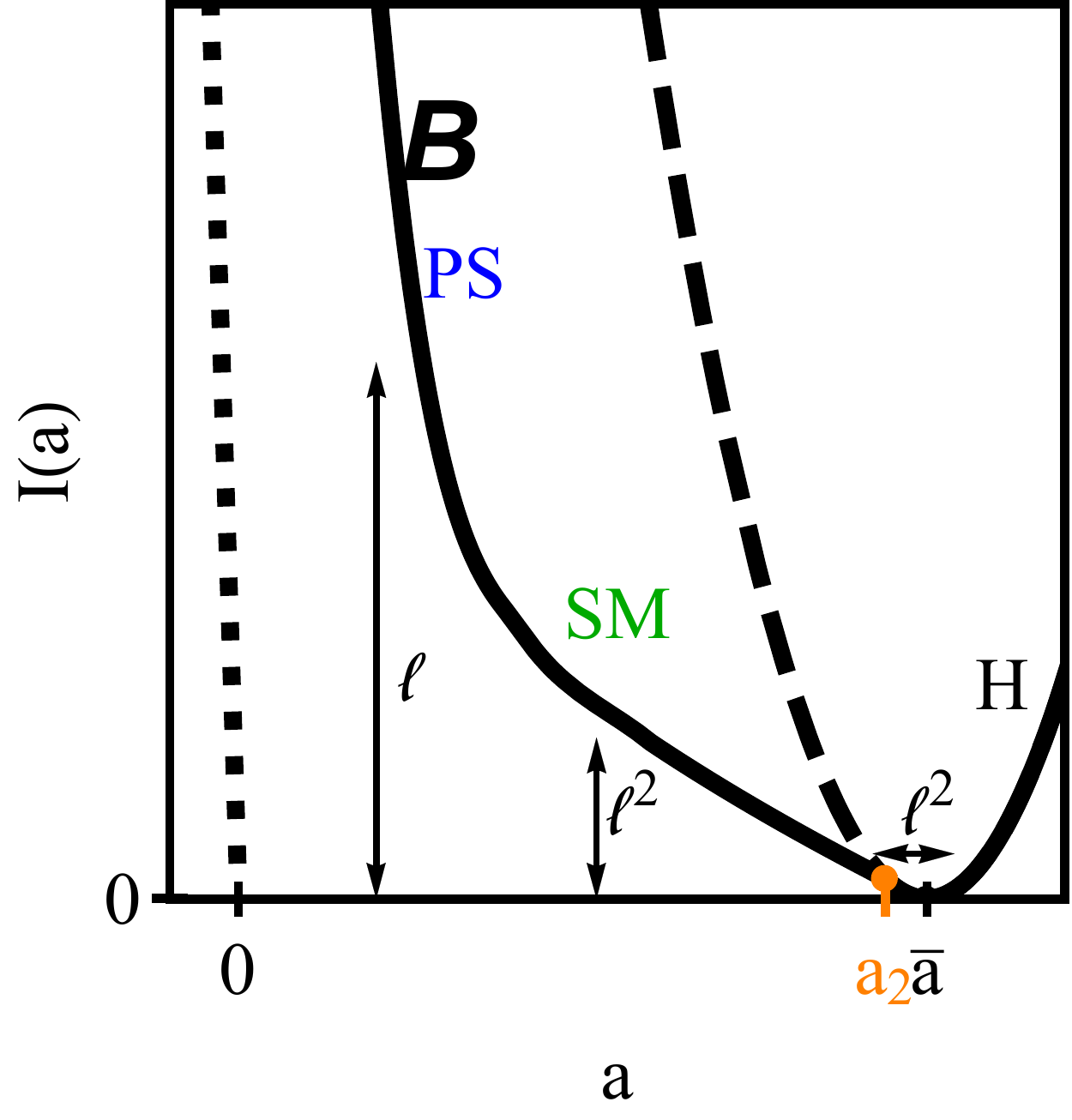}
	\end{tabular}
	\caption{(A) and the blow up panel (B) are schematic representation of the scaling behavior for the rate function $I$ for $\rho_0<0.5$ where the governing phases are H, SM and CM. The dashed curve is the upper bound given by the quadratic form in \eqref{lows}. The dotted straight line $\text{Pe}^2|a|$ marks the $\ell\to0^+$ asymptotic of $I$ that is obtained by the detailed fluctuation theorem \eqref{negext}. The corresponding kink singularity at $a=0$ is smoothed over a narrow region around $a=0$ of width $\ell$ which is characterized by a PS phase.}
	\label{schem2}	
\end{figure*}

The relationship between the inhomogeneous states of the active lattice gas and the off-lattice models is more subtle.   
The schematic Fig.~\ref{schem2} summarizes the scaling behaviour of the rate function $I(a)$ when $\rho_0<1/2$ and $\ell$ is small but finite.
Different parts of the rate function have different scalings with $\ell$: it is $O(\ell^2)$ in the SM region and $O(\ell)$ in the PS state near $a=0$.  For $a>\bar a$ it is $O(1)$, including for the CM state.
Interestingly, the arguments in~\cite{keta_collective_2021,nemoto_optimizing_2019-1} about the behaviour of the ($L$-dependent) rate function $I_m$ for the off-lattice models yield a similar picture: one has $I_m(a)=O(1)$ for $a>\bar{a}$, there is an inhomogeneous (SM) regime for reduced IEPR where the rate function $I_m(a)=O(L^{-2})$, which is equivalent to a hydrodynamic scaling (\ref{equ:Pa-hydro}).  There also is a phase separated (and dynamically arrested) regime with sharp interfaces, where the IEPR $a\approx 0$ and $I_m(a)=O(L^{-1})$.  

In other words, the behaviour of the active lattice gas and the off-lattice models appear similar if one identifies the dimensionless parameter $\ell$ in the lattice picture with the ratio $\sigma_0/L$, where $\sigma_0$ is the particle diameter, in the off-lattice models.   Recalling that $\ell=\hat{\ell}_D/L$ where $\hat{\ell}_D$ is the diffusive length that sets the natural scale for density gradients,  this correspondence is natural.  The particle diameter $\sigma_0$ sets the natural scale for density gradients in the off-lattice models, it remains fixed in the large-system limit $L\to\infty$, so one generically expects phase separation into states with sharp interfaces.  The corresponding length scale $\hat{\ell}_D$ in the active lattice gas is taken to be proportional to the system size $L$, in order to facilitate the analytic computation, but one later assumes that $\ell = \hat{\ell}_D/L\ll 1$, in order to restore the possibility of sharp interfaces.  

For this reason, the physical arguments in this work about the scalings shown in Fig.~\ref{schem2} can be similarly applied to the off-lattice models, leading to the same scalings, but with $\ell$ replaced by $\sigma_0/L$.  This is indeed consistent with~\cite{keta_collective_2021,nemoto_optimizing_2019-1}.  It shows (as expected) that the cost in probability of generating inhomogeneous density profiles is controlled by the natural scale for density gradients, while the transition to CM phase is independent of this parameter, leading to a rate function that is $O(1)$ in the relevant parameter.

Recall that in the Introduction we described the $L$-dependence chosen for the tumbling rate in the active lattice gas as `unnatural' in that it causes the magnetisation to become a slow field. However, this discussion shows that the resulting phase behaviour is nonetheless `natural', given a careful interpretation of the control parameters. Specifically, many features of the phase behaviour then echoe what is known about off-lattice models for which magnetisation relaxes much more quickly. This reinforces our view that the advantages of the active lattice gas model, in terms of what can be calculated exactly, outweigh any disadvantage in terms of its `unnaturalness'.

\section{Outlook and conclusions}\label{conc2}

We end with a few remarks on possible future directions.
As noted above, the active lattice gas behaves similarly to the off-lattice models for densities $\rho_0<1/2$, but the lattice model has complex behaviour for higher density, which has not been seen off-lattice~\cite{nemoto_optimizing_2019-1,grandpre_entropy_2021,keta_collective_2021,nemoto_optimizing_2019-1,yan_learning_2022}. Much of the complexity of the active lattice gas behavior can be traced back to the nontrivial density dependence of the expected IEPR $\bar{a}(\rho_0)$ \eqref{ent2}, whose curvature changes sign at $\rho_0=2/3$.  It would be interesting to investigate off-lattice models where the mean IEPR shows a similar property, which might well lead to novel phases such as the TB reported here, as well as to tricritical points and other interesting behaviour.  (More specifically, the TB state operates via a combination of phase separation with collective motion; it would be natural to see such combinations in a system where $\bar{a}''$ can change sign.)

With this in mind, we recall from Sec.~\ref{sec:ave-epr} that the function $\bar{a}(\rho_0)$ can also be derived directly from the microscopic dynamics, so the microscopic EPR coincides with the macroscopic IEPR. This is because, as shown in Appendix \ref{mic}, the  particles in system trajectories that contribute to IEPR traverse  macroscopic length scales over macroscopic time scales, and are thus exactly captured by the coarse-grained fluctuating hydrodynamics description. Consequently, the coarse-grained IEPR also fixes the rate at which work is expended in the system by propulsive forces.
This equivalence is in accord with Ref.~\cite{kaiser_canonical_2018} which points at a relation between the MFT large deviation functional, and the one corresponding to the underlying Markov process.

All our analysis in this work assumed that the steady state of the active lattice gas was homogeneous, and MIPS has not occurred. Nevertheless, we anticipate that to leading order at small $\ell$, most features of the problem go through to the MIPS phase.  In particular, the rate function $I$ should still be given by the convex hull of $I_H$. In all phases except the SM one, we have considered possible sharply phase-separated solutions where the optimal one among those is singled out via a convex hull construction. The MIPS state can be regarded as one such possibility, and hence we have not excluded it as part of the minimization procedure. The transition to the SM phase is more subtle as it is dominated by sub-leading corrections at small $\ell$. In this regime, the optimal profile and the nature of this phase transition might be affected by the possibility of MIPS: this could certainly be investigated further.

To conclude, in this paper we have calculated the large deviation properties of the entropy production rate in an active gas model. The special properties of this model (in which the orientational dynamics share the hydrodynamic timescales of diffusion and self-propulsion) allow such calculations to be done exactly. Despite these special features, the result show many similarities with what is known numerically from models in which the orientational dynamics represent a fast field rather than a hydrodynamic one. The resulting phase diagram shows considerable complexity and features phase separated, travelling band, and smoothly modulated states. It also shows regions of time-like phase separation in which the system spends part of its trajectory in a modulated state and part of it in a homogeneous one. Many interesting features of the phase diagram, including but not limited to tricritical behaviour, stem from the fact that the biasing variable (in this case the entropy production rate) is not everywhere a convex function of density. We expect similar features in other systems subject to bias with that property; this will be explored further in \cite{agranov_macroscopic_2022}.

\begin{acknowledgements}
We thank Jakub Dolezal for sharing with us his code for functional minimization and Sunghan Ro for helping prepare Fig.\ref{schem}. TA is funded by  Blavatnik Postdoctoral Fellowship Programme. MEC is funded by the Royal Society. Work funded in part by the European Research Council under the Horizon 2020 Programme, ERC Grant Agreement No, 740269.

\end{acknowledgements}

\newpage

	\appendix

\begin{widetext}
	
	\section{The microscopic EPR}\label{mic}
	Our aim in this section is to compute the microscopic EPR and then to express it in terms of hydrodynamic fields.
	\subsection{Preliminary: defining microscopic flows}
	The microscopic calculation of the EPR involves counting all microscopic dynamical moves that contribute to the entropy production. To this aim we now define three counting (or flow) variables:
	\begin{itemize}
		\item  Total flow: $N_{+}^T$ is the difference between the total numbers of right and left hops performed by $+$ particles up to time $T$.  The corresponding quantity for $-$ particles is $N_{-}^T$.
		\item Flow into vacant sites: $N_{+,0}^T$ is the difference between the total numbers of right and left hops into vacant sites performed by the $+$ particles up to time $T$.  Similarly $N_{-,0}^T$ for $-$ particles.
		\item  Flow due to swaps: $N_{+,-}^T$ is the difference between the total numbers of right-swaps and left-swaps between a $+$ and a $-$ particles up to time $T$, where a right-swap is a swap move in which the $+$ particle moves to the right.
	\end{itemize}
	These three variables are trivially related by 
	\begin{equation}
	N_+^T=N_{+,0}^T+N_{+,-}^T\quad,\quad N_-^T=N_{-,0}^T-N_{+,-}^T.\label{simple}
	\end{equation}
	\subsection{Computing the microscopic EPR}
	Out of the possible dynamical moves of the system, the only one which is not reversible and entails an entropy gain is when a particle  swaps positions with an adjacent vacant site. For a $+$ particle hoping to a right neighboring vacant site this move contributes an entropy gain of
	\begin{equation}
	\Delta\mathbb S=\log\left(1+\frac{\lambda}{DL}\right)=\frac{\lambda}{DL}+\mathcal O(L^{-2}).\label{ent}
	\end{equation}
	The reverse move (a $+$ particle swapping position back with a left vacant site) contributes the opposite amount $-\Delta\mathbb S$. Thus, the total entropy gain up to some time $T$ due to the $+$ particles dynamics is given by the flow into vacant sites $N_{+,0}^T$ (and similarly for $-$ particles). All in all, the total entropy gain up to time $T$ is 
	\begin{equation}
	T\mathbb S_T=(N_{+,0}^T-N_{-,0}^T)\Delta\mathbb S.\label{totact}
	\end{equation}
	We must now express the flows into vacant sites $N_{+(-),0}^T$ in terms of hydrodynamic fields.
	We will do so by first expressing the total flows $N_{+ (-)}$ and total swaps $N_{+,-}$ in terms of hydrodynamic fields, and then using \eqref{simple} to derive $\mathbb S_T$. 
	
	\subsection{Expressing the flows  in terms of hydrodynamic fields}
	
	The hydrodynamic currents $J_+(x,t)$ and $J_-(x,t)$ are directly related to the microscopic dynamics: for a short time $\Delta t$ (measured on the hydrodynamic scale), $L\Delta t J_+(x,t)$ is the number difference between right and left hops of $+$ particles across a single edge of the lattice, at position $x$. Summing over all edges on the lattice and integrating over time gives the total flow
	\begin{equation}
	N_{+}^T=L^2\int_0^Tdt\int_0^1dxJ_{+}(x,t) \; ,\label{fn}
	\end{equation}
	A similar expression holds for $-$ particles.

	The next step is to relate  the number of swaps $N_{+,-}$ to the currents.  In the absence of tumbling, the existence of counterpropagating currents determines how many swaps must take place: at leading order in $L$ one has
	\begin{equation}
	N_{+,-}^{T}=L^2\int_0^Tdt\int_0^1dx\left[J_+(x,t)\rho_-(x,t)-J_-(x,t)\rho_+(x,t)\right],\label{pm}
	\end{equation} 
	We will derive this result below, showing at the same time that it remains valid in the presence of tumbling.  (Inutuitively, the number of swap events between each tumble is large enough that the result without tumbling still applies.)
	
	The scaling of the microscopic rates with $L$ (see Sec.~\ref{model}) means that we can define a mesoscopic time scale $\Delta \hat{T}=L^{1+\delta}$ with $0<\delta<1$ (here $\Delta \hat{T}$ is measured in microscopic time units). Over such a time scale, the typical displacement of a particular particle during $\Delta\hat{T}$ will scale as $\sim\lambda L^\delta$ on account of the biased hops. This defines a mesoscopic length scale corresponding to a large number of lattice sites, so a large number of swaps must occur in this time. (Still, this scale is much smaller then the system size $L$.)
	It is on such mesoscopic length compartments that the local values of the hydrodynamic fields are defined, see Eq.\eqref{den}. In the following we will assume the large $L$ limit and express local sample means in terms of the appropriate hydrodynamic fields.
	
	It is important that a very small fraction of particles will tumble within the mesoscopic time scale.  In fact, the probability that a finite fraction of particles will tumble is even beyond the large deviation scaling for the tumbling rate probability cost \eqref{lk}, and hence can be neglected altogether, see also \cite{agranov_macroscopic_2022}. Thus, for what follows, we can neglect tumbling during $\Delta \hat{T}$.  
	
	Hence we can define an individual displacement of a $+$ particle $i$, which we denote by $n_+^i$ (here the superscript $\Delta \hat{T}$ is suppressed for ease of notation).  We relate this quantity to hydrodynamic fields. Consider a mesoscopic compartment $\Delta x$ centered at $x$ ($x\in[0,1)$ being a macroscopic variable). Now define the sample mean of particle's displacement during $\Delta \hat{T}$ over all particles in such a compartment $n_+(x,t)$. Then the contribution to the total flow $N_+^{\Delta \hat{T}}$ coming from that compartment is expressed in terms of the local hydrodynamic fields via
	\begin{equation}
	\sum_{i\in\Delta x}n_+^i=L\Delta x\rho_+(x,t)n_+(x,t)=L\Delta xJ_+(x,t)\frac{\Delta \hat{T}}{L}.\label{nav}
	\end{equation}
	Notice that here we use microscopic time. This is why there is an $L^2$ scaling difference in the right hand side, compared to \eqref{fn}.
	From \eqref{nav} we have that
	\begin{equation}
	\rho_+(x,t)n_+(x,t)=J_+(x,t)\frac{\Delta \hat{T}}{L}.\label{loc}
	\end{equation}
	Next, we can also relate the individual flows $n_+^i$ to swaps. Focus on a specific particle $i$ and denote by $\Delta_{i,j}$ its relative displacement from a specific $-$ particle labeled $j$ at time $\hat{t}=0$.
	Then the contribution to the total swaps $N_{+,-}^{\Delta\hat{T}}$ which is coming only from the particle $i$ is $N_{+,-}^{i}$, given by 
	\begin{equation}
	N_{+,-}^{i}= \sum_j\Big\lfloor\frac{n_+^i-n_-^j}{\Delta_{i,j}}\Big\rfloor,\label{sum}
	\end{equation}
	where $\lfloor\quad \rfloor$ is the floor function. Now, since $n_+^i,n_-^j\sim\lambda L^{\delta}$ are mesoscopic, all relevant $j$ particles that contribute to this sum are within the mesoscopic compartment $\Delta_{i,j}< \Delta x\sim L^{\delta}$ around particle $i$. Contribution from particles outside of this compartment is extremely rare and is beyond the large deviation scaling we consider here. In particular, we can assume the particles did not make a complete revolution around the system. 
	
	On the other hand, as this compartment contains a large (mesoscopic) number of $-$ particles, then their individual flows $n_-^j$ are self averaging and can be replaced by the sample mean $n_-(x,t)$. Thus, we have that the sum in \eqref{sum} has contributions from all $-$ particles at distance $n_+^i-n_-(x,t)$ from particle $i$; the total number of such particles is $(n_+^i-n_-(x,t))\rho_-(x,t)$, and so
	\begin{equation}
	N_{+,-}^{i}= \left[n_+^i-n_-(x,t)\right]\rho_-(x,t) \; .
	\end{equation}
	Summing over all $+$ particles in the system we have that
	\begin{equation}
	N_{+,-}^{\Delta\hat{T}}=\sum_iN_{+,-}^{i}= L\int_0^1dx\left[ n_+(x,t)- n_-(x,t)\right]\rho_-(x,t)\rho_+(x,t).
	\end{equation}
	Using the relation \eqref{loc}, switching to the hydrodynamic time scale, and integrating over time, we finally arrive at \eqref{pm}.

	\subsection{Expressing $\mathbb S_T$ in terms of the hydrodynamic fields}
	We are left with substituting the two results \eqref{fn} and \eqref{pm} into \eqref{simple} to arrive at the expression for the flow into vacant sites in terms of hydrodynamic fields to obtain
	\begin{equation}
	N_{+(-),0}=L^2\int_0^Tdt\int_0^1dx\left[J_{+(-)}(x,t)(1-\rho_{-(+)}(x,t))+J_{-(+)}(x,t)\rho_{+(-)}(x,t)\right].
	\end{equation}
	Finally, using this expression in \eqref{totact} we arrive the at the microscopic EPR expressed in terms of hydrodynamic fields
	\begin{eqnarray}
	\mathbb S_T&=&\frac{L\gamma\text{Pe}^2}{T}\,\int_0^Tdt\int_0^1dx\left[(1-2\rho_-(x,t))\frac{J_+(x,t)}{\lambda}-(1-2\rho_+(x,t))\frac{J_-(x,t)}{\lambda}\right]\\
	&=&\frac{L\gamma\text{Pe}^2}{T}\,\int_0^Tdt\int_0^1dx\left[(1-\rho(x,t))\frac{J_m(x,t)}{\lambda}+m(x,t)\frac{J_{\rho}(x,t)}{\lambda}\right],
	\end{eqnarray}
	in exact agreement with the expression \eqref{entro2} that was obtained from the coarse grained fluctuating hydrodynamics. In particular, the expected microscopic EPR coincides with the macroscopic IEPR \eqref{ent2}.
	
	\subsection{Insights from the microscopic derivation}
	The microscopic derivation helps explain two features of the IEPR. 
	First, the density dependence of the expected IEPR $\bar{a}(\rho_0)$ \eqref{ent2} can be understood. Indeed, for the expected IEPR one has $\rho_+=\rho_-=\rho_0/2$, $J_+=-J_-=\text{const}$ and $N_+^T=-N_-^T=L^2J_+$. Plugging these into \eqref{pm} one finds that the total swaps are given by $N_{+,-}^T=N_+^T\rho_0$. Using this in \eqref{simple} we find that the flow into vacant sites is a fraction $(1-\rho_0)$ of the total flow
	\begin{equation}
	N_{+,0}^T=(1-\rho_0)N_+^T\quad;\quad N_{-,0}^T=(1-\rho_0)N_-^T.
	\end{equation}
	As we detail in Sec. \ref{secentro}, the factor $(1-\rho_0)$ helps explain the density dependence in \eqref{ent2}.
	
	Secondly, the transition to collective motion is also explained by the microscopic picture. By increasing the plus minus imbalance $m$, the number of reversible swaps is reduced, while leaving the sum of total currents unchanged. By decreasing the portion of reversible moves out of the total dynamical moves that comprise the integrated current, the entropy production is increased.
	
	Indeed, assuming stationary and homogeneous density profiles and using the expressions \eqref{mean1} for the expected fluxes, one finds that the expected total swap \eqref{pm} is given by 
	\begin{equation}
	N_{+,-}^T=\frac{L^2\lambda T(1-\rho_0)}{2}(\rho_0^2-m^2).\label{pm2}
	\end{equation} 
	This expression monotonically decreases with $|m|$ until it vanishes for the completely aligned case $|m|=\rho_0$. At the same time, the sum of total flows in the direction of the self propulsion is independent of $m$
	\begin{equation}
	N_+^T-N_-^T=L^2T\lambda\rho_0(1-\rho_0)
	\end{equation}

\section{Deriving the square gradient expansion \eqref{Imin0} }\label{convex}

This appendix derives the expansion \eqref{Imin0} from the minimization problem \eqref{min}, which gives an expression for the rate function $I$ up to quadratic order in $\ell$.  As a preliminary step, we first derive a similar expression which is valid at zeroth order in $\ell$, namely
	\begin{equation}
	\lim_{\ell\to0} I(a)= \min_{\aalph(x),\rho(x)}\int_0^1dxI_H \big(\rho(x),\aalph(x)\big),\label{Imin} 
	\end{equation}
	where the minimisation is subject to the usual constraints
	$a=\int_0^1\aalph(x)dx$, and $\rho_0=\int_0^1\rho(x)dx$.
	(Recall that the rate function $I$ depends implicitly on $\rho_0,\ell$ as well as $a$.)  Building on this analysis, the full result \eqref{Imin0} is then derived in Appendix~\ref{subsec:b-gradient}.
	Since \eqref{Imin} does not involve any gradient terms, the infimum is given by the convex hull of $I_H$. 
	We will show that this is true for all $\rho_0,a$, even if the minimizer of \eqref{min} is non-stationary, as in the TB phase.

\subsection{Derivation of \eqref{Imin}}
Let us first write the Lagrangian ${\cal L}_J$ in terms of dimensionless quantities, rescaling the currents $J_\rho$ and $J_m$ by factors of $\lambda$:
\begin{equation}\label{fluxrate2}
\mathcal L_J=\frac{\gamma}{2}
\begin{bmatrix}
\text{Pe}\,\frac{ J_{\rho}}{\lambda}-\text{Pe}\,m(1-\rho)+\ell\partial_x\rho\\
\text{Pe}\,\frac{ J_{m}}{\lambda}-\text{Pe}\,\rho(1-\rho)+\ell\partial_xm
\end{bmatrix}^\dag
D\mathbb{C}(\rho,m)^{-1}
\begin{bmatrix}
\text{Pe}\,\frac{J_{\rho}}{\lambda}-\text{Pe}\,m(1-\rho)+\ell\partial_x\rho\\
\text{Pe}\,\frac{J_{m}}{\lambda}-\text{Pe}\,\rho(1-\rho)+\ell\partial_xm
\end{bmatrix}.
\end{equation}
Since all gradient terms appear with a prefactor $\ell$, they disappear as $\ell\to0$ for the computation of the action \eqref{min}. [This is still true even when the optimal fields are phase separated with a domain wall of width $\sim\ell$.  The gradient terms are $\mathcal O(1)$ in that case but the region of integration where they contribute is $\mathcal O(\ell)$, and hence their overall contribution to the functional is $\mathcal O(\ell)$, and therefore negligible.]

Next, write the dynamical equations \eqref{eq:rho} and \eqref{eq:m} as 
\begin{eqnarray}
\gamma^{-1} \partial_{ t} \rho&=& -\text{Pe}\,\ell\partial_x (J_{\rho}/\lambda)\label{eq:rho2}\\
\gamma^{-1} \partial_{ t} m&=& -\text{Pe}\,\ell\partial_x (J_{m}/\lambda) - 2K. \label{eq:m2}
\end{eqnarray} 
As $\ell\to0$, Eq.~\eqref{eq:rho2} shows that either $\rho$ must be stationary, or the gradient of $J_\rho$ must diverge.  The latter scenario can occur at sharp interfaces, leading to a TB state where the density is stationary within the bulk phases, with currents at the interfaces which cause the band to travel (see Sec. \ref{pstb}).

The same property holds for the optimal magnetization history $m$.  That is, it must also be either stationary or take the form of a traveling band. Here this does not follow immediately from \eqref{eq:m2}, and the proof is slightly more involved. One way of arriving at it is employing the Hamiltonian MFT formulation of the problem and using it to prove stationarity of $m$.  
Then from Eq.\eqref{eq:m2}, one has $K=0$ everywhere except (possibly) in the vicinity of sharp interfaces.
Recall, $K=0$ is the same value that one finds for the homogeneous solution, see Sec. \ref{hum}. 

Setting $K=0$, and neglecting the gradient terms in \eqref{fluxrate2}, the action functional \eqref{act} is of the form considered in Sec. \ref{hum}, except that the fields $\rho,m,J_{\rho},J_m$ entering the action functional are not constants.
However, in the absence of gradient terms, minimisation over the (unconstrained) fields $J_\rho,J_m$ becomes a purely local (algebraic) problem (it is important here that the IEPR constraint \eqref{entro2} does not involve gradient terms).  The computation proceeds along similar lines to those taken in the homogeneous case of Sec. \ref{hum}.

First, for stationary profiles the IEPR constraint in \eqref{min} becomes $a=\int_0^1dx\aalph(x)$ with $\alpha$ defined in \eqref{locent}. Then,
minimizing the action over $J_{\rho}(x)$ and $m(x)$, one arrives at the algebraic relations \eqref{flux} and \eqref{mag} but which are now space dependent. That is
\begin{equation}
J_\rho(x)=\frac{m(x)\aalph(x)\lambda}{\rho(x)(1-\rho(x))+m(x)^2}\quad,\quad J_m(x)=\frac{\rho(x)\aalph(x)\lambda}{\rho(x)(1-\rho(x))+m(x)^2}.\label{flux2}
\end{equation}
where $m(x)=m\left[\aalph(x),\rho(x)\right]$ is given by $m=0$ in regions where
$\aalph(x)<a_{1}(\rho(x),\text{Pe})$ with $a_1$ as in \eqref{ac1}; otherwise $m$ is obtained by solving
\begin{equation}
\aalph(x)=\left[\bar{a}(\rho(x))+(1-\rho(x))m(x)^2\right]\sqrt{1+\frac{2}{\text{Pe}^2}\frac{1}{(1-\rho(x))\sqrt{\rho(x)^2-m(x)^2}}}
\; .
\label{mag2} 
\end{equation}
Since this analysis is directly analogous to Sec.~\ref{hum}, one arrives at \eqref{Imin} with $I_H$ given by~\eqref{sto}.

\subsection{Deriving the sub-leading gradient terms}
\label{subsec:b-gradient}

We now derive the sub-leading order correction to the expression \eqref{Imin}, at small $\ell$: that is \eqref{Imin0}. 
	As noted in Sec.~\ref{opt}, this is derived under the condition  that reflection symmetry is unbroken, so it allows analysis of SM and PS states but not TB and CM.  Specifically, we assume that $J_\rho(x)=0$ (exactly) while $m(x)$ vanishes as $\ell\to0$
\begin{equation}
m(x)=\ell \tilde{m}(x)+\mathcal O\left(\ell^2\right).\label{scale}
\end{equation}
This implies that
\begin{equation}
\mathbb C^{-1}=\frac{1}{2D}\left\{
\begin{bmatrix}
\frac{1}{\rho\left(1-\rho\right)}&0\\
0&\frac{1}{\rho}
\end{bmatrix}
-\frac{\ell\tilde{m}}{\rho^2}
\begin{bmatrix}
0&1\\
1&0
\end{bmatrix}+
\frac{\ell^2\tilde{m}^2}{\rho^3}	\begin{bmatrix}
1&0\\
0&1
\end{bmatrix}\right\}+\mathcal O\left(\ell^3\right)
\end{equation}
and the mean currents appearing in  the action functional \eqref{fluxrate} are
\begin{eqnarray}
\frac{\bar{J}_{\rho}}{\lambda}&=&\ell\left[-\frac{\partial_x\rho}{\text{Pe}}+ \tilde{m}\left(1-\rho\right)\right],\nonumber\\
\frac{\bar{J}_{m}}{\lambda}&=&\rho\left(1-\rho\right)-\ell^2\frac{\partial_x\tilde{m}}{\text{Pe}}.\label{mean1b}
\end{eqnarray}
As $J_\rho=0$ we have that the local IEPR field is  \begin{equation}
	\aalph(x)=\frac{J_m\left(1-\rho\right)}{\lambda},\label{aaa}
	\end{equation} and the constraint to be satisfied by the IEPR in \eqref{min} is
\begin{equation}
a=\int_0^1 dx\, \aalph(x).
\label{a2}
\end{equation}

\subsubsection{Action functional sub-leading correction}

We now compute the correction at $O(\ell^2)$ to the action, which is to be minimised in \eqref{min}.
First, from the dynamical equation for $m$ \eqref{eq:m} we have
\begin{equation}
K=-\frac{\partial_x J_m}{2\gamma}=-\frac{\text{Pe}\,\ell}{2}\partial_x\left(\frac{\aalph}{1-\rho}\right).
\end{equation}
where the second equality is \eqref{aaa}.
Thus, to leading order in $\ell$ we find that
\begin{eqnarray}
\mathcal L_K&=&\gamma\ell^2\frac{\left[\frac{\text{Pe}}{2}\partial_x\left(\frac{\aalph}{1-\rho}\right)+\tilde{m}\right]^2}{2\rho}+\mathcal O\left(\ell^4\right)
\label{phi2}.
\end{eqnarray}
Using \eqref{mean1b}, \eqref{aaa}, and that $J_{\rho}=0$, then the other contribution to the action \eqref{fluxrate} gives
\begin{eqnarray}
\frac{\mathcal L_{J}}{\gamma}&=&I_H \left[\rho(x),\aalph(x)\right]+\ell^2\frac{\text{Pe}^2}{4}\left\{\left[\frac{\left[\partial_x\rho/\text{Pe}-\tilde{m}\left(1-\rho\right)\right]^2}{\rho\left(1-\rho\right)}+\frac{2}{\text{Pe}}\left(\frac{\aalph-\bar{a}}{1-\rho}\right)\left[\partial_x\left(\frac{\tilde{m}}{\rho}\right)+\text{Pe}\frac{\tilde{m}^2\left(1-\rho\right)}{\rho^2}+\frac{\text{Pe}}{2}\frac{\tilde{m}^2}{\rho^3}\left(\frac{\aalph-\bar{a}}{1-\rho}\right)\right]\right]\right\}\nonumber\\&+&\mathcal O(\ell^3).
\end{eqnarray}
Using integration by parts one can eliminate the derivatives of $\tilde{m}$ to arrive at
\begin{eqnarray}
\frac{\mathcal L_{J}}{\gamma}&=&I_H \left[\rho(x),\aalph(x)\right]+\ell^2\frac{\text{Pe}^2}{4}\left\{\left[\frac{\left[\partial_x\rho/\text{Pe}-\tilde{m}\left(1-\rho\right)\right]^2}{\rho\left(1-\rho\right)}+\left(\frac{\aalph-\bar{a}}{1-\rho}\right)\left[2\frac{\tilde{m}^2\left(1-\rho\right)}{\rho^2}+\frac{\tilde{m}^2}{\rho^3}\left(\frac{\aalph-\bar{a}}{1-\rho}\right)\right]-\frac{2}{\text{Pe}}\partial_x\left(\frac{\aalph-\bar{a}}{1-\rho}\right)\frac{\tilde{m}}{\rho}\right]\right\}\nonumber\\&+&\mathcal O(\ell^3).
\end{eqnarray}
Putting everything together, we have the total action is given by
\begin{eqnarray}
\frac{\mathcal{ A}}{\gamma T}&=&\int_0^1dx\left\{I_H \left[\rho(x),\aalph(x)\right]+\ell^2\left\{\frac{1}{4}\frac{\left[\partial_x\rho-\text{Pe}\,\tilde{m}\left(1-\rho\right)\right]^2}{\rho\left(1-\rho\right)}+\frac{\left[\text{Pe}\,\partial_x\left(\frac{\alpha}{1-\rho}\right)+2\tilde{m}\right]^2}{8\rho}\right\}\right\}\nonumber\\
&+&\ell^2\frac{\text{Pe}^2}{4}\int_0^1dx\left\{\left(\frac{\aalph-\bar{a}}{1-\rho}\right)\left[2\frac{\tilde{m}^2\left(1-\rho\right)}{\rho^2}+\frac{\tilde{m}^2}{\rho^3}\left(\frac{\aalph-\bar{a}}{1-\rho}\right)\right]-\frac{2}{\text{Pe}}\partial_x\left(\frac{\aalph-\bar{a}}{1-\rho}\right)\frac{\tilde{m}}{\rho}\right\}+\mathcal O(\ell^3).\label{act11}
\end{eqnarray}
This action is still to be minimized over the fields $\rho,\aalph,\tilde m$ subject to the constraints \eqref{a2} and $\rho_0 = \int_0^1 dx \rho$.
	Note that \eqref{act11} does not depend on gradients of $\tilde{m}$, and this field does not feature in either constraint. Hence, its optimal value is obtained by direct minimization of \eqref{act11}  with respect to $\tilde{m}$. Performing the  algebraic (although cumbersome) minimization we arrive at the announced result \eqref{Imin0} with the symmetric matrix $\mathbb{A}(\rho,\aalph)$ whose coefficients are given by
\begin{eqnarray}\nonumber
	\mathbb{A}_{1,1}&=&\frac{\alpha^2\text{Pe}^2+2(1-\rho)^3}{8\rho(1-\rho)^4}-\frac{\text{Pe}^2}{\frac{2}{\rho}+\text{Pe}^2(1-\rho+\frac{\alpha^2}{\rho^3(1-\rho)^2})}\nonumber\\\nonumber
	\mathbb{A}_{1,2}=\mathbb{A}_{2,1}&=&\text{Pe}^2\frac{\alpha}{8\rho(1-\rho)^3}\quad,\quad\mathbb{A}_{2,2}=\frac{\text{Pe}^2}{8\rho(1-\rho)^2}\label{fina}
	\end{eqnarray}

\subsubsection{Square gradient theory in the SM state}
In the SM state, the square gradient minimisation \eqref{Imin0} reduces to the single-field minimisation~\eqref{Imin3}.
	To see this, note that the IEPR of the SM state obeys \eqref{choise}: on plugging that result into \eqref{Imin0}, one may eventually derive~\eqref{Imin3}. However, a simpler route to the same result is to plug \eqref{choise} directly into \eqref{act11} to find that
\begin{equation} \label{A-SM}
\frac{\mathcal{ A}}{\gamma T}=\ell^2\left\{\frac{1}{4}\frac{\left[\partial_x\rho-\text{Pe}\,\tilde{m}\left(1-\rho\right)\right]^2}{\rho\left(1-\rho\right)}+\frac{\left[\text{Pe}\,\left(1-2\rho\right)\partial_x\rho+2\tilde{m}\right]^2}{8\rho}\right\}+\mathcal O\left(\ell^4\right)
\end{equation}
Performing the minimization with respect to $\tilde{m}$ now we arrive at
\begin{equation}
\tilde{m}=\frac{2\rho\text{Pe}}{2+\text{Pe}^2\,\left(1-\rho\right)}\partial_x\rho,
\end{equation}	
Plugging this expression back into \eqref{A-SM} and combining with \eqref{Imin0} finally yields \eqref{Imin3}.

\section{Finding the non-convex regions in $I_H$ \eqref{sto}}\label{convexrig}

This appendix derives the regions where $I_H$ is not convex, as illustrated in Fig.~\ref{nonzeromag}.

\subsection{Local non-convexity}
The region of local non-convexity is defined by having a negative Hessian determinant:
\begin{equation}\label{det}
\text{Det}\left[\text{Hess}(I_H)\right]=\frac{\partial ^2I_H}{\partial {\rho_0}^2}\frac{\partial ^2I_H}{\partial {a}^2}-\frac{\partial ^2I_H}{\partial {\rho_0}\partial a}<0
\end{equation}
In the region $a<a_{1}$, the expression \eqref{sto} for $I_H$  is explicit. Plugging it in \eqref{det} one finds 
\begin{equation}
\left(\frac{2}{3}-\rho_0\right)\left[a-\bar{a}(\rho_0)\right]<0.
\end{equation}
That is, in the suppressed IEPR region $a<\bar{a}$, the rate function $I_H$ is locally non-convex for all densities $\rho_0<2/3$, while for the enhanced region bounded by the critical value $\bar{a}<a<a_{1}$, it is locally non-convex for $\rho_0>2/3$.

In the region $a>a_{1}$ the rate function $I_H$ is only given in a parametric form. To find its second derivatives we use the following identities
\begin{eqnarray}
\left.\frac{\partial I_H}{\partial \rho_0}\right\vert_a&=&\left.\frac{\partial I_H}{\partial \rho_0}\right\vert_m+\left.\frac{\partial I_H}{\partial m}\right\vert_{\rho_0}\left.\frac{\partial m}{\partial \rho_0}\right\vert_a=\left.\frac{\partial I_H}{\partial \rho_0}\right\vert_m-\left.\frac{\partial I_H}{\partial m}\right\vert_{\rho_0}\frac{\left.\frac{\partial a}{\partial \rho_0}\right\vert_{m}}{\left.\frac{\partial a}{\partial m}\right\vert_{\rho_0}},\nonumber\\
\left.\frac{\partial I_H}{\partial a}\right\vert_{\rho_0}&=&\left.\frac{\partial I_H}{\partial m}\right\vert_{\rho_0}\left.\frac{\partial m}{\partial a}\right\vert_{\rho_0}=\left.\frac{\partial I_H}{\partial m}\right\vert_{\rho_0}\frac{1}{\left.\frac{\partial a}{\partial m}\right\vert_{\rho_0}}.\label{der}
\end{eqnarray}
Employing the same differentiation rules successively, one can compute the Hessian determinant in the region $a>a_{1}$:
\begin{equation}
\text{Det}\left[\text{Hess}(I_H)\right]= \chi(\rho,m) [m^2-\rho_0^2-\rho_0+1], \label{hes}
\end{equation}
where the function $\chi(\rho,m)>0$ (we do not write this function explicitly, for compactness). Thus, the determinant of the Hessian is negative for $ m^2-\rho_0^2-\rho_0+1<0$. Using the relation \eqref{mag} we find that this is the case for $a<a_{\text{non-convex}}$ with 
\begin{equation}
a_{\text{non-convex}}=(2\rho_0-1)\sqrt{(1-\rho_0)^2+\frac{2\sqrt{1-\rho_0}}{\text{Pe}^2}}.\label{noncon}
\end{equation}
Thus, beyond the critical value $a_{1}$, the rate function is locally non-convex in the region bounded by $a_{1}(\rho_0)<a<a_{\text{non-convex}}(\rho_0)$.
This concludes the mapping of the locally non-convex regions in $I_H$ that are marked in orange shading in Fig. \ref{nonzeromag} (B).

\subsection{Global non-convexity}\label{globconc}

The globally non-convex region is where $I_H$ is amenable to a two dimensional common tangent construction in the $(\rho_0,a)$ plane. This includes the SM region where tangent construction is degenerate and the convex hull is given by $I_H=0$ (see discussion at the end of Sec. \ref{sm}).
The tangent construction beyond the SM region is dominated by the point $(\rho_0=1,a=0)$ where the function $I_H$ has a pinch point singularity, see Fig. \ref{pinch}. As a consequence, the tangent construction is degenerate here as well since all tangents comprising the common tangent construction share this point as one of their ends \footnote{We could not provide an algebraic proof. We verified this by using the ConvexHullMesh command in Mathematica}. 
We are left with finding the location of the other end of all tangents which would comprise the boundary of the globally non-convex region, see Fig. \ref{tc}. Denote by $(\rho_h=1,a_h=0)$ and $(\rho_l,a_l)$ the high and low density end points of the tangents, then the tangent construction reads
\begin{equation}
\left[I_H+(1-\rho_l)\frac{\partial I_H}{\partial \rho_l}-a_l\frac{\partial I_H}{\partial a_l}\right]_{(\rho_l,a_l)}=0.\label{eqconv}
\end{equation}
For $(\rho_l,a_l)$ such that $a_l<a_{1}(\rho_l)$, the expression for $I_H$ is explicit \eqref{sto}. Substituting it into \eqref{eqconv} one finds that the boundary of the globally non-convex region is given by the vertical line $\rho_l=1/2$ for $\bar{a}\leq a\leq a_{1}(1/2)=(\sqrt{1+8/\text{Pe}^2})/8$. 

For $a>a_{1}(1/2)$ one must use the parametric representation for $I_H$ \eqref{sto}.
Plugging it into \eqref{eqconv} together with the substitution for the differentiation rules \eqref{der}, one arrives at the relation
\begin{equation}
m_l^2=\frac{\sqrt{2\rho_l-1}}{2}(1+\rho_l\sqrt{2\rho_l-1}).
\end{equation}
Plugging this back into \eqref{mag} we arrive at the curve \eqref{ac5} which marks the boundary of the globally non-convex region for $a>a_{1}(1/2)$. This concludes the mapping of the globally non-convex regions in $I_H$ that are marked in purple shading in Fig. \ref{nonzeromag} (B).

\section{Landau theory}\label{land}
We start with first identifying the critical bias $\Lambda_c$ beyond which a constant solution $\rho_0$ to the minimization problem \eqref{square} becomes linearly unstable. We then use this value to establish a Landau theory that captures the growth of the spatial excitation amplitude close to this critical value. 
\subsection{Linear stability of homogeneous solutions }\label{secondvar}
The linear stability of the constant solution $\rho_0$ is determined by examining the second variation $\delta \tilde{\psi}$ of the functional \eqref{square} with respect to small density variation $\rho=\rho_0+\delta\rho$
\begin{eqnarray}
-\delta{\psi}(\tilde\Lambda) = \int_0^1dx\left[M(\rho_0)(\partial_x\delta\rho)^2+\tilde{\Lambda}(2-3\rho_0)(\delta\rho )^2\right].\label{zeromag5}
\end{eqnarray}
The first mode to destabilize the quadratic form \eqref{zeromag5} is the principal one $\delta\rho\sim\cos(2\pi x)$. This happens as soon as:
\begin{equation}
\tilde{\Lambda}\operatorname{sign}(3\rho_0-2)>|\tilde{\Lambda}_c| \quad \text{where} \quad \tilde{\Lambda}_c=\frac{4\pi^2M(\rho_0)}{3\rho_0-2} .\label{gc2}
\end{equation}
To find the critical value of the IEPR at which this instability takes place, we consider the Legendre transform \eqref{leg} and recall that the state before the instability is homogeneous, so its SCGF is~\eqref{equ:psi-H}.  Hence, the critical value of $a$ corresponding to the critical bias $\tilde\Lambda_c$ is
\begin{equation}
a_{2}(\rho)=\left(1+\frac{2}{\text{Pe}^2}\Lambda_c\right)\bar{a}(\rho).\label{ac}
\end{equation}
Plugging \eqref{gc2} and using $\tilde\Lambda = \Lambda/\ell^2$, one arrives at \eqref{ac2}.

\subsection{Establishing the Landau theory close to $\Lambda_c$}
As we have seen in the previous Appendix \ref{secondvar}, instability of the homogeneous solution sets in via the principal mode excitation. To find the amplitude of this mode one has to go beyond the leading order expansion when close to the critical point. Such an expansion appeared in several previous works that studied similar second order transitions within the MFT framework \cite{lecomte_inactive_2012,baek_dynamical_2018,dolezal_large_2019,jack_hyperuniformity_2015}. 
The relevant small parameter is
\begin{equation}
\epsilon=\tilde{\Lambda}_c-\tilde{\Lambda} \; .
\end{equation}
The mode expansion up to the sub-leading order turns out to be
\begin{equation}
\rho=\rho_0+A\cos(2\pi x)+ B\cos(4\pi x)+\mathcal O(\epsilon^{3/2})\; ,\label{expan}
\end{equation}
where close to the critical point we have that the variational parameters scale as $A\sim\sqrt{\epsilon}$ and $B\sim \epsilon$ \cite{lecomte_inactive_2012,dolezal_large_2019}.
Plugging this expansion into  \eqref{square} and keeping terms up to order $\epsilon^2$ we find
\begin{equation}
\Psi\simeq\min_{B}\left[(16\pi^2M_0-\frac{
	\tilde{\Lambda}_c}{2}\bar{a}_0'')\frac{B^2}{2}+\epsilon \bar{a}_0'' \frac{A^2}{4}+(3\pi^2M_0'-\frac{\tilde{\Lambda}_c}{8}\bar{a}_0''')A^2B+(\frac{\pi^2}{4}M_0'')A^4\right].
\end{equation}
Here primes denote derivative with respect to the argument, and the subscript $_0$ denotes evaluating at $\rho_0$. That is, $M''_0=d^2M/d\rho^2\vert_{\rho_0}$. Plugging \eqref{gc2} we arrive at
\begin{eqnarray}\label{minb}
\Psi\simeq\min_{B}\left[\epsilon \bar{a}_0'' \frac{A^2}{4}+\frac{\mu_2}{2} B^2+\mu_3A^2B+\frac{\mu_4}{2}A^4\right],
\end{eqnarray}
with
\begin{equation}
\mu_2=12\pi^2M_0\quad;\quad \mu_3=3\pi^2M_0'-\frac{\pi^2M_0\bar{a}_0'''}{\bar{a}_0''}\quad;\quad\mu_4=\frac{\pi^2}{2}M_0''.
\end{equation}
Minimizing with respect to $B$ we find
\begin{equation}
B=-\frac{\mu_3}{\mu_2}A^2=-A^2\left[\frac{M_0'}{4M_0}-\frac{\bar{a}_0''' }{12\bar{a}_0''}\right].
\end{equation}
Plugging this relation back into \eqref{minb}, we get a Landau type of expansion with respect to the principal mode excitation amplitude $A$
\begin{equation}
\Psi\simeq\epsilon \bar{a}''_0 \frac{A^2}{4}+\frac{\beta(\rho_0,\text{Pe})}{2} A^4
\end{equation}
with
\begin{equation}
\beta=\mu_4-\frac{\mu_3^2}{\mu_2}=\frac{\pi^2}{2}M_0''-\frac{(3\pi^2M_0'\bar{a}_0''-\pi^2M_0\bar{a}_0''')^2}{12\pi^2(\bar{a}_0'')^2M_0}\label{theo}.
\end{equation}
For $\beta>0$, when crossing the critical point \eqref{gc2}, the amplitude $A$ grows as
\begin{equation}
A=\sqrt{\frac{-\bar{a}_0''\epsilon}{4\beta}}.\label{a}
\end{equation}
\subsection{Tricritical point}

As discussed in Sec.~\ref{bls}, points with $\beta=0$ are tricritical.
For any value of $\text{Pe}$, the function $\beta=\beta(\rho_0,\text{Pe})$ has at least two roots $\rho_{c1}(\text{Pe})<2/3$ and $\rho_{c2}(\text{Pe})>2/3$. As an example, for $\text{Pe=2}$ one finds $\rho_{c1}=0.600\dots$ and $\rho_{c2}=0.709\dots$, see Fig. \ref{alpha}. This signals the onset of a first order transition.

\begin{figure}[t]
	\includegraphics[scale=0.33]{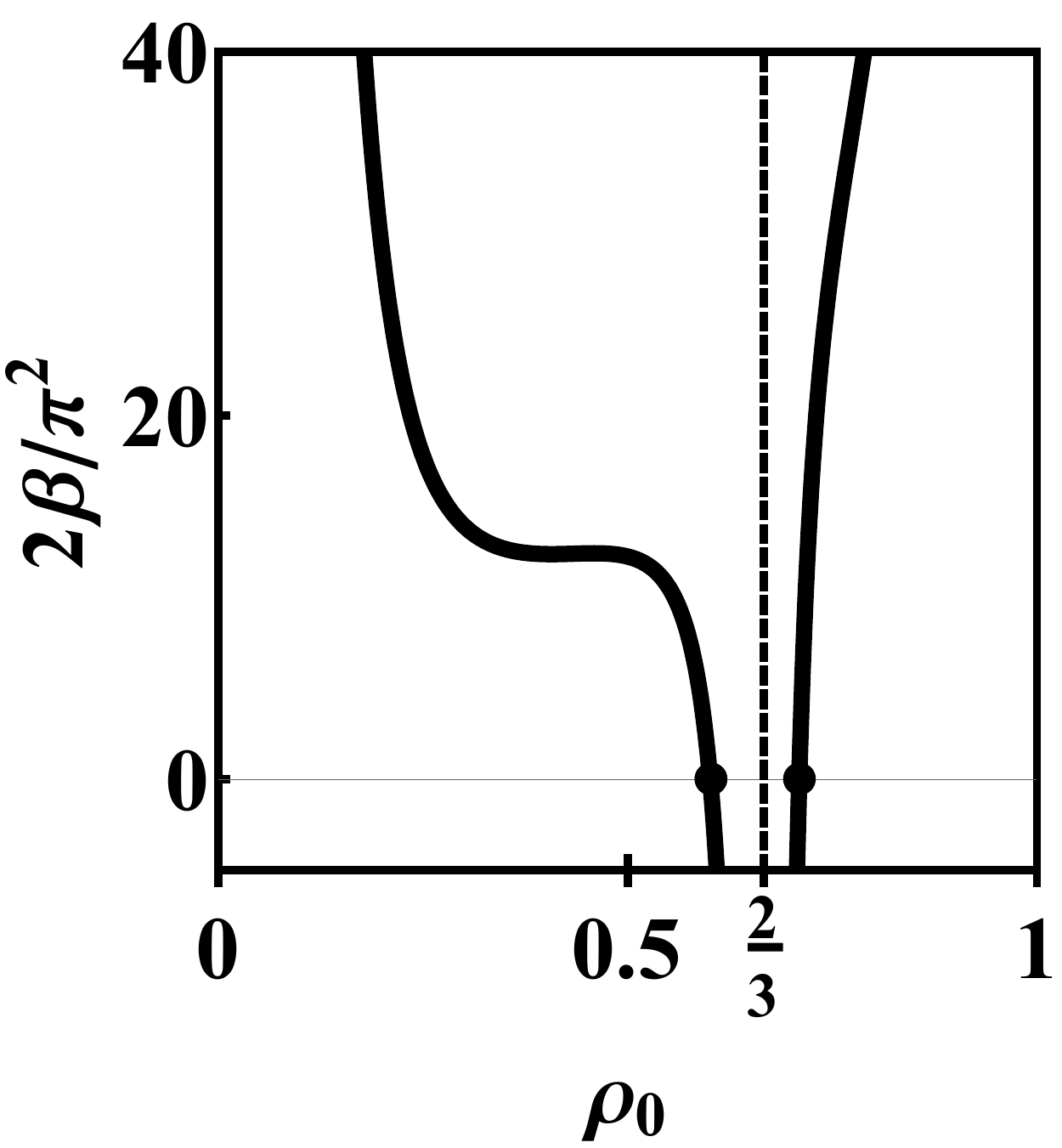}
	\caption{The function $\beta(\rho_0,\text{Pe}=2)$. The black points are its roots, marking the value at which a second order phase transition is overtaken by a first order one. The dashed vertical line at $\rho_0=2/3$ differentiates between regions where the homogeneous solution is linearly unstable for enhanced ($\rho_0>2/3$) IEPR fluctuations or suppressed ($\rho_0<2/3$) fluctuations.}
	\label{alpha}	
\end{figure}

This is verified by numerically solving the minimization problem for the CGF $\psi(\Lambda)$ \eqref{square}. As shown in Fig. \ref{delta}, once crossing the tricritical point at $\rho_c(\text{Pe})$, a first order phase transition overtakes the second order one at a value $|\tilde{\Lambda}_c^1|$ smaller than the prediction of the linear stability analysis $|\tilde{\Lambda}_c|$  \eqref{gc2}. Here the amplitude of the density modulation grows discontinuously. The magnitude of the amplitude discontinuity is increased when advancing along the first order transition line away from the tricritical point, see Fig. \ref{delta3} of the main text.

\begin{figure}[t]
	\includegraphics[scale=0.4]{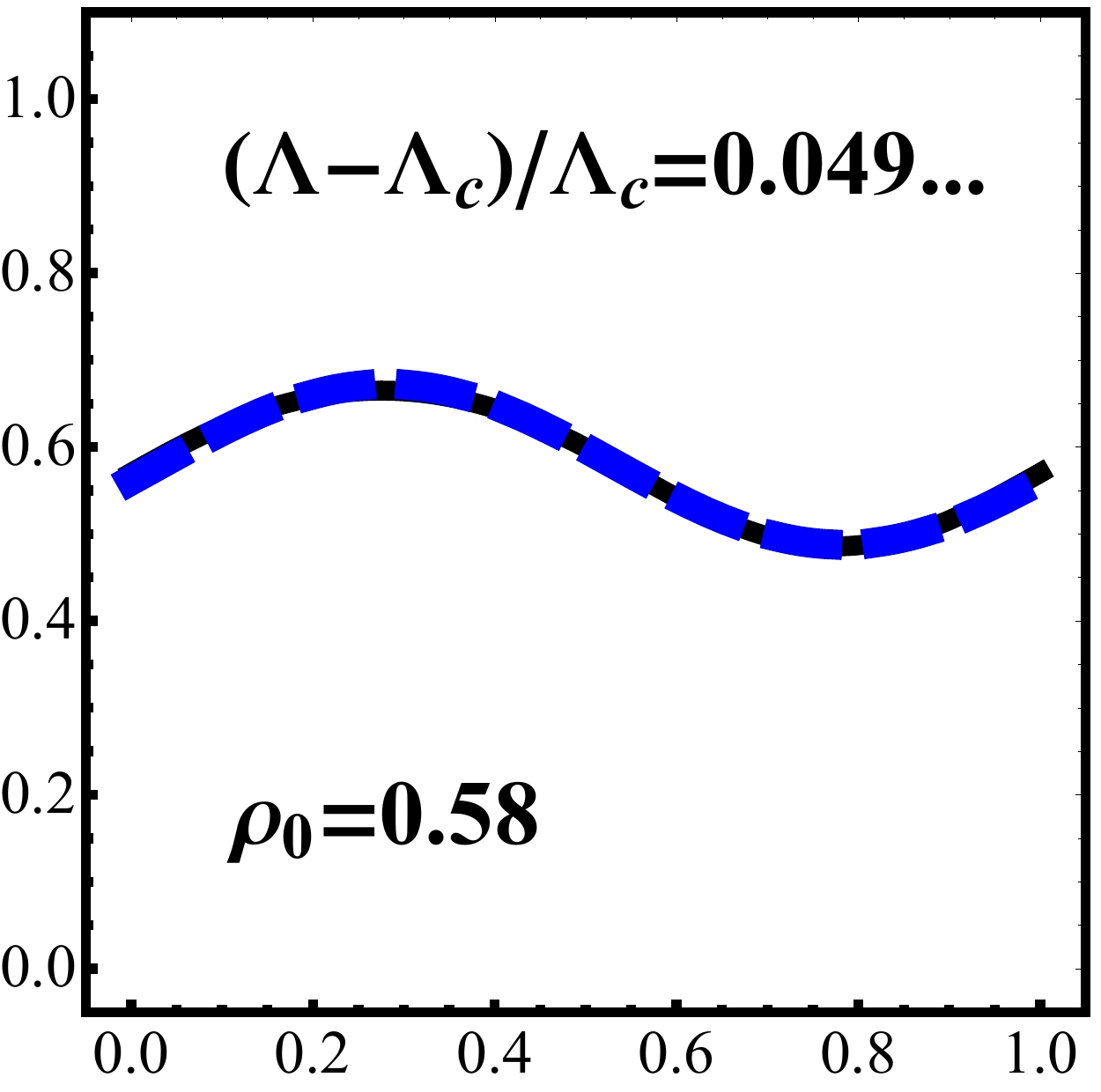}
	\includegraphics[scale=0.4]{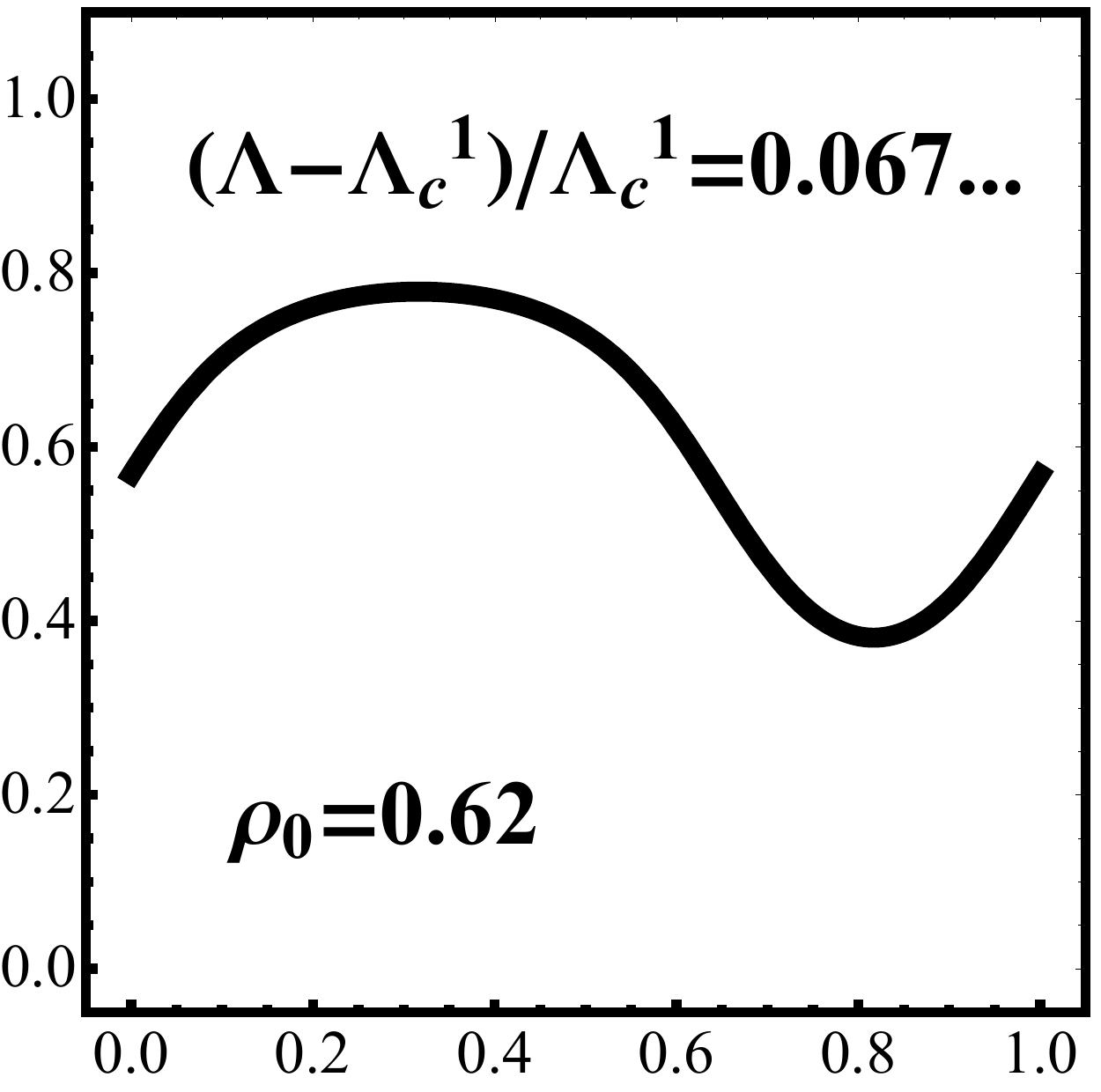}
	\caption{The optimal density profile $\rho(x)$ just after crossing the phase transition. The left panel at $\rho_0=0.58$ is below the tricritical point  $\rho_0<\rho_{c1}$. Here the phase transition is second order and corresponds to the theoretical prediction given by \eqref{lin} together with \eqref{a}. This prediction is plotted in dashed blue line (indistinguishable from the numerical black line). The right panel at $\rho_0=0.62$ is above the tricritical point  $\rho_0>\rho_{c1}$. Here the phase transition is first order, and it overtakes the  second order transition $|\tilde{\Lambda}_c^1|<|\tilde{\Lambda}_c|$. Here the amplitude of the density modulation grows discontinuously. In both panels $\text{Pe}=2$ and $\ell=0.02$.}
	\label{delta}	
\end{figure}

This behavior is consistent with the following model. Assuming the next order is accounted by a cubic term with a positive pre-factor $\gamma$
\begin{equation}
\Psi\simeq\epsilon a''(\rho_0) \frac{A^2}{4}+\frac{\beta(\rho_0,\text{Pe})}{2} A^4+\frac{\gamma}{3}A^6.
\end{equation}
Then at the tricritical point $\beta=0$ the amplitude grows with the modified power
\begin{equation}
A=\left(-\frac{\epsilon a''_0}{4\gamma}\right)^{1/4}.
\end{equation}
We found this to be consistent with our numerical computations.
Beyond the tricritical point where $\beta<0$ we have that a first order transition occurs before linear instability sets in at
\begin{equation}
\tilde{\Lambda}_c^1=\tilde{\Lambda}_c-\frac{3}{4}\frac{\beta^2}{4a''_0\gamma}
\end{equation}
where the amplitude has a jump given by
\begin{equation}
\Delta A=\sqrt{-\frac{3\beta}{4\gamma}}.
\end{equation}

\subsection{Miscibility gap at $\ell\to0$}\label{misc}
As argued in Sec.\ref{bls}, the miscibility gap for the first order transition is finite at vanishing $\ell$ as is clearly visible in the region of suppressed IEPR in Fig.~\ref{phasefig2} (A). The miscibility gap is determined in general by   $\psi_{\text{AP}}$ which is the CGF corresponding to time independent solutions for \eqref{equ:psi-min} (which are consistent with the additivity principle). In our study we solved for its leading order expansion $\psi(\Lambda)_{\text{AP}}=\ell^2\tilde{\psi}(\Lambda/\ell^2)+o(\ell^2)$.  Then denote by $a_{2}$ and $\tilde{a}_{c2}$ the two endpoints of the miscibility gap. That is, $a_{2}$ is the onset of the first order transition, marked by the orange dashed curve in Fig.~\ref{phasefig2} (A), and $\tilde{a}_{c2}$ is its other end, denoted by the magenta curve.  Then for suppressed IEPR deviations ($a<\bar{a}$), these values are given by
$a_{2}=\partial\psi_{\text{AP}}/\partial\Lambda|_{\Lambda_c^+}$ and $\tilde{a}_{c2}=\partial\psi_{\text{AP}}/\partial\Lambda|_{\Lambda_c^-}$ with $\Lambda_c$ the point of the kink singularity in $\psi_{\text{AP}}$  \cite{touchette_large_2009-1}. Then the scaling for $\psi_{\text{AP}}$ implies that these two values indeed have a finite separation at vanishing $\ell$. An equivalent statement holds for enhanced IEPR fluctuations with $\Lambda_c^+$ and $\Lambda_c^-$ interchanged between the two endpoints. Note however, that the ``slender'' shape of the SM region at enhanced IEPR fluctuations (see Fig. \ref{phasefig}) is such that this separation is bound to be small (yet finite).

\subsection{Tricritical point at MIPS criticality}\label{mipsver}
As discussed in Sec.\ref{mipsrel}, the MIPS criticality of the unbiased system is also a tricritical point that occurs in the biased ensemble when $\Lambda=0$. That is, two conditions hold. First, $\tilde{\Lambda}_c\left(\rho_0=3/4,\text{Pe}=4\right)=0$ so that the critical bias vanishes, and also $\beta\left(\rho_0=3/4,\text{Pe}=4\right)=0$ so that the MIPS criticality is also a tricritical point for the biased system. 

That these conditions are met can be read from the definition of $M(\rho)$ in \eqref{m}.  It is convenient to define
	\begin{equation}
	h(\rho_0,\text{Pe}) = 2+\text{Pe}^2\,(1-\rho)(1-2\rho)
	\end{equation} 
	such that $M \propto h^2$. From~\cite{kourbane-houssene_exact_2018-1}, 
	the spinodal instability leading to MIPS is
	$
	h(\rho_0,\text{Pe}) = 0
	$.  
	At the critical point one has additionally that $\partial h/\partial \rho_0=0$ (this is the point of minimal density on the spinodal).  Hence $M_0(\rho)\propto h^2=0$ at the MIPS critical point and similarly $M_0'(\rho) \propto h (\partial h/\partial \rho)=0$  and $M_0''(\rho) \propto (\partial h/\partial \rho)^2 + h (\partial^2 h/\partial \rho^2)=0$ there.  (Recall that $M_0$ depends implicitly on $\text{Pe}$ as well as on $\rho_0$.)  Since that $M_0$ and its first two derivatives are all vanishing at the MIPS critical point, one sees from \eqref{theo} that $\beta=0$ there too.

\end{widetext}
\bibliography{Entropy_production_active_lattice_gas}

\end{document}